\documentclass[floats,floatfix,showpacs,amssymb,prd,twocolumn,superscriptaddress,nofootinbib,reprint]{revtex4-2}

\usepackage{amssymb,amsmath,verbatim,mathtools,needspace,enumitem,etoolbox,graphicx,physics,microtype,afterpage,xspace,tabularx,lmodern,multirow}
\usepackage{siunitx}
\usepackage{gensymb}
\usepackage{ulem}  
\usepackage[dvipsnames, usenames]{xcolor}
\definecolor{linkcolor}{rgb}{0.0,0.3,0.5}
\usepackage[unicode, colorlinks=true, linkcolor=linkcolor, citecolor=linkcolor, filecolor=linkcolor, urlcolor=linkcolor, linktocpage, breaklinks]{hyperref}
\usepackage[all]{hypcap}
\usepackage[T1]{fontenc}
\usepackage[utf8]{inputenc}
\usepackage[usenames,dvipsnames]{xcolor}
\hypersetup{colorlinks=true,citecolor=romared,linkcolor=romared,urlcolor=romared}

\definecolor{romared}{RGB}{142,0,28}

\newcommand{\be}{\begin{equation}}
\newcommand{\ee}{\end{equation}}

\def\be{\begin{equation}}
\def\ee{\end{equation}}
\newcommand{\beq}{\begin{eqnarray}}
\newcommand{\eeq}{\end{eqnarray}}

\usepackage{aas_macros}
\usepackage{makecell}
\usepackage{soul}

\usepackage{lipsum}
\usepackage{orcidlink}

\newcommand{\hac}[1]{{\textcolor{black}{{{#1}}}}}

\newcolumntype{Y}{>{\centering\arraybackslash}X}

\begin{document}
\title{The First Billion Years in Seconds:\\ An Effective Model for the 21-cm Signal with Population III Stars}

\author{Hector Afonso G. Cruz \,\orcidlink{0000-0002-1775-3602}}
\email{hcruz2@jhu.edu}
\affiliation{William H. Miller III Department of Physics and Astronomy, Johns Hopkins University, 3400 N. Charles Street, Baltimore, Maryland, 21218, USA}

\author{Julian B. Mu\~noz\,\orcidlink{0000-0002-8984-0465}} 
\affiliation{Department of Astronomy, The University of Texas at Austin, 2515 Speedway, Stop C1400, Austin, Texas 78712, USA}

\author{Nashwan Sabti\,\orcidlink{0000-0002-7924-546X}}
\affiliation{William H. Miller III Department of Physics and Astronomy, Johns Hopkins University, 3400 N. Charles Street, Baltimore, Maryland, 21218, USA}

\author{Marc Kamionkowski \,\orcidlink{0000-0001-7018-2055}}
\affiliation{William H. Miller III Department of Physics and Astronomy, Johns Hopkins University, 3400 N. Charles Street, Baltimore, Maryland, 21218, USA}

\begin{abstract}
Observations of the 21-cm signal are opening a window to the cosmic-dawn epoch when the first stars formed. These observations are usually interpreted with semi-numerical or hydrodynamical simulations, which are often computationally intensive and inflexible to changes in cosmological or astrophysical effects. Here, we present an effective, fully analytic model for the impact of the first stars on the 21-cm signal, using the modular code {\tt Zeus21}. {\tt Zeus21} employs an analytic prescription of the star formation rate density (SFRD) to recover the fully nonlinear and nonlocal correlations of radiative fields that determine the 21-cm signal. We introduce the earliest Population III (Pop III) stars residing in low-mass molecular-cooling galaxies in {\tt Zeus21}, with distinct spectra from later Pop II stars. We also self-consistently model feedback in the form of $H_2$-dissociating Lyman-Werner (LW) radiation, as well as dark matter-baryon relative velocities, both of which suppress star formation in the lowest-mass halos. LW feedback produces a scale-dependence on the SFRD fluctuations, due to the long mean free path of LW photons. Relative velocities give rise to ``wiggles'' in the spatial distribution of the 21-cm signal; we present an improved calculation of the shape of these velocity-induced acoustic oscillations, showing they remain a standard ruler at cosmic dawn. Our improved version of {\tt Zeus21} predicts the 21-cm global signal and power spectra in agreement with simulations at the $\sim 10\%$ level, yet is at least three orders of magnitude faster. This public code represents a step towards efficient and flexible parameter inference at cosmic dawn, allowing us to predict the first billion years of the universe in mere seconds.
\end{abstract}

\date{\today}
\maketitle

\addtocontents{toc}{\protect\setcounter{tocdepth}{0}}
\tableofcontents

\begin{figure*}
    \centering
    \includegraphics[width=0.8\textwidth]{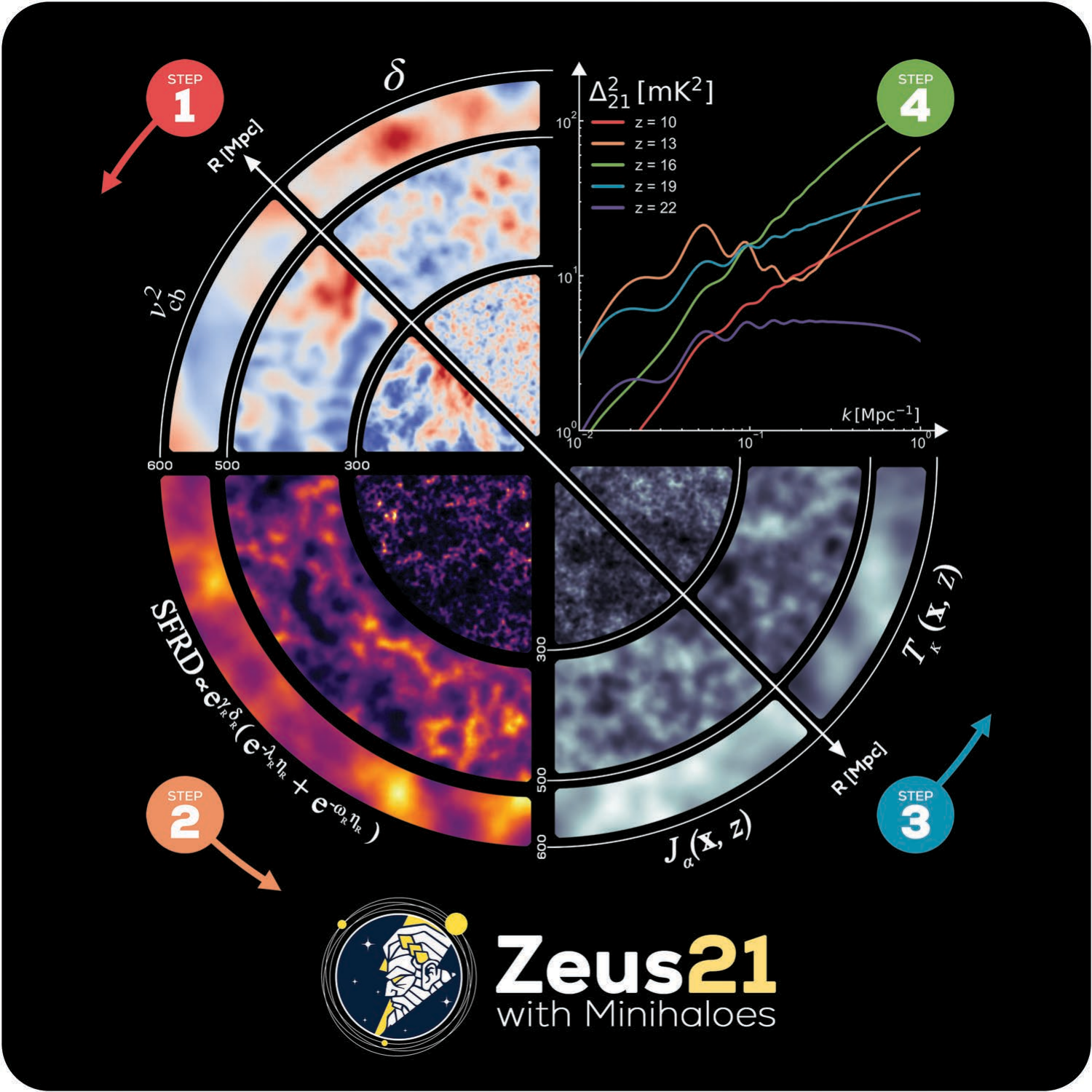}
    \caption{A schematic of how {\tt Zeus21} works, starting from the top left and moving counterclockwise. Firstly, {\tt Zeus21} calculates the behavior of the SFRD in a region of radius $R$ with overdensity $\delta_R$ and relative velocity $v_{\mathrm{cb},R}$. Secondly, we fit for the effective biases $\gamma_R$, $\lambda_R$, and $\omega_R$, which encode the lognormal and log-$\chi^2$ behavior of the SFRD against density and velocity. Thirdly, we compute the Lyman-$\alpha$ background intensity $J_\alpha$ and kinetic temperature $T_k$ of the IGM (with X-ray heating and adiabatic cooling components) as integrals of the SFRD weighted by coefficients that depend on astrophysical parameters (i.e., SEDs). Finally, these fluxes are used to compute the 21-cm global signal and fluctuations (through the power spectrum $\Delta_{21}^2$). In Step 1, the $\delta$ realization was computed assuming the standard linear matter power spectrum $P(k)$ as an input, while the $v_\mathrm{cb}^2$ realizations were created as the sum of squares of realizations of its three spatial components, each of whose input power spectrum is $P_{v_\mathrm{cb}}(k)$ as expressed in Eq.~\eqref{eq:Pvcb}. The $\mathrm{SFRD}(\delta_R, v_\mathrm{cb}^2)$ box in step 2 follows our effective model, which is then summed using astrophysical weights according to Eqs.~\eqref{eq:discreteLYA} and \eqref{eq:discreteXRAY} to create the flux realizations in step 3. As {\tt Zeus21} does not require simulation boxes, the lightcones shown above are for purely visualization purposes, and the 21-cm global signal and power spectrum in step 4 are computed analytically in seconds.
    }
  \label{fig:FIG1schematic}
\end{figure*}

\section{Introduction}

The nature of the first sources of light in the universe remains a mystery. Their emergence defines an important turning point in cosmic evolution, marking the end of the dark ages and the beginning of galaxy formation during cosmic dawn \citep{barkana01, mesinger07, bromm11, mason23}. It is thought that the pristine primordial gas within the earliest galaxies gave rise to a generation of so-called Population (Pop) III  stars with ultra-low metallicities \citep{klessen23}, which were spectrally distinct from their later Pop II counterparts \citep{stahler86a, stahler86b, schaerer02, omukai03, trenti09, maeder12, kippenhahn12, berzin21, murphy21}. While the impact of Pop II stars on galaxy evolution is relatively well understood \citep{wechsler18}, high-redshift Pop III stars have so far eluded direct detection \citep{katz19,  larkin23, trussler23}. Thus, there remain many open questions on the details of Pop III chemical enrichment \citep{trenti09, welsh19, skuladottir24}, feedback processes\citep{mebane18, liu20b, sibony22}, and initial mass functions \citep{nakamura01, clark11, parsons22, latif22}. 

We expect Pop III stars to form in small molecular-cooling galaxies (MCGs) at early times (redshifts $z\sim20-30$), hosted by dark matter (DM) halos of masses $10^5 \lesssim M_h/M_\odot \lesssim 10^7$ \citep{tegmark97, abel02, bromm04, haiman06, trenti10}. Due to their shallow potential wells, these first galaxies are highly susceptible to feedback on their star formation. In addition to the usual supernova and photo-heating feedback processes that are active at later times~\citep{draine96, barkana99, wise08, sobacchi13}, early MCGs suffer feedback from two additional sources. Firstly, Lyman-Werner (LW) radiation from the first stars can dissociate molecular hydrogen and impede further star formation in MCGs~\citep{machacek01, johnson07, oshea07, oshea08, safranek12, johnson13, visbal14, schauer17, skinner20}. Secondly, the relative (or streaming) velocities between CDM and baryons hampers gas accretion and cooling in the least massive DM halos~\citep{tseliakhovich10, greif11, naoz13, oleary12, hirano18, schauer19,schauer21, kulkarni21}. Measuring the star formation rate density (SFRD) at high redshifts holds the key to understanding the complex astrophysics that govern the nature of the first stars, as well as the underlying cosmology which seeds the initial conditions for early galaxy formation.

Perhaps the most promising way to understand the early galaxy population is through the 21-cm line of neutral hydrogen.  The 21-cm background is sensitive to the aggregate impact of all luminous sources on the intergalactic medium (IGM) \citep{furlanetto06b}, which at early times are dominated by faint galaxies hosted by small-mass DM halos. Several observatories are on the verge of a measurement of the 21-cm power spectrum~\citep{vanhaarlem13, mellema13, voytek14, deboer17, bowman18, beardsley16, greig21, mertens20, yoshiura21}, which has been proposed as a powerful tool to constrain the ionizing and heating properties of the first galaxies \citep{loeb04, santos08,parsons14,yajima14,kern17,cook24, ghara24}, the growth of large scale structure \citep{scott90, furlanetto06, cooray06, wyithe08, lidz09, villaescusa18}, and as a tool to probe fundamental physics \citep{pritchard12, barkana05,munoz20,munoz18b,berlin18, pospelov18,liu18, flitter22, qin24, flitter24, cruz24}. Yet, extracting this information is challenging owing to the complex astrophysics operating during cosmic dawn \citep{mason23b, munoz24}. The 21-cm line is a nonlinear and nonlocal tracer of galaxy formation, which traditionally has been modeled with simulations ranging from hydrodynamical~\citep{gnedin14, mutch16, ocvirk20, kannan22}, to N-body~\citep{holzbauer12, schneider21, schneider23}, or semi-analytic \citep{santos10, visbal12, fialkov13, ghara15, battaglia13, mesinger07, thomas09} (i.e., without individual galaxies) including the well-known {\tt 21cmFAST}~\citep{mesinger11, murray20}. However, simulating cosmic dawn is computationally expensive; accurate models of cosmic dawn necessitate large volumes, owing to the large mean-free paths of photons, yet fine resolutions to resolve galaxies within the first, small-mass halos. 

Recently, Ref.~\cite{munoz23} introduced a fully analytic model for the 21-cm signal (based on early works in Refs.~\cite{barkana05, pritchard06}) that can successfully capture the nonlinearity and nonlocality of cosmic dawn.
Here, we extend this model to self-consistently include the earliest Pop III stars with all their feedback effects. Our main modeling update is tracking the Pop III star formation rate density (SFRD) separately, including its behavior against density in the presence of feedback, as well as against relative velocity. We find that the SFRD is very well described as a lognormal against density, and a log-$\chi^2$ against relative velocity, for which correlation functions can be computed analytically \citep{coles91, xavier16, givans23} without having to simulate quantities on a grid.  This allows us to  predict the evolution of the cosmic 21-cm signal during {\it the first billion years in seconds}.

We implement our effective model for Pop III stars in the public code {\tt Zeus21}.\footnote{\url{https://github.com/JulianBMunoz/Zeus21}} The updated version of {\tt Zeus21} robustly and flexibly parameterizes Pop III astrophysics as user-changeable inputs, which enables fast parameter inference from the upcoming 21-cm data. We implement a novel treatment of inhomogeneous LW feedback. Overdense regions will form more galaxies, but also produce more LW photons and thus feedback. We find that this inhomogeneous LW feedback effect can be well approximated as a linear correction factor to the effective density biases $\gamma_R$, allowing us to entirely bypass simulation grids. Moreover, we find that relative velocities, which modulate the early star formation in MCGs housed in minihalos, do not affect the SFRD linearly in $v_{\rm cb}^2$, as previously assumed~\cite{dalal10,munoz19,visbal12}. Our full nonlinear treatment of the SFRD behavior against $v_{\rm cb}^2$ more precisely captures the shape of the velocity-induced acoustic oscillations (VAOs), showing an $\mathcal O(1)$ change on the amplitude and tilt of the VAO power, though no change on its acoustic structure as expected of a standard ruler~\cite{munoz19}. 

This paper is structured as follows. We begin with an overview of the 21-cm signal and star formation in Sec.~\ref{sec:II_21cm}, and elaborate upon the contributions of different stellar populations to the SFRD in Sec.~\ref{sec:III_Populations}. In Sec.~\ref{sec:IV_effectivemodel}, we introduce our effective model for cosmic dawn in terms of SFRD building blocks. With reconstructions of the averages and anisotropies of IGM quantities in Sec.~\ref{sec:V_IGM}, we construct the 21-cm signal and power spectrum in Sec.~\ref{sec:VI_21cmSignal} and compare our results against {\tt 21cmFAST}. Finally, we conclude with further discussion in Sec.~\ref{sec:VII_conclusions}. We assume a Planck 2018 cosmology \citep{planck18} (rightmost column, TT, TE, EE + low P + lensing + BAO) throughout this paper.

\section{The 21-cm Signal as a Tracer of Star Formation} \label{sec:II_21cm}

Let us begin by reviewing the basics of the 21-cm line and how it traces star formation of the first galaxies. We encourage the reader to consult the comprehensive reviews in Refs.~\cite{furlanetto06,pritchard12,liu20} for further details.

\subsection{The 21-cm Signal}

Neutral hydrogen along the line of sight to the CMB can resonantly absorb or emit (rest-frame) 21-cm photons, producing a suppression or enhancement of the apparent CMB intensity at low frequencies.
We quantify this effect through the 21-cm brightness temperature~\citep{barkana01},
    \begin{equation}
    T_{21}=\frac{T_s-T_{\mathrm{CMB}}}{1+z}\left(1-e^{-\tau_{21}}\right),
    \end{equation}
defined in terms of the temperature $T_\mathrm{CMB}$ of the radio background (in our case of the CMB with $T_\mathrm{CMB} = 2.7255(1+z)$ K).
Here, the 21-cm optical depth is expressed as
    \begin{equation}
        \tau_{21}=(1+\delta) x_{\mathrm{HI}} \frac{T_0}{T_s} \frac{H(z)}{\partial_r v_r}(1+z)\ ,
    \end{equation}
in terms of the neutral-hydrogen fraction $x_\mathrm{HI} \equiv n_\mathrm{HI} / n_\mathrm{H}$, the hydrogen density $\delta$, the Hubble parameter $H(z)$, the line-of-sight velocity gradient $\partial_r v_r$, and a normalization factor $T_0$ that depends on the cosmic baryon and total matter densities
    \begin{equation}
    T_0=34 \, \mathrm{mK} \,\left(\frac{1+z}{16}\right)^{1 / 2}\left(\frac{\Omega_\mathrm{b} h^2}{0.022}\right)\left(\frac{\Omega_\mathrm{m} h^2}{0.14}\right)^{-1 / 2}.
    \end{equation}
Finally, $T_s$ is the (non-thermodynamic) spin temperature of neutral hydrogen in the IGM. This temperature quantifies the relative occupation fraction of electrons in either hyperfine state, $n_1 / n_0 = g_1 / g_0 \exp (-hc / \lambda_{21} k_\mathrm{B} T_s)$, and can be found through a harmonic mean of $T_\mathrm{CMB}$, the color temperature $T_c$ \citep{hirata06}, and the kinetic temperature $T_k$,
    \begin{equation}
    T_s^{-1}=\frac{x_{\mathrm{CMB}} T_{\mathrm{CMB}}^{-1}+x_c T_k^{-1}+x_\alpha T_c^{-1}}{x_{\mathrm{CMB}}+x_c+x_\alpha},
    \end{equation}
weighted by the coefficients from Lyman-$\alpha$ (Ly$\alpha$) coupling through the Wouthuysen-Field effect $x_\alpha$, collisional coupling $x_c$, and radiation coupling $x_\mathrm{CMB} \equiv \left( 1-e^{-\tau_{21}}\right) / \tau_{21}$.

The 21-cm signal is then determined by the different processes that bring the spin temperature out of equilibrium with the CMB.
During the dark ages $(200 \gtrsim z \gtrsim 30)$, collisions between hydrogen atoms coupled the spin and kinetic temperatures, which are lower than $T_\mathrm{CMB}$, giving rise to an absorption signal \citep{loeb04}. 
Here, we will focus on lower redshifts, $z\lesssim 30$, where cosmic expansion renders collisions inefficient and brings $T_s$ close to $T_{\rm CMB}$, making $T_{21}$ vanish.
In that regime it is the birth of the first stars at cosmic dawn $(30 \gtrsim z \gtrsim 20)$ that gives rise to a nonzero 21-cm signal.
Stars produce a background of UV photons that can redshift into the Ly$\alpha$ transition, enabling the Wouthuysen-Field (WF, $x_{\alpha}>0$) effect \citep{wouthuysen52, field59,hirata06} to couple the spin and kinetic temperatures and give rise to absorption ($T_{21}<0$). 
Later on, X-ray photons from the first galaxies \citep{ciardi10, xu14, ewall16, sazonov17} (as well as other sources \cite{chen04, chuzhoy07}) will raise the kinetic temperature, resulting in 21-cm emission. \hac{Finally, growing reionization bubbles will overlap until the only remaining neutral hydrogen is confined to rare isolated islands, resulting in an asymptotically-decaying 21-cm signal \citep{benson06, furlanetto06b}}.

Following Ref.~\cite{munoz23}, we make a few simplifying approximations. We ignore collisional coupling ($x_c=0$), assume linear-order redshift-space distortions $\delta_v$ \citep{barkana06, mao12}, and operate in the limit of small 21-cm optical depth $\tau_{21} \ll 1$. This allows us to write
    \begin{equation} \label{eq:separableT21}
    T_{21}=T_0(z)\left(1+\delta-\delta_v\right) x_{\mathrm{HI}} \left(\frac{x_\alpha}{1+x_\alpha}\right)\left(1-\frac{T_{\mathrm{CMB}}}{T_c}\right),
    \end{equation}
which separates the cosmological contributions (sourced from the local density and bulk velocity) from the astrophysical ones (i.e., from reionization, WF coupling, and the IGM temperature).
The latter two terms in Eq.~\eqref{eq:separableT21} trace the first star formation through the Ly$\alpha$ and X-ray backgrounds through $x_\alpha$ and $T_c$, whose specific intensities are schematically given by integrals over the past lightcone of the star formation rate densities (SFRD, $\dot{\rho}$),
    \begin{equation} \label{eq:sfrdDependentFields}
        J^{(i)}_{\alpha, X}(z, R) \propto \int  c_{\alpha, X}(z, R) \dot{\rho}^{(i)}_*(z, R) \, d R, \quad i \in (\mathrm{II, III}),
    \end{equation}
of two populations of stars $i=\{\rm II, III\}$, where $R$ is the comoving radius defining the sphere over which the SFRD is averaged, and $c_{\alpha,X}(z, R)$ is a coefficient that depends on band-specific photon emissivity and propagation. 

The astrophysical components of the 21-cm brightness temperature are, thus, nonlocal tracers of the SFRD, which itself is a nonlinear function of the matter density $\delta$ and velocity $\delta_v$. This presents a technical challenge to linear approaches~\cite{barkana05, pritchard06}, but can be implemented in semi-numerical simulations~\cite{mesinger11,santos10,visbal12}. We, instead, build onto the fully nonlinear and nonlocal analytic model of Ref.~\cite{munoz23}, implemented in the public code {\tt Zeus21}. We treat the SFRD of different populations of stars as building blocks, and compute their fluctuations in density through a log-normal model, and those in velocities as a log-$\chi^2$. We show a diagram of our model in Fig.~\ref{fig:FIG1schematic}, which displays how the SFRD encodes the statistics of the matter density and velocity fields whose differential impact on Pop II and III stars can be quantified without generating a realization. Let us now describe here how we model the SFRD building blocks in detail.

\subsection{An Effective Model for the Average SFRD}

The (comoving) density at which a given population of stars $i \in (\mathrm{II, III})$ forms is given by \citep{madau96, barkana04, mirocha14, park19}
    \begin{equation} \label{eq:SFRD}
        \dot{\rho}_*^{(i)}(z) = \int \frac{dn}{dM_h}\!(M_h, z) \dot{M}_*^{(i)} dM_h,
    \end{equation}
which accounts for all halos on the halo mass function (HMF, $dn/dM_h$) weighed by their average star-formation rate (SFR)
    \begin{equation}
        \dot{M}^{(i)}_*(M_h, z) = \dot{M}_h(M_h, z) \frac{\Omega_b}{\Omega_m} f^{(i)}_{*}(M_h) f^{(i)}_\mathrm{duty}(M_h,z).
    \end{equation}
This SFR is proportional to the halo mass-accretion rate $\dot{M}_h$, with $f_*^{(i)}$ and $f^{(i)}_\mathrm{duty}$ as the (population- and mass-dependent) star formation efficiencies (SFE) and duty fractions, respectively.
We will first elaborate on the HMF and mass-accretion rate, which are sensitive only to cosmology, and return to the population-dependent SFEs and duty cycles in the next section.

We adopt the Sheth-Tormen HMF \citep{sheth01}, 
    \begin{equation} \label{eq:avgHMF}
    \frac{d n}{d M_h}=-A_{\mathrm{ST}} \frac{\rho_m}{M_h \sigma} \frac{d \sigma}{d M_h} \nu\left(1+\nu^{-2 p_{\mathrm{ST}}}\right) e^{-\nu^2 / 2},
    \end{equation}
where $\sigma^2(M_h,z)$ is the redshift-dependent variance of matter fluctuations on a scale of mass $M_h$, $\nu \equiv \sqrt{q_\mathrm{ST}} \delta_\mathrm{crit} / \sigma$ is a rescaled dimensionless peak-height of perturbations, and $\delta_\mathrm{crit} \equiv 3(12\pi)^{2/3}/20 \simeq1.686$. Here, the normalization is fixed to be $A_\mathrm{ST} = 0.32218 \sqrt{2/\pi}$, and the parameters $q_\mathrm{ST} = 0.707$ and $p_\mathrm{ST} = 0.3$ are chosen to correct the abundance of small-mass halos and fit high-redshift simulations \cite{iliev12, schneider18,schneider21}.

The mass accretion rate at high redshifts can be modeled as a simple exponential of the form $M_h(z) \propto e^{\alpha_\mathrm{acc} z}$ with a parameter $\alpha_\mathrm{acc} = 0.79$ shown to provide a good fit to simulations \citep{schneider21}. Additionally, the extended Press-Schechter (EPS) formalism provides an alternative accretion rate \cite{neistein06}, derived from an empirical fit to halo merger trees \citep{fakhouri10, trac15}. It can be shown that the EPS model, as well as other prescriptions of $\dot M_h$, can be approximately recast as different mass-dependent values of $\alpha_\mathrm{acc}$, and that these differences can be reabsorbed into the low- and high-mass slope of $f_{\star}$ vs $M_h$ (see Appendix A of \citep{munoz23b} for more details.) We proceed with the exponential accretion for our fiducial model.

The remaining two components are the SFE ($f_{\star}$) and duty cycle ($f_{\rm duty}$), which characterize how efficiently stars form in halos of mass $M_h$ given the different feedback and cooling processes.
Let  us now show how we model Pop II and III star formation at high redshifts.

\begin{figure}
    \centering
    \includegraphics[width=\columnwidth]{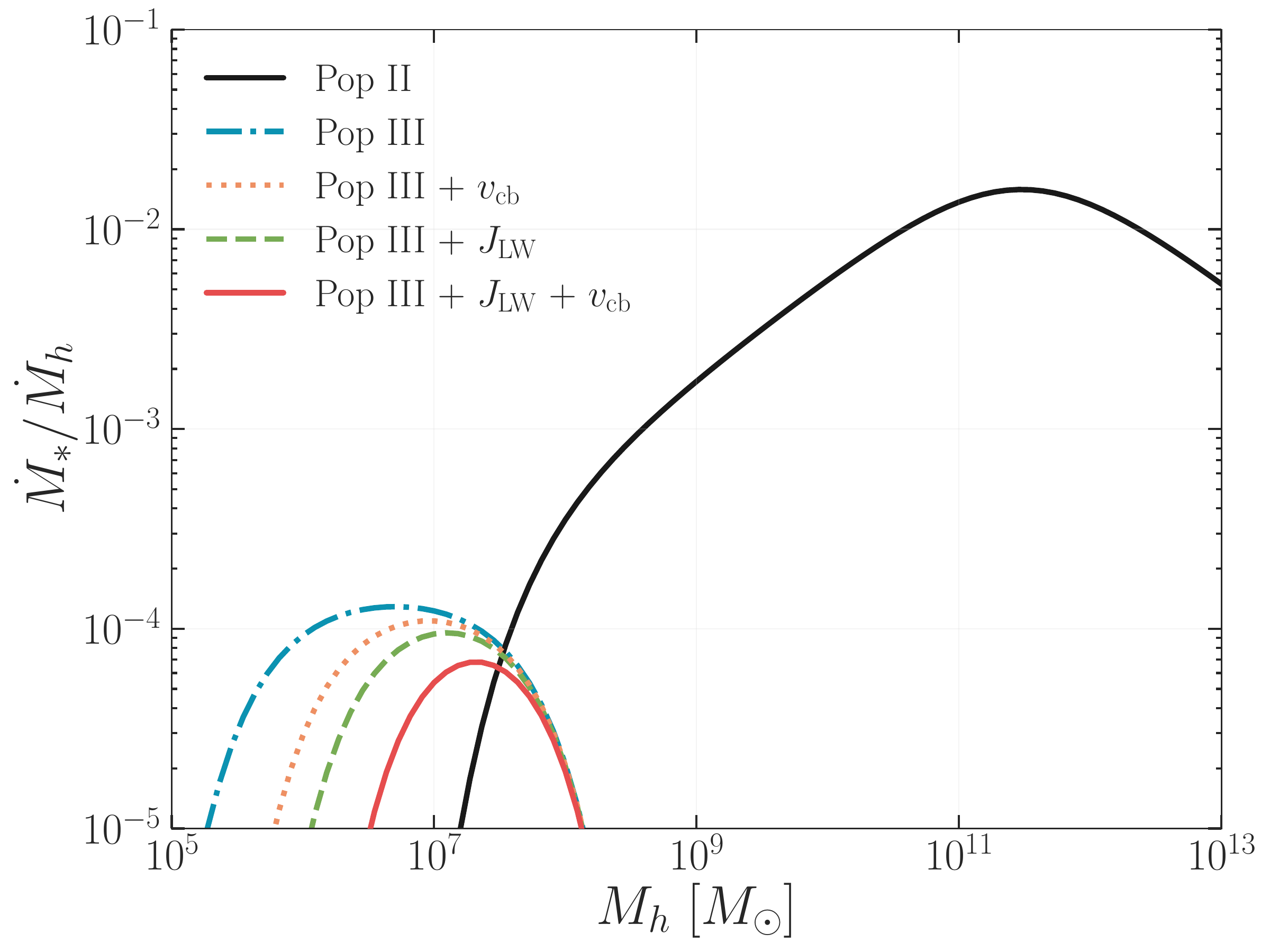}
    \caption{
    A plot of the $z = 15$ star-formation efficiencies $\dot{M}_* / \dot{M}_h = f_b f_* f_\mathrm{duty}$ for galaxies hosting Pop II stars (black) and Pop III stars (colors for different feedback assumptions) as a function of their halo mass $M_h$. The blue dot-dashed line shows the SFE for Pop III stars in the absence of feedback, whereas orange and green  show the result when including either relative-velocity or Lyman-Werner feedback, both of which suppress star formation at small $M_h$.
    The solid red line shows the Pop III SFE with both sources of feedback included.}
  \label{fig:FIG2fstar_vs_m}
\end{figure}

\section{Modeling Successive Populations of Stars} \label{sec:III_Populations}

The first (Pop III) stars are hypothesized to form around $z\sim 30$ from nearly metal-free hydrogen and helium gas.
Over time, the interstellar gas inside galaxies will self enrich, transitioning to form the more common and better-understood Pop II stars. 
As structure formation proceeds hierarchically, and stellar material is recycled, we follow Refs. \citep{xu16, mebane18} and model Pop III stars as residing in small molecular-cooling galaxies (MCGs, with halo masses between $M_h \approx 10^6 - 10^8 \, M_\odot$ ), whereas the subsequent Pop II stars form in atomic-cooling galaxies (ACGs, with $M_h \gtrsim 10^8 \, M_\odot$ \citep{oh02, tegmark09}).
This toy model allows us to keep track of two distinct populations of stars\footnote{Note that we do not need Pop III stars to be strictly metal free, but to have different SEDs on average than the later Pop II stars.} without anisotropically evolving the metallicity of the ISM and IGM in detail.

The onset of Pop III star formation is expected to kickstart the epoch of Ly$\alpha$ coupling at $z\sim 20$, and thus to produce 21-cm absorption.
Nevertheless, Pop II stars are likely the ones driving reionization at $z\sim 5-10$.
As such, it is imperative to robustly and flexibly quantify the transition between both populations of stars \citep{schneider02}. In this section, we review the formalism for this transition, following Refs.~\cite{park19,qin21,munoz22,tacchella18,mason15,sabti22}, and demonstrate our usage of population-averaged quantities in computations of stellar properties and feedback processes. The reader familiar with the details of Pop II and III modeling in 21-cm calculations may want to glance at Table~\ref{tab:tableParams} for the full list of parameters and skip to Sec.~\ref{sec:IV_effectivemodel}.

\newcolumntype{C}{>{$}c<{$}}
\begin{table*}
\caption{\label{tab:tableParams} A table of all astrophysical parameters used in this work. The left two columns represent the parameters used in our fiducial {\tt Zeus21} model, while the right two columns are the parameters used in our comparisons with {\tt 21cmFAST}.}
\begin{ruledtabular}
\begin{tabular}{C|CCCCC}
 &\multicolumn{2}{c}{{\tt Zeus21} (fiducial)}&\multicolumn{2}{c}{{\tt 21cmFAST} (code comparison)}\\
  & \textrm{Pop II} & \textrm{Pop III} & \textrm{Pop II} & \textrm{Pop III} \\ 
  \hline\hline \\
 \alpha_* & 0.5 & 0.0 & 0.5 & 0.0 \\
 \beta_* & -0.5 & 0 & -0.5 & 0 \\
 \epsilon_* & 10^{-1} & 10^{-3} & 10^{-1.25} & 10^{-1.25} \\
 M_p & 3\times 10^{11} \ M_\odot & 10^7 \ M_\odot & 10^{10} \ M_\odot & 10^7 \ M_\odot \\
 f_{\mathrm{esc},0} & 10^{-1}  & 10^{-1.35} & 0.0 & 0.0 \\
 M_\mathrm{esc} & 10^{10} \, M_\odot &  10^7 \, M_\odot &  10^{10}\, M_\odot &  10^7 \, M_\odot \\
 \alpha_\mathrm{esc} & 0.0 & -0.3 & -0.3 & -0.3 \\
 L_{40} & 10^{0.5} & 10^{0.5} & 10^{0.5} & 10^{0.5} \\
 E_{0,X} & 500 \, \mathrm{eV} & 500 \, \mathrm{eV} & 500 \, \mathrm{eV} & 500 \, \mathrm{eV} \\
 E_{\mathrm{max},X} & 2000 \, \mathrm{eV} & 2000 \, \mathrm{eV} & 2000 \, \mathrm{eV} & 2000 \, \mathrm{eV} \\
 \alpha_X & -1.0 & -1.0 & -1.0 & -1.0 \\
 N_\alpha & 9690 & 17900 & 11825 & 5080 \\
 N_\mathrm{LW} & 6200 & 12900 & 3030 & 560 \\
 A_\mathrm{LW} & 2.0 & 2.0 & 2.0 & 2.0 \\
 \beta_\mathrm{LW} & 0.6 & 0.6 & 0.6 & 0.6 \\
 A_{v_\mathrm{cb}} & 1.0 & 1.0 & 1.0 & 1.0 \\
 \beta_{v_\mathrm{cb}} & 1.8  & 1.8  & 1.8  & 1.8 \\
\end{tabular}
\end{ruledtabular}
\end{table*}

\subsection{Population II Stars}

We begin with the (simpler) Pop II stars. 
We assume that halos can host Pop II star-forming galaxies if their mass is above a turnover threshold $M_\mathrm{turn}^{\rm II}(z) = M_\mathrm{atom}(z) = 3.3\times 10^7 \, M_\odot [(1+z)/21]^{-3/2}$,  corresponding to a virial temperature of $T_\mathrm{vir} \sim 10^4 \, \si{K}$ \citep{oh02}, which we parameterize through the duty-cycle fraction,
\begin{equation}
    f^\mathrm{\rm II}_{\text {duty }}=\exp \left[-M^\mathrm{II}_{\mathrm{atom}}(z)/ M_h\right].
\end{equation}
Throughout this work we will focus on $z\gtrsim 10$, and thus neglect feedback from the photo-evaporation of halos during reionization, which at later times halts star formation depending on the local ionizing background \citep{thoul96, noh14, sobacchi14}. 

Other feedback processes, including radiative and supernova~\citep{wyithe13, dayal14, yung19}, as well as AGN~\citep{furlanetto17, mirocha20}, will shape the star formation efficiency $\dot{M}_* / \dot{M}_h$ even if they do not produce a strict turnover~\citep{mason15, sun16, mirocha17}.
Following Ref.~\cite{sabti21} we capture these effects by modeling the SFR of both star populations $i=\{\rm II,III\}$ as
    \begin{equation} \label{eq:SFE}
    f^{(i)}_*(M_h)=\frac{2 \epsilon^{(i)}_*}{\left(M_h / M^{(i)}_p\right)^{\alpha^{(i)}_*}+\left(M_h / M^{(i)}_p\right)^{\beta^{(i)}_*}},
    \end{equation}
designed to peak at a halo mass $M^{(i)}_p$ with an amplitude $\epsilon^{(i)}_*$, and to decrease towards the fainter end with a power-law index $\alpha_*^{(i)}$ and towards the bright end with a different $\beta_*^{(i)}$. 

The same SFE that we use for 21-cm ought to fit the UV luminosity function (UVLF) data from the Hubble and James Webb Space Telescopes~\cite{park19,mirocha20}.
We therefore adopt values consistent to the MCMC search in Ref.~\cite{munoz23b}, with $\epsilon_*^\mathrm{II} = 10^{-1}$, $\alpha_*^\mathrm{II} = -\beta_*^\mathrm{II} = 0.5$, and $M_\mathrm{pivot}^\mathrm{II} = 3 \times 10^{11} \, M_\odot$, summarized in Table~\ref{tab:tableParams}. 
This parameter set could be enhanced to include stochasticity as well as time dependence~\cite{behroozi19,mirocha21}, which we will examine in future work.

We show in Fig.~\ref{fig:FIG2fstar_vs_m} the Pop II SFE at $z=15$ as a function of halo mass $M_h$.
The amplitude $\epsilon_*^\mathrm{II}$ and peak mass $M_\mathrm{pivot}^\mathrm{II}$ determine the height of the curve, whereas the
faint-end slope $\alpha_*^\mathrm{II}$ contributes heavily to setting the SFRD. The SFE drops sharply below $M_{\rm atom}$ due to the exponential duty cycle. 
Even though the bigger halos $M_h\sim 10^{12}$ form stars more efficiently, at  $z \gtrsim 15$ the HMF drops exponentially with increasing mass, so that faint galaxies are the main contributors to the Pop II SFRD.

\begin{figure*}
    \centering
    \includegraphics[width=\textwidth]{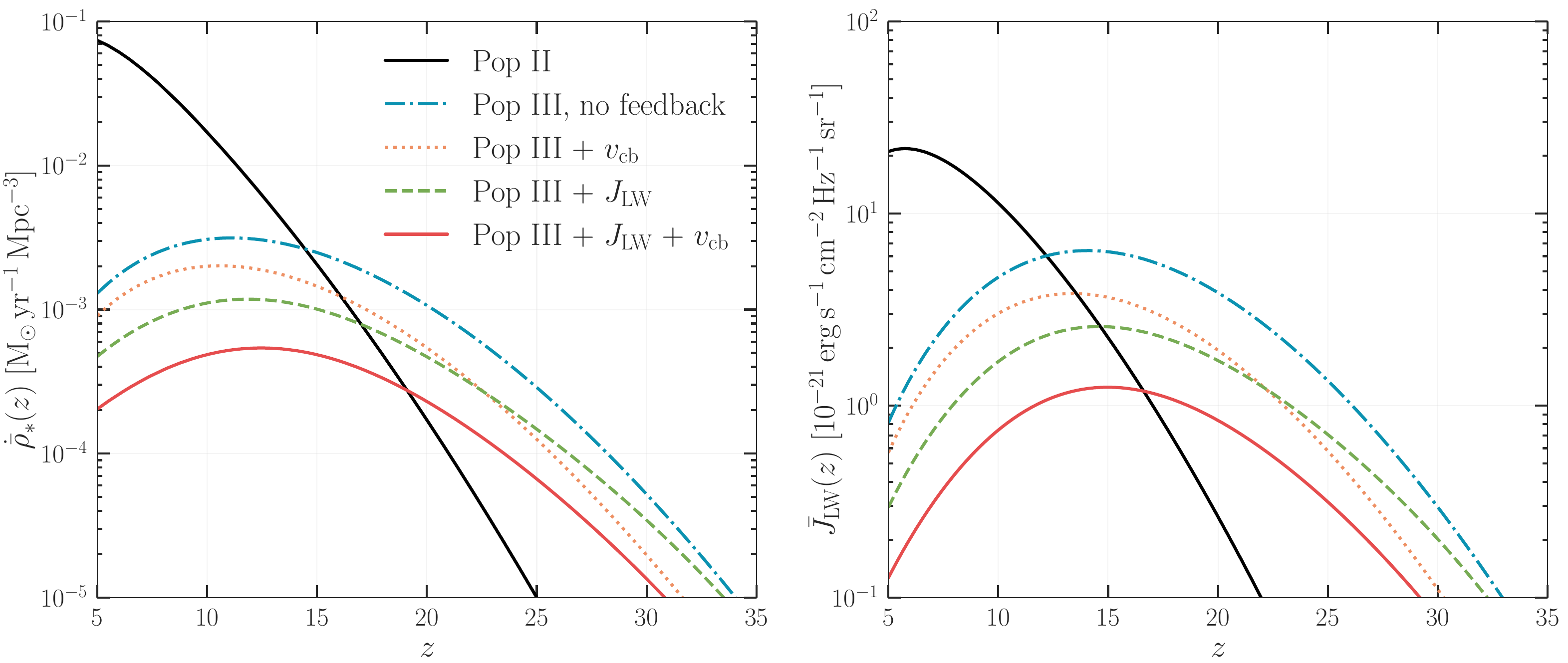}
    \caption{
    Plots of the average SFRD ({\bf left}) and LW background ({\bf right}) against redshift \hac{contributed by Pop II and III stars separately.} 
    For Pop II stars (black) the SFRD keeps growing towards lower $z$ while $J_\mathrm{LW}$ peaks at around $z \sim 5$.
    The situation is different for Pop III stars, shown as different colored lines.
    \hac{From $\dot{\bar{\rho}}_*^\mathrm{III}(z)$ with no feedback (blue), to LW alone (orange dotted), $v_{\rm cb}$ alone (green) and both forms of feedback together (red solid), the SFRD of Pop III stars is reduced in a $z$-dependent way.}
    For our choice of fiducial parameters, the rise of Pop II stars within atomic-cooling galaxies at lower redshifts will produce enough LW photons that the Pop III SFRD will decline for $z\lesssim 10$. 
    \hac{The LW flux and SFRD depend on each other as shown in Eq.~\eqref{eq:averageJLW}, which we self-consistently model through a simple recursive iteration process delineated in Sec. \ref{sec:LWfeedback}.}}
  \label{fig:FIG3_sfrd_j21_vs_z}
\end{figure*}

\subsection{Population III Stars}
Due to a paucity of observations of the earliest generation of stars, the formation and feedback mechanisms that govern the evolution of molecular-cooling galaxies are still uncertain. 
We will use a flexible parameterization, similar to the  Pop II case above, but enhanced to account for new sources of feedback.

We follow the same functional form of the SFE from Eq.~\eqref{eq:SFE}, but assuming a flat SFE with the parameters $\epsilon_*^\mathrm{III} = 10^{-3.0}$, $\alpha_*^\mathrm{III} = \beta_*^\mathrm{III} = 0.0$, and $M_\mathrm{pivot}^\mathrm{III} = 10^{7} \, M_\odot$. 
These are summarized in Table~\ref{tab:tableParams}.
To ensure a smooth transition between Pop III formation (in MCGs with halos of mass below $M_\mathrm{atom}$) and Pop II (in ACGs above $M_\mathrm{atom}$), we implement a double exponential duty cycle
    \begin{equation}
    f_\mathrm{duty}^\mathrm{III} = \exp\left(-\frac{M_\mathrm{turn}^\mathrm{III}(z)}{ M_h}\right) \exp\left(-\frac{M_h}{M_\mathrm{turn}^\mathrm{II}(z)}\right),
    \end{equation}
in terms of some turnover mass for haloes containing MCGs $M_\mathrm{turn}^\mathrm{III}(z) = M_\mathrm{mol}(z)$ where we again neglect the effects of photo-heating and photo-ionization in $M_\mathrm{mol}^\mathrm{III}$, as we did for ACGs. 
Because MCGs reside in less-massive halos, they suffer from additional feedback effects that suppress star formation. 
In particular, molecular cooling is highly sensitive to  Lyman-Werner radiation and to the relative (or streaming) velocity between cold dark matter and baryons \citep{schauer21, kulkarni21}. 
These two effects are expected to be cumulative, where we can model~\cite{munoz22,fialkov13} 
    \begin{equation}\label{eq:Mmol}
    M_\mathrm{mol}^\mathrm{III}=M_0(z) f_{v_{\mathrm{cb}}} f_{\mathrm{LW}},
    \end{equation}
given feedback functions $f_\mathrm{LW}$ and $f_{v_\mathrm{cb}}$.
Here, the zero-feedback molecular-cooling threshold $M_0(z) = 3.3 \times 10^7 (1+z)^{-3/2}\, M_\odot$ corresponds to a virial temperature of $T_\mathrm{vir} = 10^3 \, \si{K}$. 
Let us now describe the functional forms of both feedback functions, which highlight their dependence on the relative velocity $v_\mathrm{cb}$ and Lyman-Werner background $J_\mathrm{LW}$.

\subsubsection{Lyman-Werner Feedback} \label{sec:LWfeedback} 

Hydrogen molecules are highly susceptible to photo-dissociation by Lyman-Werner (LW) radiation (of energies between $11.2 - 13.6 \, \si{eV}$) through the Solomon process \citep{stecher67}. This precludes gas reservoirs in the IGM from cooling sufficiently to form stars in MCGs if the LW flux is high enough \citep{haiman97, tegmark97, abel02, bromm04, haiman06, wise07, oshea07, trenti10}. As the regulatory influence of LW photons at any location depends on luminous sources in its surroundings, LW feedback is both a nonlinear and nonlocal process. A full computation of the total impact of the LW photons must consider the complicated processes of photon recycling within the Lyman-series cascade \citep{pritchard06, qin20} and self-shielding \citep{skinner20}. 
The resulting threshold mass for a galaxy to be able to cool gas can be approximated through a simple power-law~\citep{machacek01, munoz22}.

For a population of stars $i$, we quantify the spatially averaged LW specific intensity (in units of $\si{erg/s/cm^2/Hz/sr}$) as an integral \hac{over the past spatially-averaged SFRD lightcone \citep{visbal14, mebane18} in Eq.~\ref{eq:SFRD}},
    \begin{equation} \label{eq:averageJLW}
    \bar{J}_{\mathrm{LW}}^{(i)}(z)=\frac{c(1+z)^2}{4 \pi} \int_z^{z_{\mathrm{m}}} \epsilon^{(i)}_\mathrm{LW}\left(z^{\prime}\right) \dot{\bar{\rho}}_*^{(i)}(z')  \frac{dz'}{H(z')},
    \end{equation}
where, instead of considering frequency-dependent attenuation \citep{haiman97}, we follow the prescription from Ref.~\cite{visbal14} that Lyman-series photons approximately redshift by at most 4\% before falling into a Lyman-resonance via the relation
    \begin{equation}
    \frac{1+z_{\mathrm{m}}}{1+z}=\frac{\nu_i}{\nu_{\mathrm{obs}}} \approx 1.04.
    \end{equation}
Here, we assume an average LW emissivity of the form,
    \begin{equation} 
        \epsilon_\mathrm{LW}^{(i)}(z) = \frac{N^{(i)}_{\mathrm{LW}} E_{\mathrm{LW}}}{\Delta \nu_{\mathrm{LW}}m_p} ,
    \end{equation}
where $m_p$ is the proton mass, $N^{(i)}_\mathrm{LW}$ is the total number of LW photons per baryon produced in Pop $i$ stars, $E_\mathrm{LW} = 11.9 \, \si{eV}$ is the average LW photon energy, and $\Delta \nu_\mathrm{LW} = 5.8 \times 10^{14} \, \si{Hz}$ is the width of the Lyman-Werner band. For Pop II stars we assume $N_\mathrm{LW}^\mathrm{II } = 6200$, following the prescription that Pop II stars emit 9690 photons per baryon over the full Lyman-series energy range \cite{barkana05}. We similarly assume $N_\mathrm{LW}^\mathrm{III} = 12900$, such that Pop III stars produce 17900 Lyman-series photons per baryon \citep{gesseyjones22, klessen23}. In reality, these latter values are dependent on the initial mass functions of early stars; we leave an IMF-flexible parameterization of the LW emissivities for future work.

Given a LW flux, the turnover mass for a halo to host an MCG is increased by a factor~\cite{munoz22},
    \begin{equation}
        f_\mathrm{LW} = 1 + A_\mathrm{LW}\bar{J}_{21}^{\beta_\mathrm{LW}},
    \end{equation}
fitted to simulations of Pop III stars in Refs.~\citep{schauer21, kulkarni21}. Here, $\bar{J}_\mathrm{21} = \bar{J}_\mathrm{LW} / 10^{21} \ \mathrm{erg \, s^{-1} \, cm^{-2} \, Hz^{-1} \, sr^{-1}}$, and we take fiducial parameters $A_\mathrm{LW} = 2.0$ and $\beta_\mathrm{LW} = 0.6$ as in Ref.~\cite{munoz22}. 
It is noteworthy that the turnover mass depends on the LW background, which itself computed over past SFRD lightcones that depend on the turnover mass. \hac{To solve for both quantities, we recursively iterate both $\bar{J}_{21}(z)$ and $\dot{\bar{\rho}}_*^\mathrm{III}(z)$ until the iterations stabilize in {\tt Zeus21}, which executes on the order of $10^{-1}$ seconds.} We will see how this anisotropic feedback affects the 21-cm fluctuations in Sec.~\ref{sec:LWcorrection1} and \ref{sec:LWcorrection2}, but for now we illustrated the average $J_{21}(z)$ in the rightmost panel of Fig.~\ref{fig:FIG3_sfrd_j21_vs_z}. While ACGs contribute a monotonic increase in the LW background, MCGs show peak contribution at around $z \sim 15$ followed by a steady decline at later times.
The transition from Pop III to II is apparent by the decrease of the Pop III SFRD towards lower $z$, as Pop II stars become the main contributors to the cosmic SFR.

\subsubsection{Relative Velocity Feedback} \label{sec:avgVCBfeedback}

The absence of  baryon acoustic oscillations (BAOs) for dark matter gives rise to a supersonic relative velocity between it and baryons, which can impede star formation in the lowest mass halos. This velocity is set at kinematic decoupling $z_\mathrm{kin} \approx 1060$ and only redshifts as $(1+z)$ thereafter, with its amplitude keeping a $\chi^2$ distribution with three degrees of freedom (for each of the spatial directions), defined by an RMS $\sigma_\mathrm{rms} \approx 29.23 \ \mathrm{km/s}$ and average value $v_\mathrm{avg} \approx 0.92\,\sigma_\mathrm{rms}$~\citep{tseliakhovich10}. The statistics of the relative velocity field are encapsulated by its two-point correlation defined as
    \begin{equation} \label{eq:Pvcb}
        \left\langle v_\mathrm{cb}^{m}(\mathbf{k}) v_\mathrm{cb}^{*n}\left(\mathbf{k}^{\prime}\right)\right\rangle=\frac{k^{m} k^n}{k^2} P_{v_\mathrm{cb}}(k)(2 \pi)^3 \delta^3\left(\mathbf{k}-\mathbf{k}^{\prime}\right),
    \end{equation}
for spatial components $m,n$. Here, we define the power spectrum through
    \begin{equation}
        P_{v_\mathrm{cb}}(k) = A_s \left(\frac{k}{k_\mathrm{pivot}} \right)^{n_s - 1} \left[\frac{\theta_b(k)-\theta_c(k)}{k}\right]^2 \frac{2\pi^2}{k^3},
    \end{equation}
where $\theta_b$ and $\theta_c$ are the baryon and CDM transfer functions respectively, and $k_\mathrm{pivot} = 0.05 \, \si{Mpc}^{-1}$.

This relative velocity affects the formation of the first galaxies. 
Large values of $v_\mathrm{cb}$ inhibit star formation in galaxies residing in the lowest-mass halos, chiefly MCGs, in three main ways. Firstly, patches of the universe containing large relative velocities exhibit suppressed matter fluctuations on small scales \citep{tseliakhovich10, naoz12, bovy13}, which impacts the abundance of the low-mass end of the halo mass function \citep{munoz19}. Secondly, relative velocities tend to disrupt the inner, dense cores of protogalaxies, which interferes with its cooling efficiency \citep{dalal10, greif11, hirano18, schauer19}. Lastly, halos residing in areas with large relative velocity accrete and cool gas less efficiently \citep{dalal10, tseliakhovich11, stacy11, oleary12, naoz13}. The latter two effects have been shown to be the most dominant in its impact on the Pop III SFRD \citep{mcquinn12, fialkov12}.  

The hydrodynamical simulations from Refs.~\citep{schauer21, kulkarni21} found that both the LW feedback and relative velocities increase the necessary halo mass for Pop III star formation by a factor, fitted by~\cite{munoz23}
    \begin{equation}
    f_{v_{\mathrm{cb}}}=\left(1+A_{v_{\mathrm{cb}}} \frac{v_{\mathrm{cb}}}{\sigma_{\mathrm{rms}}}\right)^{\beta_{v_{\mathrm{cb}}}},
    \end{equation} 
with best-fit parameters $A_{v_{\mathrm{cb}}} = 1$ and $ \beta_{v_{\mathrm{cb}}} = 1.8$. 
To compute the spatially-averaged suppression factor, we use the global average velocity $v_\mathrm{cb} = v_\mathrm{avg} $.

\subsection{Joint Effect}

Before finishing this section, let us briefly explore the joint effect of LW and velocity feedback on Pop III star formation.
The leftmost panel of Fig.~\ref{fig:FIG2fstar_vs_m} shows the Pop III SFE under different feedback effects at $z = 15$. 
Even with no feedback, the requirement of molecular cooling restricts the range of halo masses allowed to form MCGs to around $10^6 \lesssim M_h \lesssim 10^8 \, M_\odot$.
Adding each of the feedback effects increases the low-mass turnover of the SFE, and together they rise it by more than an order of magnitude.

The halos hosting MCGs are less massive, and thus far more abundant than those hosting ACGs.
Consequently, at high redshifts they can dominate the SFRD even with mdoest SFEs. 
This is clear in Fig.~\ref{fig:FIG3_sfrd_j21_vs_z}, where we show the redshift evolution of the SFRD when including different feedback processes. With influences from both LW photons and relative velocities, MCGs dominate the SFRD above redshifts $z \gtrsim 16.5$. Including relative velocities reduces the Pop III SFRD by a factor of four by $z \sim 10$, whereas LW feedback impacts the SFRD by a factor of two. 
This highlights the need to model the joint feedback effects on Pop III star formation.

\section{An Effective Model for Feedback-Dependent Anisotropies}
\label{sec:IV_effectivemodel}

In previous sections, we have deconstructed the cosmic-dawn 21-cm signal into its different astrophysical components, using the SFRD as a building block to find the early radiation fields. Let us now build a purely analytical framework that extends beyond average quantities (i.e., one-point statistics) and into fluctuations. 
Finding the 21-cm fluctuations requires knowing the distribution of the radiation fields sourced by Pop II and III stars, which also source the anisotropic feedback that regulates star formation. 
We will follow the effective approach of~Ref.~\cite{munoz23} for density fluctuations, extending it to relative-velocities, which affect Pop III star formation. 
This effective model relies on approximating the dependence of SFRDs against $\delta$ and $v_\mathrm{cb}$ so that arbitrary power spectra can be computed analytically. 
In this section, we introduce our perturbative approach to the SFRD in terms of anisotropies in density and velocity, correspondingly decomposed into lognormal and log-chi-squared building blocks from which we construct the inhomogenous 21-cm signal. 

\begin{figure*}[t!]
    \centering
    \includegraphics[width=\textwidth]{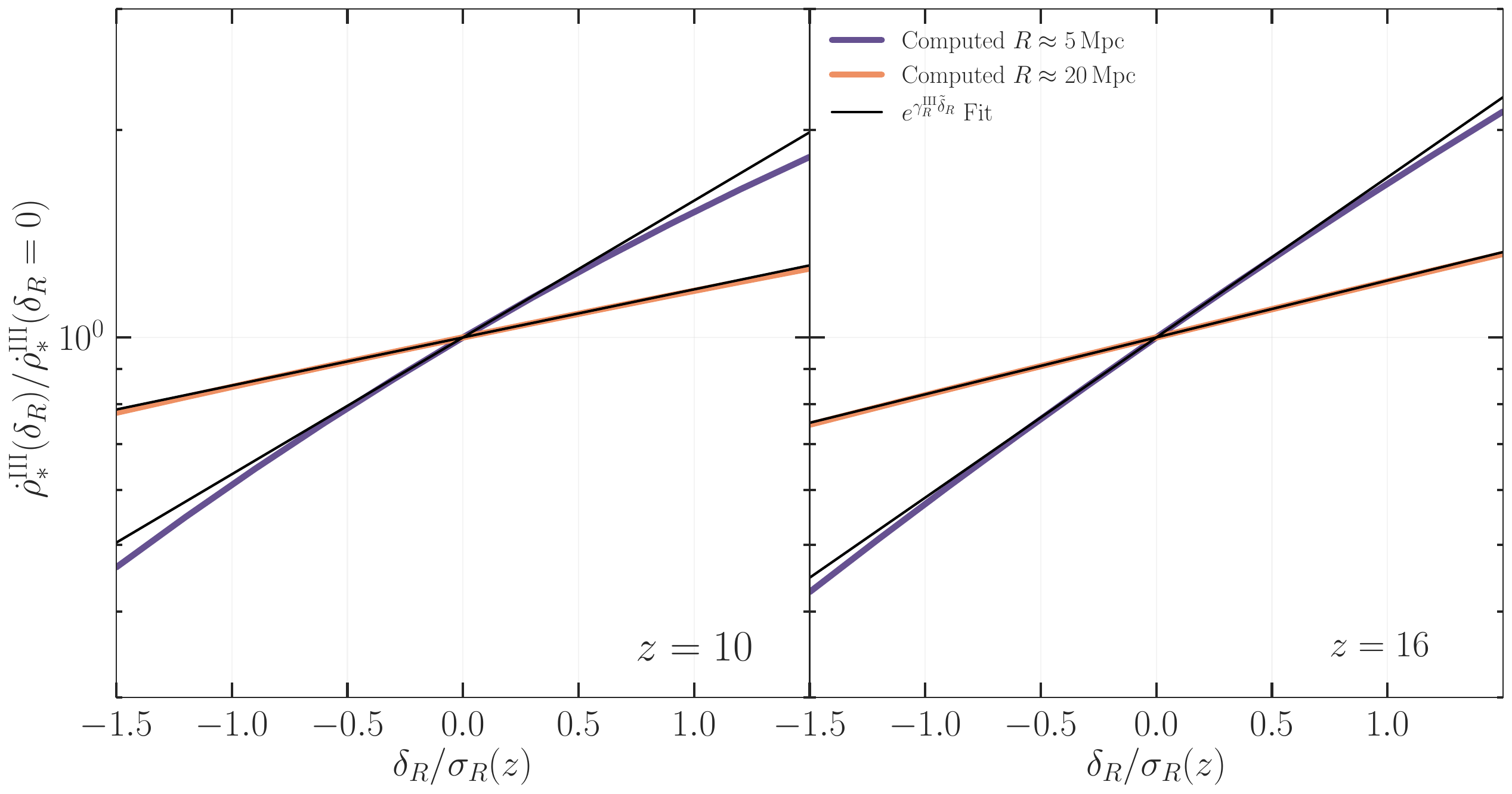}
    \caption{A depiction of the Pop III SFRD (normalized to the $\delta = 0$ case) vs. overdensity $\delta$ according to Eq. \eqref{eq:perturbedSFRD}, at two cases of redshift $z = 10$ (\textbf{left}) and $z=16$ (\textbf{right}). In orange and purple, we plot the result using a tophat smoothing function of radius $5$ and $20 \, \mathrm{Mpc}$ respectively. Increasing the smoothing radius will decrease the matter variance $\sigma_R(z)$ and therefore decrease anisotropies in the SFRD, resulting in shallower slopes. In solid black, we show our log-linear fits $\exp(\gamma^\mathrm{III}_R \delta_R)$ for both cases. In later sections, we elaborate over how Lyman-Werner radiation hampers Pop III star formation, which homogeneizes the SFRD, and therefore decreases the $\gamma^\mathrm{III}_R$ slopes in a scale-dependent way as described in Secs.~\ref{sec:LWcorrection1} and \ref{sec:LWcorrection2}.}
  \label{fig:FIG4sfrdIII_vs_delta}
\end{figure*}

For notational clarity, we will follow the \citet{Peebles:1980yev} convention of defining fluctuations on a quantity $q$ to be $\delta q(\mathbf{x}) = q(\mathbf{x}) - \langle q \rangle$, such that correlations can be described as a mean-normalized covariance of the field $q(\mathbf{x})$
    \begin{equation} \label{eq:peeblesCorrFunc1}
    \xi(r) \equiv \frac{\big\langle\left[q(\mathbf{x}_1)-\langle q\rangle \right]\left[q(\mathbf{x}_2)-\langle q\rangle \right]\big\rangle}{\langle q\rangle^2}.
    \end{equation}
Additionally, we will define correlation functions with lowercase $\xi_q (r)$ and when correlating quantities $q$. To avoid redundancy, we refer to $\xi_\delta(r)$ as the matter correlation function $\langle \delta (\mathbf{x}_1) \delta (\mathbf{x}_2)\rangle$.

\subsection{The SFRD Building Block}
We begin by expanding upon our model of the globally averaged SFRD (in Eq.~\eqref{eq:SFRD}) to regions of comoving radius $R$ with density $\delta_R$~\citep{barkana05} and relative velocity $v_{{\rm cb},R}$~\cite{tseliakhovich10},
    \begin{align} \label{eq:perturbedSFRD}
        \dot{\rho}^{(i)}_*(z | \delta_R,v_{{\rm cb},R}) &= (1 + \delta_R) \int \dfrac{dn}{dM_h}(z | \delta_R)  \dot{M}_h(M_h,z_R) \nonumber \\
        &\times f^{(i)}_{*}(M_h) f_b  f^{(i)}_\mathrm{duty}(v_{{\rm cb},R}) \, dM_h ,
    \end{align}
where the factor of $(1 + \delta_R)$ accounts for the conversion between Lagrangian and Eulerian space \citep{mesinger11}. Technically, the HMF depends on $v_{{\rm cb},R}$~\cite{tseliakhovich11}, though we ignore this effect here. We assume that the local overdensity $\delta_R$ only impacts cosmology through the HMF (i.e., we  neglect assembly bias \citep{wechsler02}), which is modulated as \citep{press74, bond91, barkana05}
    \begin{equation}
        \frac{dn}{dM_h}(z | \delta_R) = \frac{dn}{dM_h}(z) \frac{dn^{\rm PS}/dM_h(\delta_R)}{\langle dn^{\rm PS}/dM_h(\delta_R)\rangle },
    \end{equation}
whose behavior follows \citep{lacey93}
    \begin{equation}
        \frac{dn^{\rm PS}/dM_h(\delta_R)}{\langle dn^{\rm PS}/dM_h(\delta_R)\rangle } = \mathcal C \frac{\tilde \nu}{\nu_0} \frac{\sigma^2}{\tilde \sigma^2} e^{a_{\rm EPS} (\tilde \nu^2 - \nu_0^2)/2},
    \end{equation}
where we use $a_\mathrm{EPS} = q_\mathrm{ST}$ and define
\begin{equation}
    \begin{array}{ll}
        \tilde{\nu} = \tilde{\delta}_\mathrm{crit} / \tilde{\sigma}, &  \nu_0 = \delta_\mathrm{crit} / \sigma,  \\
        \tilde{\delta}_\mathrm{crit} = \delta_\mathrm{crit} - \delta_R,  &  \tilde{\sigma}^2 = \sigma^2 - \sigma_R^2,  
    \end{array}
\end{equation}
which use the rescaled variables for a region of overdensity $\delta_R$ and variance $\sigma_R^2$. The amplitude $\mathcal{C}$ is found numerically.
The relative velocities affect the SFE through the duty cycle, as shown in Sec.~\ref{sec:avgVCBfeedback}, making the SFRD inherit the $v_{\rm cb}$ fluctuations.

We take the ansatz that the inhomogeneous SFRD can be expressed as the globally average SFRD modulated separately by densities and velocities,
    \begin{equation} \label{eq:sfrdBuildingBlocks}
        \dot{\rho}^{(i)}_* \left(z \big| \delta_R, v^2_\mathrm{cb}\right) =  \mathcal{V}^{(i)}\left( z | v_{\mathrm{cb}, R}^2\right) \mathcal{D}^{(i)}\left( z | \delta_R \right)\dot{\bar{\rho}}^{(i)}_*(z),
    \end{equation}
in which we have defined
    \begin{align}
        \mathcal{V}^{(i)}\left( z | v_{\mathrm{cb}, R}^2\right)  & \propto \frac{\dot{\rho}^{(i)}_* \left(z \big| \delta_R, v^2_{\mathrm{cb},R}\right)}{\dot{\rho}^{(i)}_* \left(z \big| \delta_R, v^2_\mathrm{avg}\right)}, \\
        \mathcal{D}^{(i)}\left( z | \delta_R \right) &\propto  \frac{\dot{\rho}^{(i)}_* \left(z \big| \delta_R, v^2_\mathrm{avg}\right)}{\dot{\rho}^{(i)}_* \left(z \big| \delta_R=0, v^2_\mathrm{avg}\right)},
    \end{align}
that isolate the density $\mathcal{D}$ and relative velocity $\mathcal{V}$ dependencies which will be normalized such that $\langle \mathcal{D} \rangle = \langle \mathcal{V} \rangle = 1$. Hereafter, we will refer to these as the density and velocity ``building blocks'' of the perturbed SFRD. Since we model relative velocities to affect only Pop III stars, $\mathcal V$ only affects the Pop III SFRD and we thus drop its $(i)$ population-specifying superscript for notational simplicity. In the remainder of the section we will first describe these $\mathcal D^{(i)}$ and $\mathcal V$ building blocks, and then compute their correlations to build the 21-cm power spectrum.

\begin{figure}[t!]
    \centering
    \includegraphics[width=\columnwidth]{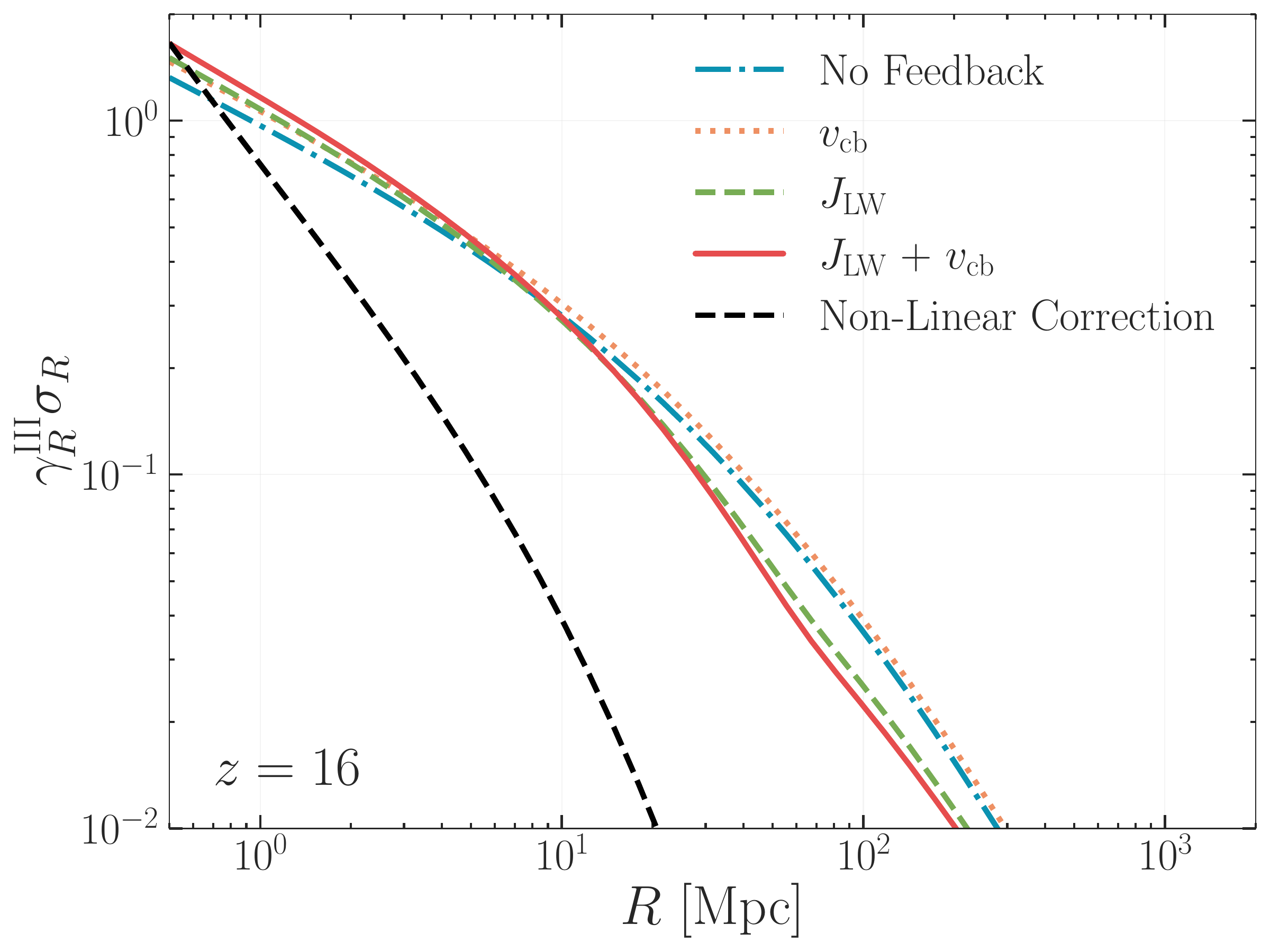}
    \caption{
    A prediction for our effective bias parameter $\gamma_R^\mathrm{III}$ as a function of smoothing scale $R$, at $z=16$, multiplied by $\sigma_R$ for normalization purposes. 
    Different colored lines show different feedback assumptions, where LW feedback reduces $\gamma_R^\mathrm{III}$ at large scales but increases it at small $R$.
    In dotted black we show the size of the correction due to non-linearities for all cases, defined as $\left[ \exp (\gamma_R^2 \sigma_R^2) - 1 \right] / \gamma_R^2 \sigma_R^2$. Our lognormal model thus predicts overall higher power than the linear prediction at small scales, thus boosting the high-$k$ power spectrum of different astrophysical quantities.}
  \label{fig:FIG5gammaRsigmaR_vs_R}
\end{figure}

\subsection{Effective lognormal model for $\delta$ anisotropies}

Over- and under-densities strongly modulate the abundance of halos in the early universe, and thus the SFRD. \citet{munoz23} showed that for Pop $i = \mathrm{II}$ stars, the behavior of the SFRD against the (Gaussianly distributed) smoothed density $\delta_R$ is approximately exponential
\begin{equation}
    \mathcal{D}^{(i)}\left( z | \delta_R \right) \approx e^{ \gamma_R^{(i)} \tilde{\delta}_R },
\end{equation}
with an effective bias $\gamma_R^{(i)}$, provided that $\delta_R \ll \delta_\mathrm{crit}$, valid for the scales and redshifts of interest. We have recast the smoothed overdensity using $\tilde{\delta}_R=\delta_R-\gamma_R \sigma_R^2 / 2$ such that the exponential quantity $\langle \exp (\gamma_R \tilde{\delta}_R )\rangle = 1$ is normalized appropriately. Such a lognormal parameterization for the SFRD density dependence allows for analytic calculations of its two point functions. 
We will now extend this computation to include the Pop III SFRD. In Fig.~\ref{fig:FIG4sfrdIII_vs_delta}, we demonstrate how the behavior of the Pop III SFRD against $\delta_R$ also follows a log-linear fit, whose slope is given by a fitted effective bias $\gamma_R^\mathrm{III}$. At $| \delta_R | \ll 1$ (which corresponds to large $R \gg 1 \, \si{Mpc}$), we recover the linear bias from Refs.~\citep{barkana05, pritchard06}, while at higher $|\delta_R|$, the lognormal model deviates from the true SFRD by no more than $\sim 3\%$. Thus, our lognormal model still captures precisely the nonlinearities of density-dependent SFRD fluctuations toward the tails of the distribution in density. 

In Fig.~\ref{fig:FIG5gammaRsigmaR_vs_R} we show our prediction for these effective biases as a function of smoothing scale $R$ at $z=16$. We plot $\gamma_R^\mathrm{III}\sigma_R$ to account for the scale of matter fluctuations that multiplies the effective bias in the exponent of $\mathcal D$. Smaller $\gamma_R^\mathrm{III}$ (at large $R$) translate into linear-like behavior; as expected from the expansion $\exp(\gamma_R \delta_R) \sim 1 + \gamma_R \delta_R$. 
The opposite is true for small $R$, where the abundance of structure, and thus of emitting sources, behaves nonlinearly with overdensity. Fig.~\ref{fig:FIG5gammaRsigmaR_vs_R} shows that, in the absence of feedback, the SFRD of Pop III stars can be captured through a log-normal approximation, as done for Pop II in Ref.~\cite{munoz23}.

Before turning to feedback, we note that the (position-space) two-point correlation of the SFRD of two populations of stars $i$ and $j$ will have a structure \citep{coles91, xavier16},
    \begin{align}
        \xi^{(i)\times(j)}_{\mathcal{D}}(r, z) &\equiv \left\langle e^{\gamma^{(i)}_{R_1} \tilde{\delta}_{R_1}} e^{\gamma^{(j)}_{R_2} \tilde{\delta}_{R_2}}\right\rangle - \left\langle e^{\gamma^{(i)}_{R_1} \tilde{\delta}_{R_1}}\right\rangle \left\langle e^{\gamma^{(j)}_{R_2} \tilde{\delta}_{R_2}}\right\rangle \nonumber \\
        &=e^{\gamma^{(i)}_{R_1} \gamma^{(j)}_{R_2} \xi_\delta^{R_1, R_2}(r, z)}-1,
    \end{align}
where the smoothed position-space matter correlation function is
    \begin{equation}
    \xi_{\delta}^{R_1, R_2}(r, z)=\int P_m(k, z) W_{R_1}(k) W_{R_2}(k) e^{i \mathbf{k} \cdot \mathbf{r}} \frac{d^3k}{\left(2 \pi \right)^3},
    \end{equation}
given the matter power spectrum  $P_m(k,z)$, and $W_R(k)$ are window functions conventionally described with a spherical tophat
    \begin{equation}
    W_R(k)=\frac{3\left( \cos kR -kR \sin kR \right)}{(k R)^3} .
    \end{equation}
In practice, {\tt Zeus21} numerically computes the perturbed SFRD in Eq.~\eqref{eq:perturbedSFRD} and fits for the $\gamma_R^{(i)}$ for each redshift $z$ and smoothing radius $R$.

\begin{figure*}[t!]
    \centering
    \includegraphics[width=\textwidth]{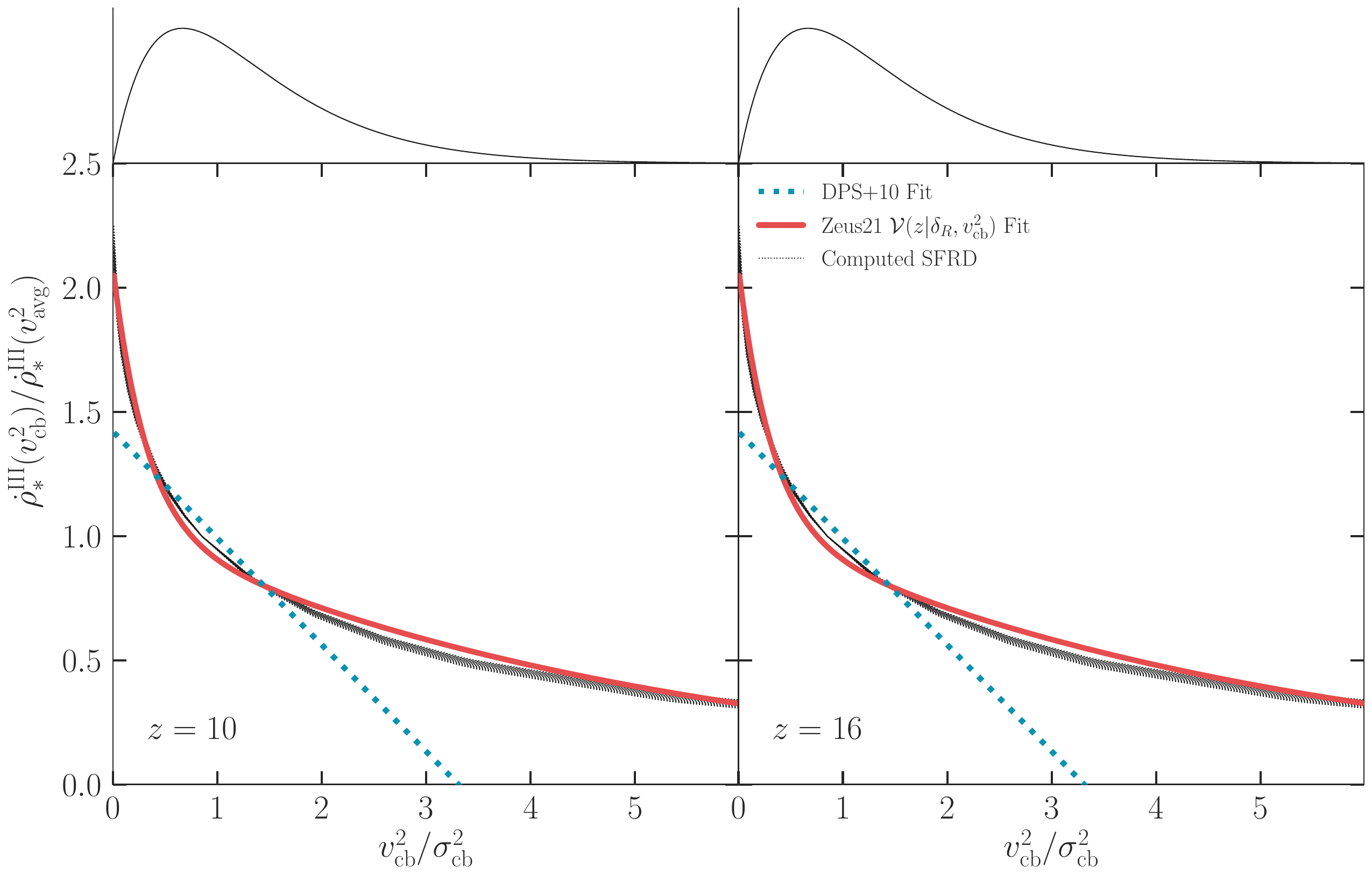}
    \caption{Behavior of the Pop III SFRD against $v_\mathrm{cb}^2$ (normalized to the $v_\mathrm{avg}^2$ case) at $z=10$ ({\bf left}) and $z=16$ ({\bf right}). In dotted black we show a family of SFRDs computed at different fractional densities $-\sigma_R \leq \delta_R \leq \sigma_R$ for $R \sim 3 \, \mathrm{Mpc}$, showing that the density and velocity behaviors are independent (as argued in Eq.~\ref{eq:sfrdBuildingBlocks}). In red we show the best-fit sum of two exponentials, as used in the velocity building block in Eq.~\eqref{eq:logchisquareBlock}, while in blue is the best linear fit following Ref.~\cite{dalal10}. A linear fit does not sufficiently capture the behavior of the SFRD, making the two-exponentials necessary. 
    Top panels show the unnormalized probability distribution functions ($\chi^2_{k=3}$, or Maxwell-Boltzmann) of $v^2_{\rm cb}$. 
    Such a fit recovers the decaying trend of the SFRD, and matches the full numerical calculation better than $\sim 10\%$ across the central $95\%$ of the velocity distribution. }
  \label{fig:FIG6sfrdIII_vs_vcb} 
\end{figure*}

\subsubsection{Lyman-Werner Anisotropic Modifications} \label{sec:LWcorrection1}

Pop III star formation depends sensitively on the intensity of the local LW background.
Regions of higher density will form more galaxies but also generate more LW photons, making the strength of LW feedback depend on density.
Moreover, LW photons can travel a (comoving) distance of $R_\mathrm{max} \sim 50-100 \, \si{Mpc}$ before being absorbed (see Eq.~\eqref{eq:averageJLW}), making the effect of LW feedback highly nonlocal.
Rather than keep track of the local value of $J_{21}$ in a simulation grid (or $\log (M_{\rm turn}^{\rm III})$, as in Refs.~\cite{qin20,munoz22,qin21b}), we will capture the effect of LW feedback into a shift of the effective biases $\gamma_R^{\rm III}$.
That is, we will write
    \begin{align} \label{eq:DIII_LW}
        \mathcal{D}^\mathrm{III}(z|\delta_R) &= \exp \Big[ \gamma_R^\mathrm{III} \tilde{\delta}_R  \\
        &+ \frac{\partial \ln \left(\dot{\rho}_{*}^\mathrm{III}(z|\delta_R)\right)}{\partial J_\mathrm{LW}}\bigg|_{\bar{J}_\mathrm{LW}(z)} \!\! \delta J_\mathrm{LW} (z|\delta_R) \Big] \nonumber 
    \end{align}
to linear order in $\delta J_{\rm LW}$. 
We will expand this expression in Sec.~\ref{sec:LWcorrection2}, along with our IGM treatment.

\subsection{Effective log-$\chi^2$ model for $v_\mathrm{cb}$ anisotropies} \label{sec:effectiveLNCHImodel}

Let us now compute how the Pop III SFRD depends on $v_\mathrm{cb}$.
Unlike LW feedback, the effect of relative velocities on star formation is more straightforward as it is strictly local. 

\begin{figure*}[t!]
    \centering
    \includegraphics[width=\textwidth]{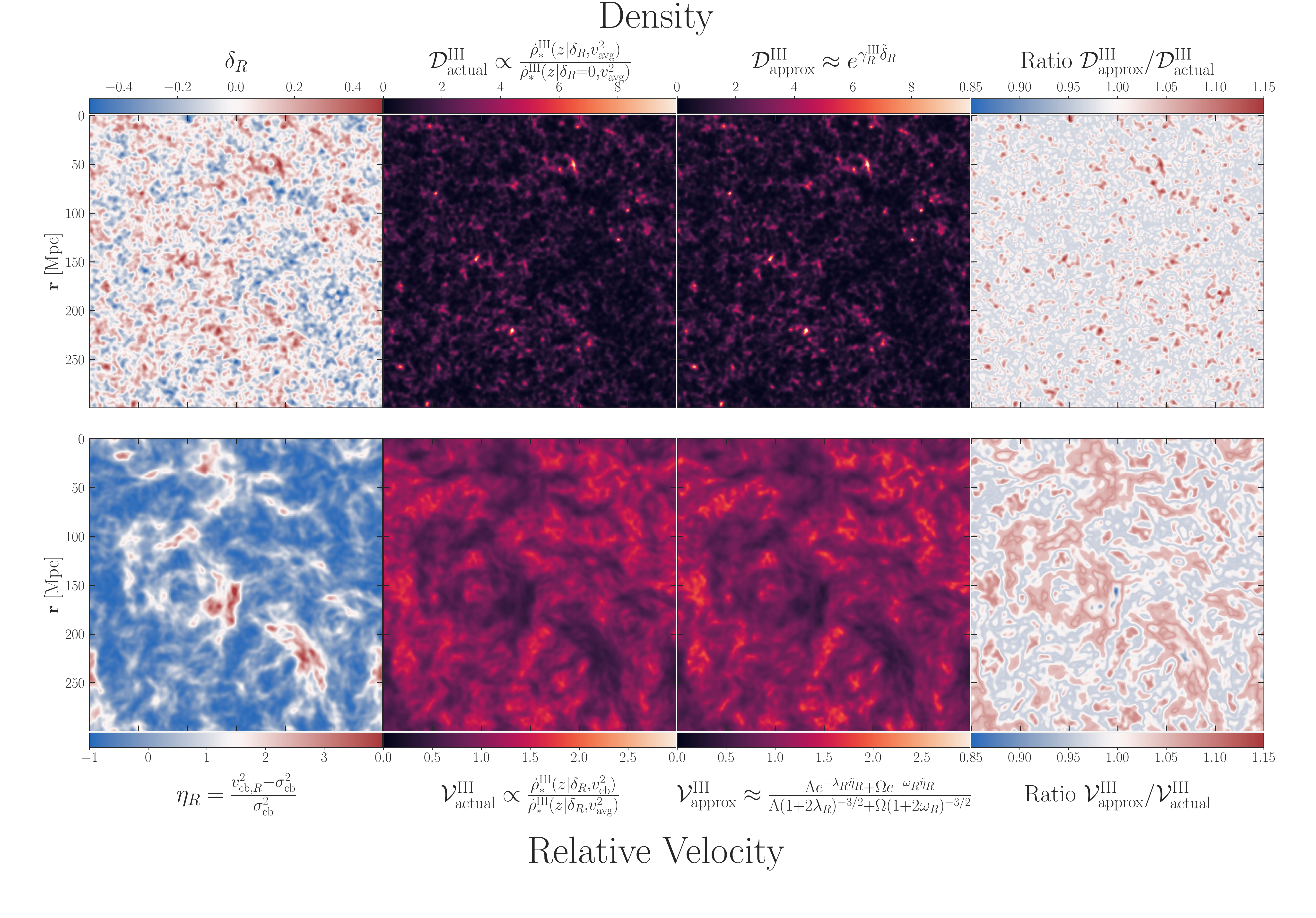}
    \caption{Slices of different quantities dependent on density (\textbf{top row}) and velocity (\textbf{bottom row}) at $z = 16$ across a simulated box with length $600 \, \mathrm{Mpc}$ and resolution $3 \, \mathrm{Mpc}$. Top left is the density $\delta_R$ smoothed with a $5 \, \mathrm{Mpc}$ tophat. Top middle panels show the remarkable similarity between the actual SFRD (middle left) and our lognormal treatment to the density dependence (middle right). The rightmost top panel portrays their ratio, which deviates from unity by less than $\sim 5\%$ nearly everywhere, and by $\sim 10 \%$ at the most nonlinear overdense regions. 
    Bottom left shows the velocity field $\eta_R$ in the same realization. Bottom middle panels show the likewise nearly indistinguishable actual SFRD (middle left) and our log-$\chi^2_3$ treatment (middle right). The rightmost bottom panel portrays their ratio, deviating by $\sim 5-7\%$ from unity across most of the simulated space.}
  \label{fig:FIG7SFRDRealizations} 
\end{figure*}

Higher values of $v_{\rm cb}$ will suppress the SFR in early galaxies, and thus the SFRD. This is clear in Fig.~\ref{fig:FIG6sfrdIII_vs_vcb}, which shows how the Pop III SFRD decreases as a function of $v_{\rm cb}$. As hinted by our construction of the separable SFRD building blocks in Eq. \eqref{eq:sfrdBuildingBlocks}, we plot in Fig.~\ref{fig:FIG6sfrdIII_vs_vcb} the family of $\dot{\rho}_*^\mathrm{III} (v_\mathrm{cb}^2) / \dot{\rho}_*^\mathrm{III} (v_\mathrm{avg}^2)$ curves at different $\delta_R$ to show that the velocity-induced suppression to the SFRD is unaffected by density. By rotational invariance, physical quantities can only depend on $v_\mathrm{cb}^2$, so we define the quantity
    \begin{equation} \label{eq:velBuildingBlock}
            \eta \left(\mathbf{r}\right) = \frac{v^2_\mathrm{cb}\left(\mathbf{r},z\right)}{\sigma^2_\mathrm{cb}(z)} - 1,
    \end{equation}
normalized to have zero mean and to be approximately redshift-independent \citep{ferraro12}.
\hac{Through trial and error we have found that the SFRD can be well fit through the approximation 
    \begin{equation} \label{eq:logchisquareBlock}
        \mathcal{V}\left( z | v_{\mathrm{cb},R}^2 \right) \approx \frac{\Lambda_R e^{-\lambda_R \tilde{\eta}_R} + \Omega_R e^{-\omega_R \tilde{\eta}_R}}{\Lambda_R(1+2\lambda_R)^{-3/2} + \Omega_R(1+2\omega_R)^{-3/2}},
    \end{equation}
whose functional form is motivated below. Firstly, we see by eye that the trend in Fig.~\ref{fig:FIG6sfrdIII_vs_vcb} resembles exponential decay-like behavior. To replicate this decaying behavior, we find it necessary to use the sum of two exponentials in $v_\mathrm{cb}^2$, using normalizations $\Lambda_R$ and $\Omega_R$ and effective biases $\lambda_R$ and $\omega_R$ to fit its shape and curvature. Here, the velocity term has been recast as
    \begin{equation}
        \tilde{\eta}_R \equiv 3\left( \eta_R + 1 \right) = \frac{3v_{\mathrm{cb},R}^2}{\sigma_\mathrm{cb}^2},
    \end{equation}
which is conveniently a $\chi_3^2$-distributed variable since each component of the smoothed relative velocity $v^j_{\mathrm{cb},R}(\mathbf{r}), j \in (x, y, z)$ is Gaussianly-distributed with mean zero. 
Both a linear fit (as in DPS+10~\cite{dalal10}) and a single exponential underestimate the suppression due to $v_{\rm cb}$ by a factor of two, though as we show in Fig.~\ref{fig:FIG6sfrdIII_vs_vcb} the full effect is well captured by two exponentials.
This functional form can, of course, return to a single exponential by setting one of the amplitudes to zero, and further to linear by setting the effective biases to be small. The terms in the denominator are normalization factors that arise from the expectation $\langle \exp (-3\lambda_R v_{\mathrm{cb},R}^2/\sigma_\mathrm{cb}^2 ) \rangle = (1+2\lambda_R)^{-3/2} $, which must be divided out of the velocity building block such that the expectation $\langle \mathcal{V}(z | v_{\mathrm{cb},R}^2) \rangle = 1$. }

The main benefit of our log-$\chi^2$ parameterization for the $\mathcal{V}$-building block is that its correlation function can be analytically found. Its spatial two-point function is
    \begin{align} \label{eq:xi_velBlock}
        \xi_{\mathcal{V}}(r, z) &\equiv \left\langle \mathcal{V}\left( z |  v_{\mathrm{cb},R_1}^2 \right) \mathcal{V}\left( z | v_{\mathrm{cb},R_2}^2 \right)\right\rangle \nonumber \\
        &- \left\langle \mathcal{V}\left( z |  v_{\mathrm{cb},R_1}^2 \right) \right\rangle \left\langle \mathcal{V}\left( z |  v_{\mathrm{cb},R_2}^2 \right)\right\rangle,
    \end{align}
but leave its derivation and lengthy functional form for Appendices~\ref{sec:logchisquareCorrFunc} and \ref{sec:correlationsInV}. The two-point correlation function of the $\mathcal{V}$-building block $\xi_\mathcal{V}(r,z)$ is written in terms of the smoothed spatial correlation function in $\eta$, defined as
    \begin{equation}
        \xi^{R_1, R_2}_{\eta}(r)=\int P_\eta(k) W_{R_1}(k) W_{R_2}(k) e^{i \mathbf{k} \cdot \mathbf{x}} \frac{d^3k}{\left(2 \pi \right)^3}.
    \end{equation}
Here, $P_\eta(k)$ is the power spectrum of $\eta$,
derived using Eq.~\eqref{eq:Pvcb} and Wick's theorem~\cite{dalal10},
    \begin{equation}
        P_\eta(k)=4 \pi \int d r r^2 j_0(k r)\left(6 \psi_0(r)^2+3 \psi_2(r)^2\right),
    \end{equation}
where $\psi_0(r), \psi_2(r)$ are defined as
    \begin{align}
    \psi_0(r) &=\frac{1}{3\sigma_\mathrm{cb}^2} \int \frac{k^2 d k}{2 \pi^2} P_{v b c}(k) j_0(k r), \\
    \psi_2(r) &=-\frac{2}{3\sigma_\mathrm{cb}^2} \int \frac{k^2 d k}{2 \pi^2} P_{v b c}(k) j_2(k r),
    \end{align}
in terms of spherical Bessel functions $j_\ell(x)$.

\begin{figure}[t!]
        \centering
        \includegraphics[width=\columnwidth]{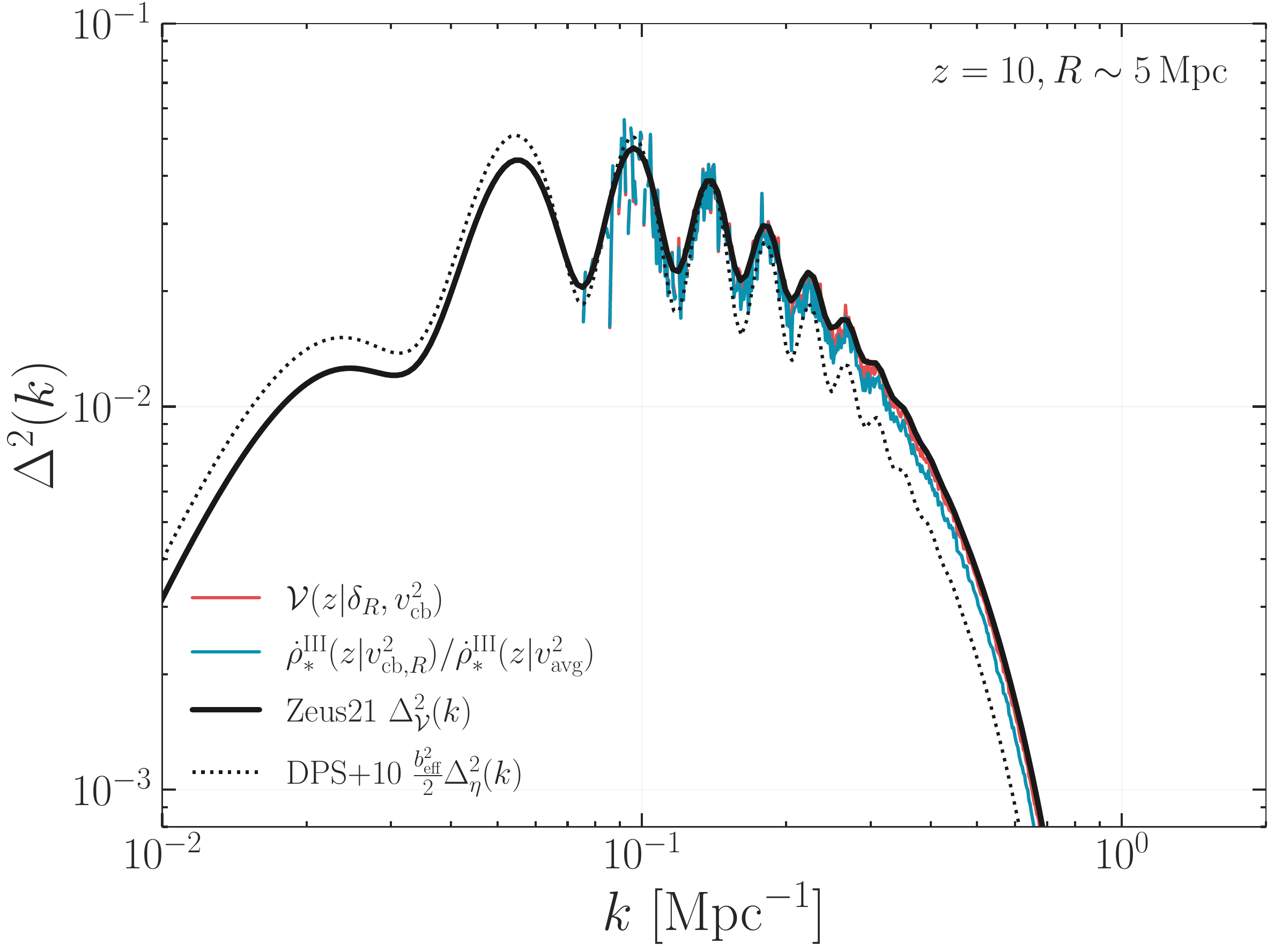}
        \caption{
        The velocity contribution to the SFRD power spectrum at $z=10$, smoothed by a spherical tophat of radius $R \sim 5 \, \mathrm{Mpc}$. In cyan we show the power spectrum of the actual velocity-modulated SFRD, while in red we show the power spectrum of our $\mathcal{V}$ approximation, using the sum of two exponentials (equivalent to the bottom middle-left and -right panels of Fig.~\ref{fig:FIG7SFRDRealizations}, respectively). 
        In solid black, we show the analytic power spectrum of the $\mathcal{V}$ building block, whereas we portray the \citep{dalal10} linearized power spectrum in dotted black. We show that the linearized assumption using an effective bias recovers the position of the acoustic features in the power spectrum, but may over (under)estimate the power at large (small) scales by $\sim 10-100\%$. }
      \label{fig:FIG8_PS_SFRD}
    \end{figure}
    
\subsection{Combining Density and Velocity} \label{sec:DandVtogether}

Using the fits in Figs.~\ref{fig:FIG4sfrdIII_vs_delta} and \ref{fig:FIG6sfrdIII_vs_vcb}, we have shown that both the velocity and density dependence of the SFRD can be recreated using the $\mathcal{D}$ and $\mathcal{V}$ building blocks. 
We now test this formalism against realizations, compare it with existing prescriptions in the literature, and show that we can combine the two building blocks.

To test that our fits adequately represent star formation during during cosmic dawn, we show in Fig.~\ref{fig:FIG7SFRDRealizations} a slice of a realization of the Pop III SFRD at $z=16$ (see Ref.~\cite{munoz23} for the Pop II case). We simulate a Gaussian linear density field $\delta_R$ and a $\chi^2_3$ relative-velocity field $v_{{\rm cb},R}$, both smoothed with a spherical tophat of radius $R=5 \, \mathrm{Mpc}$, and compare a simulated slice of the SFRD using our fitted $\mathcal{D}$ and $\mathcal{V}$ approximations against the actual SFRD from Eq.~\eqref{eq:perturbedSFRD}. 
The two slices are visually identical, and have most galaxies within the densest regions, or in patches of low relative velocity, dominating the SFRD.
Fig.~\ref{fig:FIG7SFRDRealizations} also shows the ratio of our $\mathcal{D}$ and $\mathcal{V}$ approximations to the actual SFRD, which deviate by at most $5\%$ across the majority of the simulated space (with the $\mathcal{D}^\mathrm{III}_\mathrm{approx}/\mathcal{D}^\mathrm{III}_\mathrm{actual}$ ratio exhibiting an RMS of 1.0065, and the $\mathcal{V}_\mathrm{approx}/\mathcal{V}_\mathrm{actual}$ exhibiting an RMS of 0.9928.) These differences shrink to less than percent level for $R \gtrsim 10 \, \mathrm{Mpc}$, which corresponds to the $k$-scales most readily observable by current and upcoming 21-cm experiments.

Our $\mathcal{V}$ building block is more complex than introduced in Ref.~\cite{dalal10} and used in e.g.~Refs.\cite{munoz19,Hotinli:2021vxg,Sarkar:2022dvl,Sarkar:2022mdz,Zhang:2024pwv}.
\citet*{dalal10} argued that the correlation function $\xi_{f}$ of a function $f$ of the relative velocity can be well approximated in terms of a linear effective bias, i.e., $\xi_{f}(r) \approx b^2_\mathrm{eff} \xi_\eta(r)$. This is equivalent to a linear fit around $v_\mathrm{avg}$, shown in Fig.~\ref{fig:FIG6sfrdIII_vs_vcb}, which we show does not adequately fit the SFRD behavior for both the low and high velocity tails. 
Moreover, this misfit propagates to the power spectrum, as shown in Fig.~\ref{fig:FIG8_PS_SFRD}.
The power spectrum of a realization of the full velocity-modulated SFRD (equivalent to the third panel from the left on the bottom row of Fig.~\ref{fig:FIG7SFRDRealizations}) is well matched by our $\mathcal V$ double-exponential approximation, but not by the linear approximation introduced in~Ref.~\cite{dalal10}. 
The peaks of the acoustic features in both treatments are aligned, however, showing that VAOs remain a standard ruler.
Yet, power-spectrum amplitude can differ by $\mathcal O(1)$ at low and high $k$, which could bias inferences of astrophysical parameters dependent on velocity. 

The exponential building blocks $\mathcal D$ and $\mathcal V$ allow us to compute correlation functions of the (nonlinear and nonlocal) SFRD without realizations on a grid. 
We have updated {\tt Zeus21} to numerically calculate $\dot{\rho}_*(z | \delta_R, v_{\mathrm{cb}, R}^2)$ and fit for the parameters $\gamma_R, \Lambda_R, \lambda_R, \Omega_R, \omega_R$, which form the base of our effective model of the SFRD during cosmic dawn.
The total SFRD can then be constructed as the sum of its Pop II and III contributors,
    \begin{align}
        \dot{\rho}_*(z| \delta_R, v_{\mathrm{cb}, R}^2) &= \dot{\bar{\rho}}_*^\mathrm{II}(z) \mathcal{D}^\mathrm{II} \left(z|\delta_R\right)\\
        &+ \dot{\bar{\rho}}_*^\mathrm{III}(z) \mathcal{V} \left(z|v_{\mathrm{cb}, R}^2\right) \mathcal{D}^\mathrm{III} \left(z|\delta_R\right) , \nonumber
    \end{align} 
whose ``unitfull'' correlation function can be expressed in terms of a sum over those of the building blocks,
    \begin{equation}
        \xi_{\dot{\rho}_*}(r, z) = \sum_{i, j} \dot{\bar{\rho}}_*^{(i)}(z) \dot{\bar{\rho}}_*^{(j)}(z) \xi^{(i) \times (j)}_{\mathcal{D}/\mathcal{V}}(r, z),
    \end{equation}
where $\xi^{(i) \times (j)}_{\mathcal{D}/\mathcal{V}}(r, z)$ is a general correlation function in terms of $\xi^{(i) \times (j)}_\mathcal{D}$ and/or $\xi_\mathcal{V}$, whose full expression can be found in App.~\ref{sec:apx_sfrdCorrelation}.

\section{The IGM during Cosmic Dawn} \label{sec:V_IGM}

In previous sections, we introduced our treatment of the SFRD, which encoded its dependence on the density and velocity fields. To connect this prescription to the 21-cm signal, we construct in this section each of the components of the $T_{21}$ signal in Eq.~\eqref{eq:separableT21} through these building blocks. The 21-cm signal is sensitive to the thermal and ionization history of the IGM, and therefore to the radiative fields that cause coupling and energy injection. Our analytical approach replaces the pixel-by-pixel calculations of fluxes in simulations with simple integrals of the SFRD over the past lightcone, weighted by the properties of stars and the IGM that regulate the radiation background. In particular, the Lyman-$\alpha$ and X-ray backgrounds are determined by integrals of the spectral energy distributions (SEDs) of the first galaxies, accounting for the opacity of the IGM. 
In this section, we compute how these ingredients fold into our computations of the early astrophysics at cosmic dawn, first through spatially averaged quantities and then anisotropies. 
Subsec.~\ref{sec:LWcorrection2} deals with the LW background, and its feedback effect on the Pop III SFRD, whereas the rest of the section follows Ref.~\cite{munoz23} but including the effect of the different SEDs in Pop II and III stars.

    \begin{figure*}[t!]
        \centering
        \includegraphics[width=\textwidth]{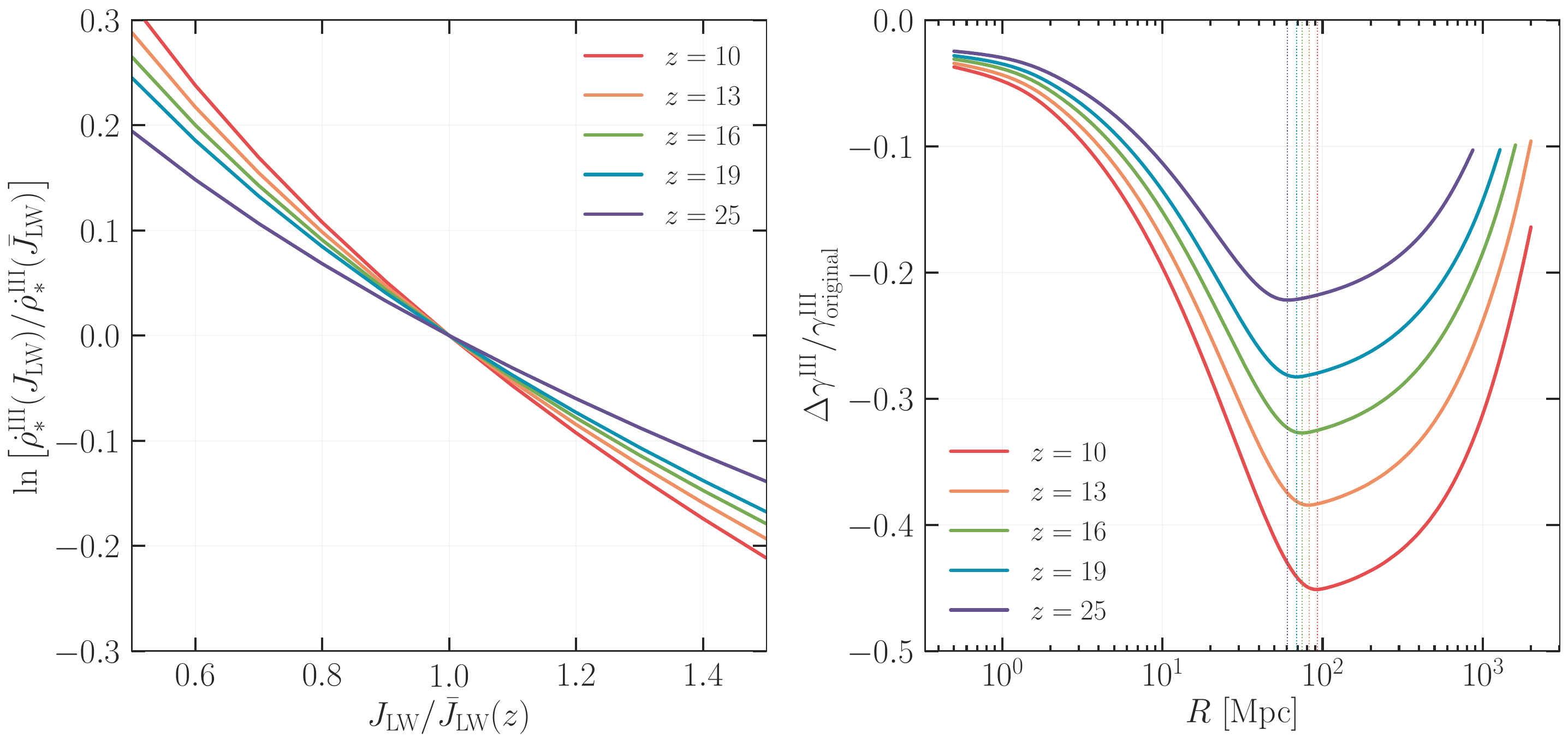}
        \caption{
        Plots of the behavior of the SFRD in the presence of a Lyman-Werner background. (\textbf{Left}): A plot of the Pop III SFRD (normalized to the average $\bar{J}_\mathrm{LW}(z)$ case) vs. the fractional change $J_\mathrm{LW} / \bar{J}_\mathrm{LW}(z)$ in LW intensity at redshift $z$, shown at five different redshifts. For reference, the 1$\sigma$ deviation of $J_\mathrm{LW}$ is never more than $18\%$ across redshifts $10 \gtrsim z \gtrsim 19$ and never more than $25\%$ at redshifts greater than $z \gtrsim 19$. As expected, higher LW background intensities result in lower Pop III SFRD. The slopes of each line are approximately constant near the 2$\sigma$ window around the mean of $J_\mathrm{LW}$, which motivates our linear correction to $\gamma_R^\mathrm{III}$ presented in Eqs.~\eqref{eq:LWcorrectionToGamma1} and \eqref{eq:LWcorrectionToGamma2}. (\textbf{Right}): A plot of the fractional change in the effective bias $\gamma_R^\mathrm{III}$ over distance $R$ at different redshifts. In dotted lines, we depict the $4\%$ redshifting limit at each $z$ that LW photons can travel before falling into a Lyman-series resonance. This distance limit very nearly defines the maximum correction factor to $\gamma_R^\mathrm{III}$, estimated to be nearly $40\%$ at redshifts $z \lesssim 13$ which falls to $20\%$ at redshifts above $z \gtrsim 25$.
        }
      \label{fig:FIG8b_LWcorrections}
    \end{figure*}

\subsection{Lyman-Werner Background} \label{sec:LWcorrection2}

As the first stars are born, they will begin producing an early LW background, which will suppress $H_2$ cooling and thus impede further star formation. The LW flux will inherit the spatial distribution of the first galaxies, so we must model not only its average, but also its anisotropies. As foreshadowed in Sec.~\ref{sec:LWcorrection1}, our goal will be to find the correction to the effective biases $\gamma_R^\mathrm{III}$ that modulate the Pop III SFRD. Let us describe how.

Firstly, we need to know the fluctuations on $J_\mathrm{LW}$.
Converting the redshift integral on Eq.~\eqref{eq:averageJLW} into a discrete sum over spherical shells of comoving radius $R$ yields the relation,
\begin{align}
    J_\mathrm{LW}(z | \delta_R) &= J^\mathrm{II}_\mathrm{LW}(z | \delta_R) + J^\mathrm{III}_\mathrm{LW}(z | \delta_R) \nonumber \\
    &= c_{1, \mathrm{LW}} \sum_R c^\mathrm{II}_{2, \mathrm{LW}}e^{\gamma^\mathrm{II}_R \tilde{\delta}_R} + c^\mathrm{III}_{2, \mathrm{LW}}e^{ \gamma^\mathrm{III}_R \tilde{\delta}_R} ,
    \label{eq:defJLWfull}
\end{align}
where, at redshift $z_R$ (corresponding to a comoving distance $R$ away from $z$), the coefficients are defined as
\begin{align}
    c_{1, \mathrm{LW}} &\equiv \frac{(1+z)^2}{4\pi}, \\
    c^{(i)}_{2, \mathrm{LW}} &\equiv \frac{N_\mathrm{LW}^{(i)} E_\mathrm{LW}}{m_p \Delta \nu_\mathrm{LW}} \dot{\bar{\rho}}^{(i)}_{*}(z_R) w(R) \Delta R. \nonumber
\end{align}
Here, $w(R)$ is a tanh weight, designed to zero out the contribution from photons that redshift beyond the 4\% limit from Ref.~\cite{visbal14}. Fluctuations in the LW background can be written as
\begin{align}\label{eq:defJLWlin}
	\delta J_\mathrm{LW}(z|\delta_R) &= J_\mathrm{LW}(z|\delta_R) - \bar{J}_\mathrm{LW}(z)  \\
	&= c_{1, \mathrm{LW}} \sum_R c^\mathrm{II}_{2, \mathrm{LW}}\left(e^{\gamma^\mathrm{II}_R \tilde{\delta}_R}-1\right) \nonumber \\
 &+ c^\mathrm{III}_{2, \mathrm{LW}}\left(e^{\gamma^\mathrm{III}_R \tilde{\delta}_R}-1\right). \nonumber 
\end{align}

Secondly, given this $\delta J_\mathrm{LW}$ we can compute the correction to the Pop III SFRD inhomogeneities. From the left panel in Fig.~\ref{fig:FIG8b_LWcorrections}, the Pop III SFRD decreases quasi-linearly with increasing LW background. For reference, we calculate that the LW background integrated over all scales deviates from the average by $\sigma_{J_{21}}/J_{21} \sim 0.16$ at $z=10$, $\sim 0.18$ at $z=16$, and $\sim 0.24$ at $z=25$. This suggests that our $\partial \log \dot{\rho}_* / \partial J_\mathrm{LW}$ correction factor is a safe assumption to make around 2$\sigma$ deviations from the the average LW background intensity. Using the structure from Eq.~\eqref{eq:DIII_LW} we can see how the ``new'' effective bias on the density building block $\mathcal D^{\rm III}(\delta)$ will be
\begin{equation}
    \gamma^\mathrm{III}_\mathrm{new}\tilde{\delta} = \gamma^\mathrm{III}_\mathrm{old}\tilde{\delta} + \frac{\partial \ln \left(\dot{\rho}_{*}^\mathrm{III}(z|\delta_R)\right)}{\partial J_\mathrm{LW}}\bigg|_{\bar{J}_\mathrm{LW}(z)} \delta J_\mathrm{LW}.
\end{equation}
Plugging the linearized fluctation $\delta J_\mathrm{LW}$ from Eq.~\eqref{eq:defJLWlin} yields the correction to the Pop III effective bias
\begin{equation}
    \Delta \gamma^\mathrm{III} \tilde{\delta} = \frac{1}{\dot{\bar{\rho}}_{*}} \frac{\partial \dot{\bar{\rho}}_{*}}{\partial J_\mathrm{LW}}\bigg|_{\bar{J}_\mathrm{LW}} c_{1, \mathrm{LW}}\sum_R \sum_i c^\mathrm{(i)}_{2, \mathrm{LW}}\left(e^{\gamma^\mathrm{(i)}_R\tilde{\delta}_R} -1\right),
\end{equation}
where as usual, quantities with a subscript $R$ are averaged with a window function using a spherical tophat of radius $R$. 
The astute reader may notice that this equation still depends on densities, both unsmoothed ($\tilde \delta$) and smoothed ($\tilde \delta_R$).
In order to obtain an statistical average, we smooth both sides of the equation by window functions $W_r(\textbf{x})$ of different radii $r$, and approximate a  density perturbation that has been smoothed twice as $\tilde{\delta}_{\max(r, R)}$.\footnote{In practice this introduces additional nonlocality, as photons that arrive to a point from a radius $R$ can come from $R+r$ away. As we will see this is a small effect, since the SFRD grows exponentially so there are more photons closer to the source.}\ \ Multiplying by $\tilde{\delta}_{r}$ on both sides and taking the ensemble average yields
\begin{align} \label{eq:LWcorrectionToGamma1}
    \Delta \gamma^\mathrm{III}_r \xi_\delta^{rr}(0) &= \frac{1}{\dot{\bar{\rho}}_{*}} \frac{\partial \dot{\bar{\rho}}_{*}}{\partial J_\mathrm{LW}}\bigg|_{\bar{J}_\mathrm{LW}} c_{1, \mathrm{LW}} \sum_R \left(  c^\mathrm{II}_{2, \mathrm{LW}}\gamma^\mathrm{II}_R +  c^\mathrm{III}_{2, \mathrm{LW}} \gamma^\mathrm{III}_R  \right) \nonumber \\
    &\times \xi_\delta^{\max(r,R), r}(0),
\end{align}
in which $\xi^{R_1 R_2}_\delta(0)\equiv \sigma^2_{R_1 R_2}$ denotes the zero-lag autocorrelation function smoothed across spherical tophats of radii $R_1, R_2$. 
This relation allows us to find the correction to the Pop III effective bias $\Delta \gamma^\mathrm{III}$ at different radii $r$, and thus the full exponent of $\mathcal D^{\rm III}$, including feedback, as 
\begin{equation} \label{eq:LWcorrectionToGamma2}
    \gamma_R^\mathrm{III} \rightarrow \gamma_R^\mathrm{III} + \Delta \gamma_R^\mathrm{III}.
\end{equation}

We show how LW feedback affects the Pop III effective biases $\gamma_R^\mathrm{III}$ in the right panel of Fig.~\ref{fig:FIG8b_LWcorrections}.
At all $z$ we find that the correction to $\gamma_R^\mathrm{III}$ is negative, as expected from LW feedback, which tends to homogenize the SFRD. 
The relative correction $\Delta \gamma_R^\mathrm{III}/\gamma_R^\mathrm{III}$ is largest at $R\sim 100$ Mpc, corresponding to the maximum (comoving) distance that LW photons are expected to travel before falling into a Lyman-series resonance (approximated as a 4\% redshift). This correction peaks at a $40\%$ reduction of $\gamma_R^\mathrm{III}$ at $z\sim 10$, which shrinks to a maximum $20\%$ shift at $z\sim 25$. 
We note that, technically, there should be fluctuations in the LW from relative velocities (as $J_{\rm LW}$ should be modulated by them), which would necessitate a higher-order correction factor that we ignore in this work.

\subsection{Wouthuysen-Field Coupling}

    \begin{figure*}[t!]
        \centering
        \includegraphics[width=\textwidth]{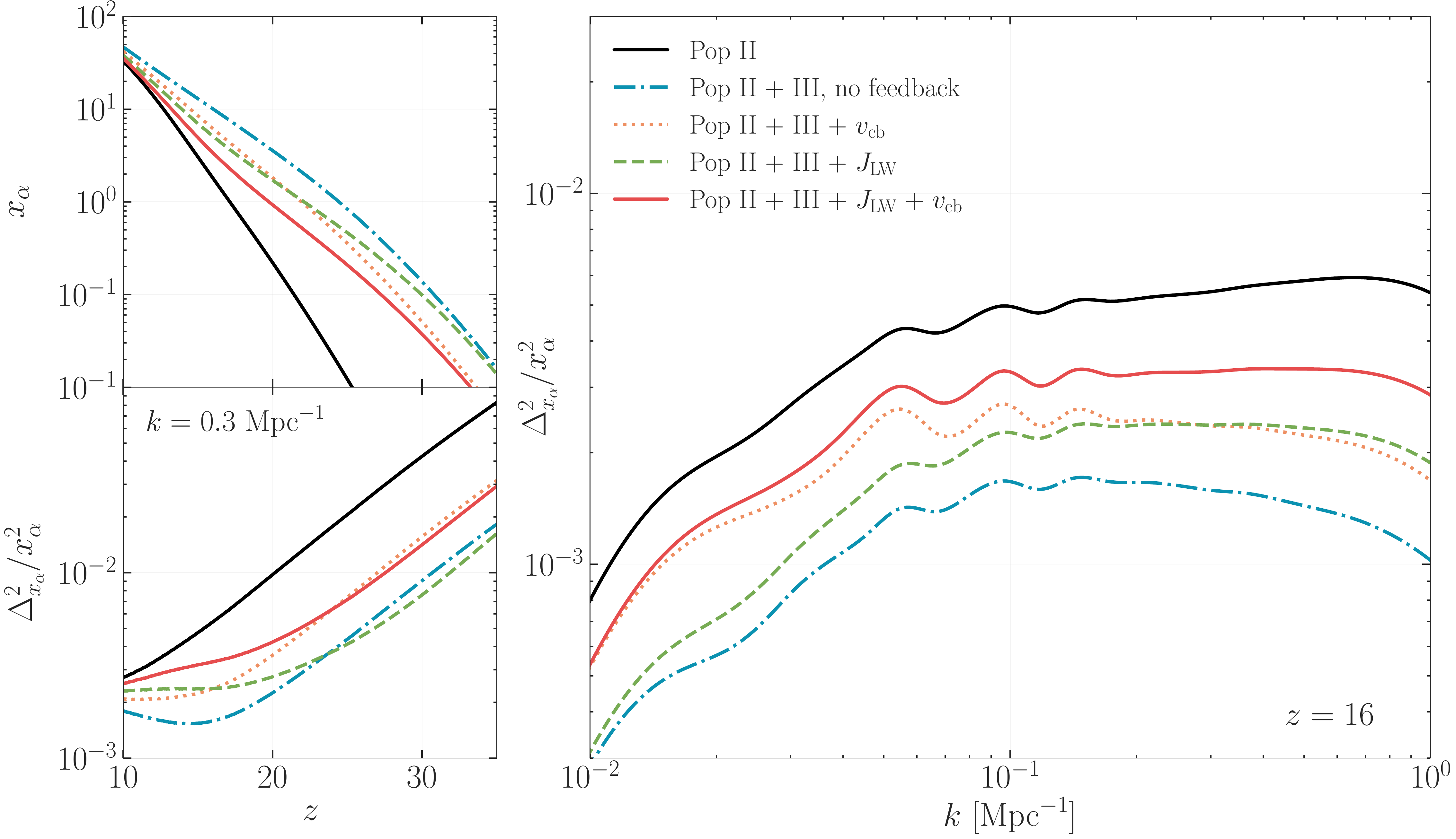}
        \caption{
        Plots of the evolution of averages and anisotropies of the Lyman-$\alpha$ background, using the same color scheme for different stellar populations and feedback effects as Figs.~\ref{fig:FIG2fstar_vs_m} and \ref{fig:FIG3_sfrd_j21_vs_z}. {\bf Top left}:  Average coupling parameter $x_\alpha$ vs. redshift $z$. {\bf Bottom left}: Unitless power spectrum of $x_\alpha$ against $z$, at a fixed $k = 0.3 \, \mathrm{Mpc}^{-1}$. 
        {\bf Right}: Unitless $x_\alpha$ power spectrum now against scale $k$, at fixed $z=16$.  Efficient 21-cm absorption begins when $x_\alpha \gtrsim 0.1$, which for our fiducial model with Pop III stars (in red) occurs for $z \sim 25$, whereas with Pop II stars only (black) takes longer, until $z\sim 20$.
        Including $v_{\rm cb}$ feedback creates large-scale power in the form of wiggles, which we will describe in Sec.~\ref{sec:VAOs}.}
      \label{fig:FIG9_xa}
    \end{figure*}
    
The early ultraviolet background from the first galaxies will produce Wouthuysen-Field coupling of the spin and kinetic temperatures~\citep{wouthuysen52, field59}, which will produce 21-cm absorption at high redshifts (where $T_k$ is lower than $T_{\rm CMB}$). 
The strength of the WF coupling depends on the Lyman-$\alpha$ specific intensity (in units of $\mathrm{erg \, s^{-1}\, cm^{-2} \, Hz^{-1} \, sr^{-1}}$), which for a population $i$ of stars can be computed as \citep{barkana05}
    \begin{equation} \label{eq:LyAIntegral}
        J^{(i)}_\alpha(\mathbf{x}, z) = \frac{(1+z)^2}{4 \pi} \int_0^\infty \dot{\rho}^{(i)}_* (\mathbf{x}, R) \epsilon_\alpha^{\mathrm{tot},(i)} (\nu') dR,
    \end{equation}
where we sum over the possible transitions that give rise to Lyman-$\alpha$ photons
    \begin{equation}
        \epsilon_\alpha^{\mathrm{tot},(i)}\left(\nu^{\prime}\right) \equiv \epsilon_\alpha^{(i)}\left(\nu^{\prime}\right) \sum_{n=2}^{n_{\max }} f_{\mathrm{rec}}(n) w_\alpha(n),
    \end{equation}
  given the recycling fractions $f_\mathrm{rec}(n)$ from Ref.~\cite{pritchard06} (truncated at $n_\mathrm{max} = 23$), and the Pop $i$ SED $\epsilon_\alpha^{(i)}$, evaluated at the redshifted frequency $\nu' = \nu [1+z'(R)] / (1+z)$. 
The weighting functions $w_\alpha(n)$ equal unity for $z < z_\mathrm{max}(n)$ and zero above, with $[1+z_\mathrm{max}(n)] / (1+z) = [1 - (1+n)^{-2}]/(1-n^{-2})$,
to account for the maximum redshift that photons can travel before entering a resonance.

The difference between the two populations of stars is their intrinsic spectrum, which in both cases we approximate as a piecewise power-law via
    \begin{equation}
    \epsilon^{(i)}_\alpha(\nu)=\frac{N^{(i)}_\alpha}{\mu_b} \mathcal{A}^{(i)}\left(\frac{\nu}{\nu_\beta}\right)^{\alpha^{(i)}}
    \end{equation}
for frequencies between Lyman-$\alpha$ and the Lyman-limit, with $\mu_b$ being the mean baryonic mass. 
The power-law indices are
    \begin{equation}
    \alpha^\mathrm{II} = \begin{cases}
    0.14 & \nu_\alpha < \nu < \nu_\beta \\
    -8.0 & \nu > \nu_\beta
    \end{cases}
    \quad 
    \alpha^\mathrm{III} = \begin{cases}
    1.29 & \nu_\alpha < \nu < \nu_\beta \\
    0.2 & \nu > \nu_\beta
    \end{cases},
    \end{equation}
which follow the prescription of Ref.~\cite{munoz23} for Pop II stars (with a total number $N_\alpha^\mathrm{II} = 9690$ of Lyman-series photons per baryon),  and for Pop III stars follow the prescription of Refs.~\citep{barkana05, klessen23, gesseyjones22} (with $N_\alpha^\mathrm{III} = 17900$). The population-specific amplitudes $\mathcal{A}^{(i)}$ normalize the SED between certain frequencies, and are defined as
    \begin{equation}
    \mathcal{A}^\mathrm{II} = \begin{cases}
    0.68 & \nu_\alpha < \nu < \nu_\beta \\
    0.32 & \nu > \nu_\beta
    \end{cases}
    \quad 
    \mathcal{A}^\mathrm{III} = \begin{cases}
    0.56 & \nu_\alpha < \nu < \nu_\beta \\
    0.44 & \nu > \nu_\beta
    \end{cases}.
    \end{equation}

We will compute the influence of Lyman-$\alpha$ photons on the spin temperature $T_s$ in terms of the dimensionless coupling coefficient \citep{furlanetto06},
    \begin{equation}
        x_\alpha = S_\alpha \frac{J_\alpha^\mathrm{II} + J_\alpha^\mathrm{III}}{J_\alpha^c},
    \end{equation}
which uses a numerical constant defined as 
    \begin{equation}
        \frac{1}{ J_\alpha^c} = \frac{1.811 \times 10^{11}}{(1+z)\left(2.725 \mathrm{~K} / T_{\mathrm{CMB}}^{(0)}\right) } \ \mathrm{cm^2 \, s \, Hz \, sr},
    \end{equation}
with a correction factor $S_\alpha$ that depends on the kinetic temperature $T_k$ and the free-electron fraction $x_e$, which is computed iteratively \citep{hirata06}.
We show the redshift evolution of $x_\alpha$ in the top-left panel of Fig.~\ref{fig:FIG9_xa}. The Lyman-$\alpha$ background expectedly grows with redshift, following a behavior similar to the average redshift evolution of the SFRD in Fig.~\ref{fig:FIG3_sfrd_j21_vs_z}, with population-specific weighting factors. Either quantity traces the other; therefore, a measurement of $x_\alpha$ inferred from 21-cm observations can help estimate the early SFRD \citep{madau96}, which is expected to saturate (i.e., $T_s\approx T_k$) when $J_\alpha \sim J_\alpha^c$ or when $x_\alpha$ grows to be of order unity.
The addition of Pop III stars increases $x_\alpha$ significantly at $z\gtrsim 15$, and as we will see it kickstarts the 21-cm absorption era.
However, by the end of our calculations ($z\sim 10$) Pop II stars dominate the SFRD and thus set the value of $x_{\alpha}$, as expected.

\subsubsection*{Lyman-$\alpha$ Anisotropies}
Our description of the Lyman-$\alpha$ background in Eq.~\eqref{eq:LyAIntegral} shows that fluctuations in the SFRD will give rise to fluctuations in $x_\alpha$. 
We now follow the procedure we did for $J_{\rm LW}$ and construct the $x_\alpha$ fluctuations using our density and velocity building blocks by discretizing the integrals, 
\begin{align}\label{eq:discreteLYA}
        x^\mathrm{II}_\alpha(z) &= c_{1, \alpha}(z) \sum_R c_{2, \alpha}^\mathrm{II}(z, R) \mathcal{D}^\mathrm{II}(z | \delta_R), \\
        x^\mathrm{III}_\alpha(z) &= c_{1, \alpha}(z) \sum_R c_{2, \alpha}^\mathrm{III}(z, R) \mathcal{V}(z | v_{\mathrm{cb},R}^2) \mathcal{D}^\mathrm{III}(z | \delta_R), \nonumber
    \end{align}
where
    \begin{equation}
    c_{1, \alpha}(z)=\frac{(1+z)^2}{4 \pi} \frac{S_\alpha}{J_\alpha^c}
    \end{equation}
is a coefficient independent of $R$, and
    \begin{equation}
        c_{2,\alpha}^\mathrm{i} = \dot{\bar{\rho}}_*^{(i)}(R) \epsilon_\alpha^\mathrm{tot} \Delta R
    \end{equation}
includes all $R$-dependent factors and the discretized differential $\Delta R$. 
Using these formulae we can find the autocorrelation function of $x_\alpha$ as
\begin{equation} \label{eq:xiAlpha1}
    \xi_\alpha(r, z) =  \xi^\mathrm{II}_\alpha(r, z) + 2\xi^\mathrm{II \times III}_\alpha(r, z) + \xi^\mathrm{III}_\alpha(r, z),
\end{equation}
where each autocorrelation function (between populations of stars $i$ and $j$) is described by
    \begin{align} \label{eq:xiAlpha2}
        \xi^{(i) \times (j)}_\alpha(r,z) &= c_{1,\alpha}^2(z) \sum_{R_1, R_2} c^{(i)}_{2, \alpha}(z, R_1)c^{(j)}_{2, \alpha}(z, R_2) \nonumber \\
        &\times \xi^{(i) \times (j)}_{\mathcal{D}/\mathcal{V}}(r, z),
    \end{align}
where $\xi^{i \times j}_{\mathcal{D}/\mathcal{V}}(r, z)$ is a general correlation function in terms of $\xi^{i \times j}_\mathcal{D}$ and/or $\xi_\mathcal{V}$ dependent on the correlated populations $i, j$. The lengthy explicit form for each of the terms in Eq.~\eqref{eq:xiAlpha1} is in App.~\ref{sec:allCorrelations}.
We compute the $x_\alpha$ power spectrum by simply Fourier transforming $\xi_\alpha(r, z)$. 

We show in the bottom left and right panels of Fig.~\ref{fig:FIG9_xa} the power spectrum of $x_\alpha$ fluctuations over different wavenumbers $k$ and redshifts. 
Compared to the Pop II case only, adding Pop III stars gives rise to acoustic features in the power spectrum, the so-called VAOs (larger and off phase compared to the BAOs that appear for Pop II only). 
We note that a linear calculation of the $x_\alpha$ power spectrum would underestimate power by $\mathcal O(1)$ at observable scales ($k\gtrsim 0.1$ Mpc$^{-1}$), highlighting the need for our lognormal and log-$\chi_3^2$ nonlinear building blocks $\mathcal{D}$ and $\mathcal{V}$.
This is in agreement with the results from previous work~\cite{munoz23,santos10, mesinger11}

\subsection{Kinetic Temperature of the IGM}

    \begin{figure*}[t!]
        \centering
        \includegraphics[width=\textwidth]{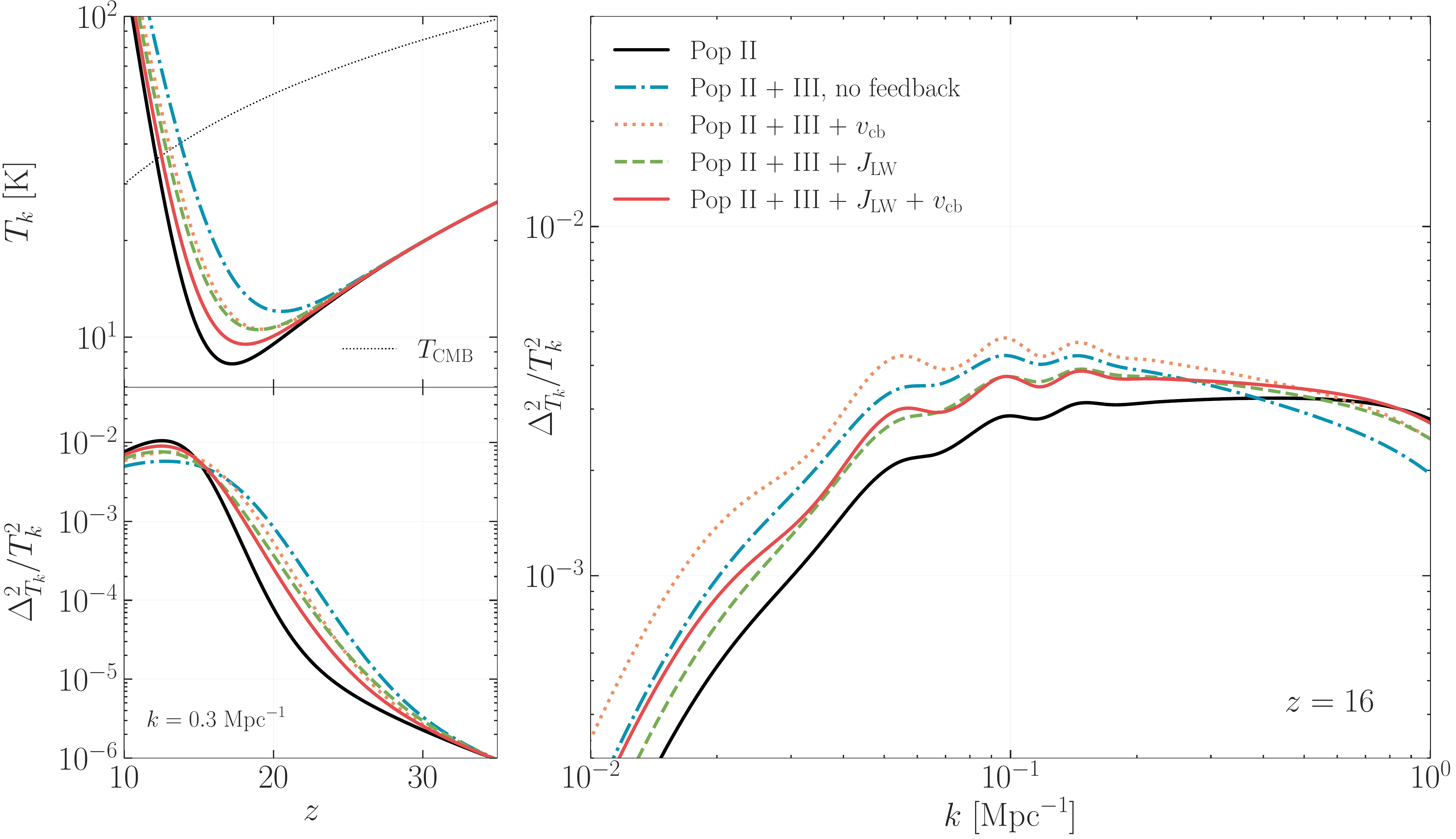}
        \caption{Evolution of averages and anisotropies of the IGM kinetic temperature, with the same color scheme as Figs.~\ref{fig:FIG2fstar_vs_m} and \ref{fig:FIG3_sfrd_j21_vs_z}. {\bf Top left:}  Average kinetic temperature $T_k$ vs.~redshift. {\bf Bottom left:} Unitless power spectrum of $T_k$ against $z$, at a fixed $k = 0.3 \, \mathrm{Mpc}$. {\bf Right:} Unitless $T_k$ power spectrum as a function of scale $k$ at fixed $z=16$. 
        Heating takes over earlier when including Pop III stars, though in our fiducial model, with all sources of feedback included, the process is expected to be dominated by Pop II stars. }
      \label{fig:FIG10_Tk}
    \end{figure*}
    
The gas kinetic temperature $T_k$, which determines the strength of the 21-cm signal, is defined by the competing effects of X-ray heating (from the first galaxies) and adiabatic cooling (due to cosmic expansion). 
At each point $T_k$ evolves as
    \begin{equation}
    \frac{3}{2} T_k^{\prime}=T_k \frac{d \log \rho}{d z}-\frac{\Gamma_C}{H(z)(1+z)}-\frac{\Gamma_X}{H(z)(1+z)},
    \end{equation}
where the first term on the RHS accounts for adiabatic cooling, $\Gamma_c(z)$ is the Compton heating rate to the CMB (see \citep{ma95}), and $\Gamma_X(z)$ includes the heating due to X-rays. For ease of calculation, we will split the kinetic temperature as
    \begin{equation}
        T_k(z) = T_\mathrm{cosmo}(z) + T_X(z),
    \end{equation}
where the first ``cosmology'' term accounts for adiabatic expansion and Compton coupling,
    \begin{equation}
    \frac{3}{2} T_\mathrm{cosmo}^{\prime}=T_\mathrm{cosmo} \frac{d \log \rho}{d z}-\frac{\Gamma_C}{H(z)(1+z)},
    \end{equation}
obtained self-consistently using {\tt CLASS}, whereas the ``astrophysics'' term includes X-ray heating via
    \begin{equation} \label{eq:TxDiffEQ}
    \frac{3}{2} T_\mathrm{X}^{\prime}=T_\mathrm{X} \frac{3}{1+z}-\frac{\Gamma_X}{H(z)(1+z)}.
    \end{equation}
where we have approximated that $\rho \propto (1+z)^3$ (its $\delta$ dependence will be expanded upon in the adiabatic fluctuations discussion of Sec. \ref{sec:LSS}). 
This split allows one to separately treat the contributions from Pop II and III stars to Eq.~\eqref{eq:TxDiffEQ}, so $T_X = T_X^{\rm II} + T_X^{\rm III}$, with
    \begin{equation} \label{eq:XrayIntegral}
    T^{(i)}_X(z)=(1+z)^2 \int_z^{\infty} \frac{1}{\left(1+z^{\prime}\right)^2}\left[-\frac{2}{3} \frac{\Gamma^{(i)}_X(z')}{H\left(z^{\prime}\right)\left(1+z^{\prime}\right)}\right] d z^{\prime},
    \end{equation}
for a population of stars $i$.

The heating rate $\Gamma_X$ of the IGM by X-rays can be described as a local integral over the value of the X-ray flux $J_X^{(i)}$ (over all photon frequencies, which we will define below) \citep{pritchard07}
    \begin{align}
        \Gamma^{(i)}_X(\mathbf{x},z') &=(4 \pi) f_{\text {heat }}(z') \int  J^{(i)}_X(\mathbf{x}, z', \nu) \\
        & \times \sum_{k=\mathrm{HI}, \mathrm{HeI}} f_k \sigma_k(\nu) \left(\nu-\nu_{\text {ion }}^{k}\right)   d \nu, \nonumber
    \end{align}
in terms of the number fraction of each species $f_k$ for $k=\{$HI, HeI\}, their ionization energies $\nu_{\text {ion }}^{(k)}$, and their ionization cross sections $\sigma_k(\nu)$ \citep{verner96}. During the epochs of time considered, we safely neglect the second helium ionization. We approximate the heat deposition efficiency as $f_\mathrm{heat} \approx \bar{x}_e^{0.225}$ \citep{schneider21} (ignoring its fluctuations on this first work), and compute the average free-electron fraction $\bar{x}_e$ through
    \begin{equation}
    \bar{x}_e=\bar{x}_e^{\operatorname{cosmo}}+\bar{x}_e^X,
    \end{equation}
where again we have divided into a ``cosmological'' term $\bar{x}_e^{\operatorname{cosmo}}$ (residual from recombination and obtained from {\tt CLASS} using {\tt Hyrec} \citep{alihamoud11, lee20}), and an ``astrophysical'' term, estimated as an integral over the average ionization rate \citep{mirocha14}, given by
    \begin{equation}
        \bar{x}_e^X = \frac{3}{2} f_\mathrm{ion} \int  \frac{\Gamma_X^\mathrm{II} + \Gamma_X^\mathrm{III}}{\langle \nu_\mathrm{ion} \rangle} dz,
    \end{equation}
with $\nu_\mathrm{ion}$ the average ionization frequency of HI and HeI, and use the estimated ionization fraction fit $f_\mathrm{ion} = 0.4 \exp (-\bar{x}_e / 0.2)$ for the fraction of energy imparted by X-ray photons into secondary ionization \citep{furlanetto10}.

We can now return to the X-ray specific intensity $J_X$ (in units of $\si{erg/s/cm^2/Hz/sr}$), which  is determined as an integral over the past SFRD lightcone
    \begin{equation}
    J^{(i)}_X(\mathbf{x}, \nu)=\frac{(1+z)^2}{4 \pi} \int_0^{\infty}  \dot{\rho}^{(i)}_* (\mathbf{x}, R) \epsilon^{(i)}_X\left(\nu^{\prime}\right) e^{-\tau_X} d R,
    \end{equation}
weighted by a frequency-dependent opacity term given by the optical depth
    \begin{equation}
    \tau_X(\nu, z, R)=\int_z^{z(R)}  n_b \sum_{k=\mathrm{HI}, \mathrm{HeI}} f_k \sigma_k\left(\nu^{\prime}\right) \, d R,
    \end{equation}
using the average baryon density $n_b$ and the redshifted frequency $\nu'$. Finally, $J^{(i)}_X$ depends on the (population-specific) X-ray SED
    \begin{equation}
    \epsilon^{(i)}_X(\nu)=L^{(i)}_{40} \times \frac{10^{40} \mathrm{erg} / \mathrm{s}}{\left(M_{\odot} / \mathrm{yr}\right)} \frac{I^{(i)}_X(\nu)}{\nu}.
    \end{equation}
Here, we use an arbitrary amplitude $L^{(i)}_{40}$,
and model the shape of the X-ray spectrum as a power-law $I^{(i)}_X(\nu) \propto \nu^{\alpha^{(i)}_X}$,  normalized to integrate to unity over the frequency band $E_{0,X} \lesssim h\nu \lesssim E_{\mathrm{max},X}$ (where $E_{0,X} = 0.5 \, \mathrm{keV}$ and $E_{\mathrm{max},X} = 2.0 \, \mathrm{keV}$), in order to compare to past literature (e.g.,~\citep{greig15,mirocha14}).
For both Pop II and III stars, we use the fiducial values $\alpha^\mathrm{II}_X = \alpha^\mathrm{III}_X= -1$ for the power-law slope and $L^\mathrm{II}_{40} = L^\mathrm{III}_{40} = 10^{0.5}$ for the normalization, inspired by studies of low-metallicity high-mass X-ray binaries \citep{fragos13, gilfanov04}.

We show the average kinetic temperature of the IGM in the upper left panel of Fig.~\ref{fig:FIG10_Tk}. As expected, adding Pop III stars heats the IGM through the extra X-rays. Unlike the average evolution of $x_\alpha$, the kinetic temperature $T_k$ does not increase monotonically with redshift. This is due to the adiabatic cooling of the IGM at high redshifts.
The X-rays heat up the IGM above the CMB at $z\sim 12-15$ across all models regardless of feedback, showing that for our fiducial models the process of reionization (at $z\lesssim 10$) will not be cold, as pointed by early 21-cm observations~\cite{hera23}.

\subsubsection*{X-ray Anisotropies}

For X-ray fluctuations we will mirror the formalism for $x_\alpha$, though in this case the IGM temperature captures temporal nonlocality, as the gas heating depends on the integrated heating rate $\Gamma_X$ at all previous times.
Then, the population-specific X-ray temperature integral in Eq.~\eqref{eq:XrayIntegral} can be discretized as
    \begin{align}\label{eq:discreteXRAY}
    T^\mathrm{II}_X(z) &=(1+z)^2 \sum_{z^{\prime} \geq z} c_{1, X}\left(z^{\prime}\right) \sum_R c^\mathrm{II}_{2, X}\left(z^{\prime \prime}, R\right) \mathcal{D}^\mathrm{II}(z | \delta_R) \nonumber \\
    T^\mathrm{III}_X(z) &=(1+z)^2 \sum_{z^{\prime} \geq z} c_{1, X}\left(z^{\prime}\right) \sum_R c^\mathrm{III}_{2, X}\left(z^{\prime \prime}, R\right) \\
    &\times \mathcal{V}(z | v_{\mathrm{cb},R}^2) \mathcal{D}^\mathrm{III}(z | \delta_R)\nonumber,
    \end{align}
given a sum over earlier times ($z'>z$), with the coefficients written as
    \begin{equation}
        c_{1,X}(z') = \frac{-2 f_\mathrm{heat}}{3 H(z')(1+z')} \Delta z',
    \end{equation}
and
    \begin{align}
        c_{2, X}^{(i)} &= \dot{\bar{\rho}}_*^{(i)}(R) \epsilon^{(i)}(\nu'') e^{-\tau_X(\nu'')} \\
        &\times \left(\sum_{\nu'} \sum_{k=\mathrm{HI}, \mathrm{HeI}} \sigma_j (\nu'-\nu_{\text{ion }}^{\prime k}) \Delta \nu' \right) \Delta R. \nonumber
    \end{align}
With these formulas, analytic computations of correlations of $T_X$ become straightforward:
    \begin{equation} \label{eq:xi_X1}
        \xi_X(r, z) =  \xi^\mathrm{II}_X(r, z) + 2\xi^\mathrm{II \times III}_X(r, z) + \xi^\mathrm{III}_X(r, z),
    \end{equation}
where each correlation function is described by
    \begin{align} \label{eq:xi_X2}
        \xi^{(i)\times (j)}_X(r, z) &= (1+z)^4 \sum_{z_1', z_2' \geq z }c_{1, X}(z_1')c_{1, X}(z_2')  \\
        & \times \sum_{R_1, R_2} c_{2, X}^{(i)} (z_1', R_1) c_{2, X}^{(j)} (z_2', R_2) \xi^{(i) \times (j)}_{\mathcal{D}/\mathcal{V}}(r, z). \nonumber 
    \end{align}
We can also define the cross-power spectrum between $x_\alpha$ and $T_X$ as
    \begin{align} \label{eq:xi_alphaX1}
        \xi_{\alpha X}(r, z) &= \xi^\mathrm{II}_{\alpha X}(r, z) + \xi^{\mathrm{II}\times \mathrm{III}}_{\alpha X}(r, z)  + \\
        &+ \xi^{\mathrm{III}\times \mathrm{II}}_{\alpha X}(r, z)  + \xi^\mathrm{III}_{\alpha X}(r, z),  \nonumber
    \end{align}
where each term can be described by
    \begin{align} \label{eq:xi_alphaX2}
        \xi^{(i)\times (j)}_{\alpha X}(r, z) &= (1+z)^2 c_{1, \alpha}(z)\sum_{z_2' \geq z }c_{1, X}(z_2') \\
        & \times \sum_{R_1, R_2} c_{2, \alpha}^{(i)} (z, R_1) c_{2, X}^{(j)}(z_2', R_2) \xi^{(i) \times (j)}_{\mathcal{D}/\mathcal{V}}(r, z). \nonumber
    \end{align}
The full explicit forms for Eqs. \eqref{eq:xi_X1}, \eqref{eq:xi_X2}, \eqref{eq:xi_alphaX1}, and \eqref{eq:xi_alphaX2} can be found in App.~\ref{sec:allCorrelations}. 

As fluctuations in the IGM kinetic temperature are sourced from adiabatic fluctuations in addition to anisotropies in X-ray heating, correlation functions of $T_k$ must include an adiabatic component. Since adiabatic fluctuations are dependent on the local matter density $\delta$, we defer discussion on this topic until the next section.

\subsection{Large-Scale Structure} \label{sec:LSS}
Let us now study the contribution from the density and redshift-space distortion terms in Eq.~\eqref{eq:separableT21}. These cosmological fluctuations are ``Eulerian'', as they are given by the local environment, as opposed to $x_\alpha$ and $T_X$, which are ``Lagrangian'' in that they depend on extrapolated and evolved initial conditions. 
We refer to them as  large-scale-structure (LSS) terms, and our approach will follow the standard {\tt Zeus21} model, though adding cross correlations with Pop III fluctuations.
That is, we assume that the non-linear density terms also follow lognormal behavior, with  $\Delta = \exp(\tilde{\delta}_{R=0}) $~\citep{coles91}, so that
\begin{align}
    \xi_\Delta(r, z) &= e^\xi - 1, \\
    \xi^{(i)}_{\Delta, \alpha}(r, z) &= c_{1, \alpha}(z) \sum_R c^{(i)}_{2, \alpha}(z, R)\left(e^{\gamma_R \xi^{R, 0}}-1\right) \nonumber, \\
    \xi^{(i)}_{\Delta, X} (r, z) &= (1+z)^2\sum_{z^{\prime} \geq z} c_{1, X}(z) \sum_R c^{(i)}_{2, X}\left(e^{ \gamma_R \xi^{R, 0}}-1\right). \nonumber
\end{align}
The lognormal density model is included as a flag that can be turned on or off in {\tt Zeus21}, and will be improved upon in future versions through perturbation theory in Eulerian \citep{shaw08} or Lagrangian \citep{crocce06} space.
Densities are expected to be approximately linear at the epochs and scales of interest ($z\gtrsim 10$ and $k \lesssim 1 \, \mathrm{Mpc^{-1}}$), so this model is expected to be accurate at the $\sim 5-10\%$ level. 

We account for redshift-space distortions with a simple linear model (see~\cite{mao12, ross21, shaw23} for more sophisticated treatments). We take the linear relation,
    \begin{equation}
    \delta_v = -\mu_\mathrm{RSD}^2 \delta, 
    \begin{cases}
        &\mu_\mathrm{RSD} = 0.0, \quad \textrm{no RSD},\\
        &\mu_\mathrm{RSD} = 0.6, \quad \textrm{spherical RSD},\\
        &\mu_\mathrm{RSD} = 1.0, \quad \textrm{line-of-sight RSD},
    \end{cases}
    \end{equation}
where $\mu = k_\parallel / k$ is the line-of-sight (LoS) cosine. Spherical RSDs are the standard for simulations, whose value is chosen such that $(1+\mu)^2 \approx 1.87$. The value for the foreground-avoided RSD model was chosen such that runs of our code can properly simulate an interferometric observation outside the foreground ``wedge.'' Unless specified otherwise, we will proceed assuming spherical RSDs to match past literature.

As hinted in the previous section, the above large scale structure terms also impact the kinetic temperature of the IGM. The upper left panel of Fig.~\ref{fig:FIG10_Tk} shows that the global evolution of $T_k$ is dominated by adiabatic cooling prior to the epoch of X-ray heating. Though the average effect of adiabatic cooling is determined by the cosmic expansion rate, anisotropies in adiabatic cooling are sourced from the local matter density $\delta$, and therefore source fluctuations in $T_k$. Compton coupling to the CMB further complicates the picture, as electrons scattering off the post-recombination photon bath imprint their trace on the IGM thermal evolution. We follow the prescription in Ref.~\cite{munoz23} to find the effective cosmological adiabatic index including Compton heating, and use it to compute the $T_k$ fluctuations. We plot in Fig.~\ref{fig:FIG10_Tk} the power spectrum of the kinetic temperature across redshift and scale, including both the X-ray and adiabatic terms. In the lower left panel, the amplitude of $\Delta_{T_k}^2$ bends sharply at a redshift $z\sim22$ in the Pop II-only model and at $z\sim25$ in the fiducial model (with both stellar populations and all feedback processes). This point marks the epoch where heating by X-rays begin to dominate over adiabatic fluctuations. After X-rays become dominant, $T_k$ will roughly trace the SFRD and the adiabatic component of the temperature fluctuations will diminish. In the rightmost panel, we show the presence of acoustic features similar to those in Fig.~\ref{fig:FIG9_xa} imprinted onto the $T_k$ power spectrum from relative velocities.

    \begin{figure*}[t!]
        \centering
        \includegraphics[width=\textwidth]{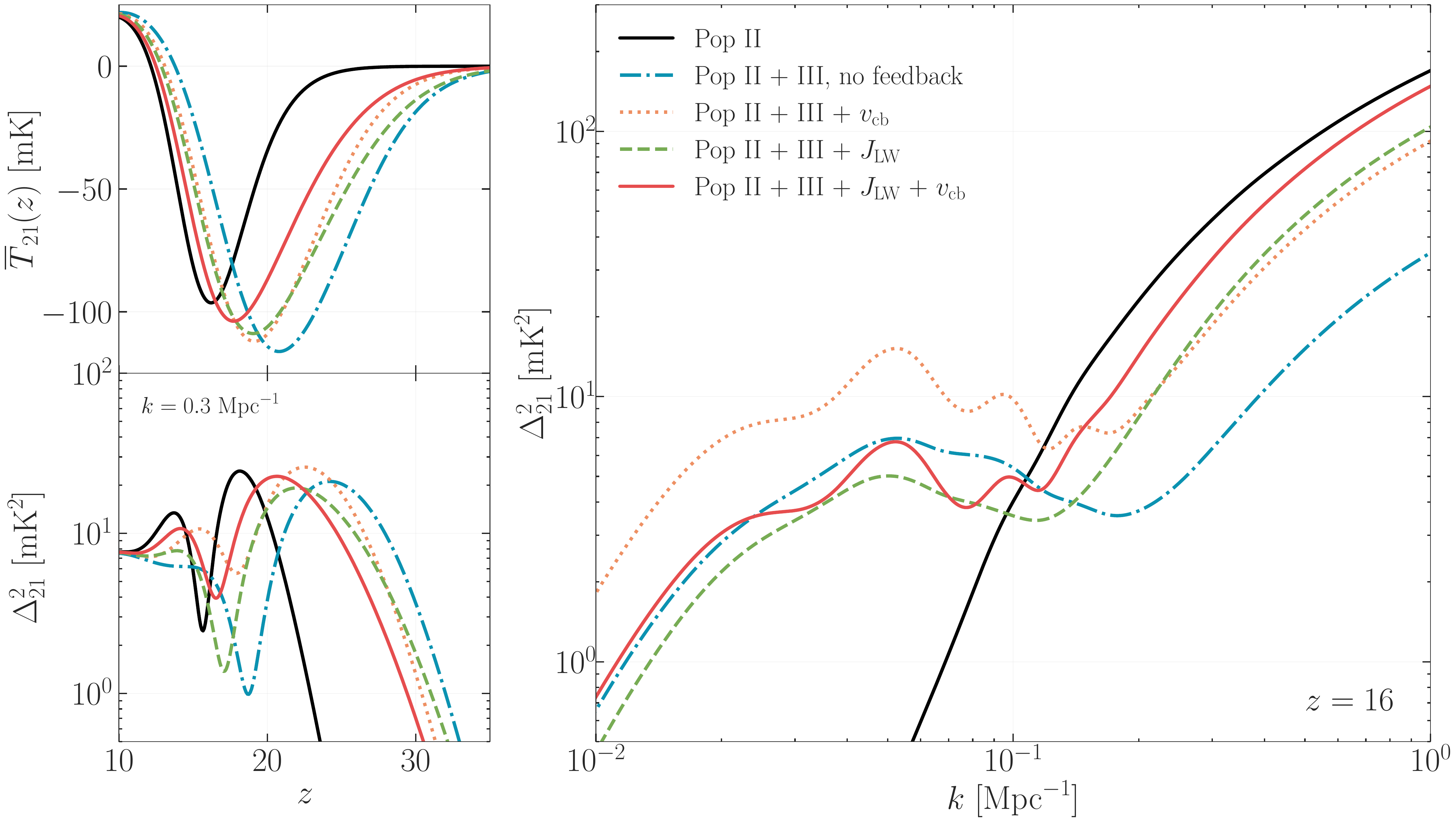}
        \caption{Plots of the 21-cm global signal and power spectrum across redshift and scale, following the same color scheme as Figs.~\ref{fig:FIG2fstar_vs_m} and \ref{fig:FIG3_sfrd_j21_vs_z}. {\bf Top left:} Global signal $\overline T_{21}$ vs.~$z$ across the five models we consider.
        Adding Pop III stars hastens the cosmic-dawn landmarks by introducing early Lyman-$\alpha$ and X-ray backgrounds, which shift the position of the 21-cm absorption trough towards higher $z$. Our fiducial model (red, with all sources of feedback) exhibits a global minimum at around $z \sim 17$.  {\bf Bottom left:} The 21-cm power spectrum at fixed $k = 0.3 \, \mathrm{Mpc}^{-1}$ across redshift, showing that adding Pop III stars spreads the characteristic shape of the Pop II only (black) power spectrum towards higher redshifts. {\bf  Right:} A plot of the 21-cm power spectrum vs.~scale $k$ at fixed redshift $z = 16$. As opposed to the Pop II only model, those including Pop III stars exhibit large-scale power and acoustic features in their spectrum imprinted by relative-velocity fluctuations.}
      \label{fig:FIG11_T21}
    \end{figure*}
    
\subsection{Reionization}
Successive generations of stars will reionize the intergalactic medium, so the cosmic neutral hydrogen fraction $x_\mathrm{HI}$ will decay from near unity at cosmic dawn to zero by $z \sim 5$ \citep{becker15}, dragging the 21-cm brightness temperature along with it \citep{furlanetto16}. 
A comprehensive analytic treatment of reionization through the  $x_\mathrm{HI}$ term in Eq.~\eqref{eq:separableT21} is beyond the scope of this work, so we will follow Ref.~\cite{munoz23} and proceed with the average evolution only, and leave anisotropies in $x_\mathrm{HI}$ (e.g., through perturbative~\citep{mcquinn18, qin22, sailer22} or excursion-set methods \citep{furlanetto04}) for future work. 

For a population of stars $i$, we calculate the evolution of the neutral fraction through the ionizing emissivity \citep{mason19},
    \begin{equation}    \dot{N}^{(i)}_{\mathrm{ion}}= \frac{N_\mathrm{ion}^{(i)}}{\rho_{b,0}}\int d M_h\left(\frac{\partial \dot{\rho}_*^{(i)}(z)}{\partial M_h}\right) f^{(i)}_{\mathrm{esc}}\left(M_h\right),
    \end{equation}
where $N_\mathrm{ion}^{(i)}$ is the population-dependent number of ionizing photons produced per stellar baryon and $\rho_{b,0}$ is the baryon mass density at the present time. Additionally, we parameterize the fraction $f_\mathrm{esc}^{(i)}$ of ionizing photons that escape from its original galaxy into the IGM as a population-dependent power law defined as
    \begin{equation}
        f_\mathrm{esc}^{(i)} = f_{\mathrm{esc},0}^{(i)} \left(\frac{M_h}{M_\mathrm{esc}^{(i)}} \right)^{\alpha_\mathrm{esc}^{(i)}},
    \end{equation}
whose parameters are defined in Table~\ref{tab:tableParams}. This simply enhances the model in {\tt Zeus21} by separating the Pop II and III contributions, with different power-law $f_{\mathrm{esc}}^{i}$ (see also \citep{park19}). \hac{As we do not yet have a model for ionized bubbles, we restrict our study to $z \gtrsim 10$ and will expand {\tt Zeus21} towards the conclusion of reionization in a forthcoming update. } Since Pop II + III stars are expected to ionize less than $1\%$ of the neutral hydrogen in the IGM by redshift $z \sim 10$, we will simply set $x_\mathrm{HI} = 1$ throughout the rest of the paper unless specified otherwise.
This will allow us to more easily compare against other results which would include $x_{\rm HI}$ fluctuations.

\section{A Full Analytical Calculation of the 21-cm Signal from the First Stars} \label{sec:VI_21cmSignal}

We have devoted previous sections to delineate the different puzzle pieces that factor into of the 21-cm brightness temperature $T_{21}$. In this section, we show how {\tt Zeus21} combines these pieces to form the 21-cm global signal and power spectrum. 

\subsection{The {\tt Zeus21} Global Signal}

We begin with the redshift evolution of the globally averaged $\overline{T}_{21}$. As a first step, we neglect anisotropies in the kinetic temperature $T_k$ and the Lyman-$\alpha$ coupling parameter $x_\alpha$, and set the local density and RSD to $\delta = \delta_v = 0$ in Eq.~\eqref{eq:separableT21}.

We plot in the top left panel of Fig.~\ref{fig:FIG11_T21} the global signals from models including Pop II or II + III stars, when considering different feedback mechanisms. Each global signal exhibits a characteristic shape marked by three phases.
First, $\overline{T}_{21}$  becomes negative from WF coupling when the first Pop III stars form. 
As they begin producing X-rays, which heat the IGM, the global signal begins to increase until it eventually shows in emission.
Eventually, $\overline{T}_{21}$ will decay as the hydrogen is ionized, at redshifts below $z \lesssim 10$ (which we remind the reader we do not model in this work).

As expected, adding two populations of stars hastens the onset of Lyman-$\alpha$ coupling, seen as a shift towards higher $z$ of the 21-cm absorption trough. LW and relative velocity feedback both inhibit the amount of WF photons, and together they yield a fiducial trough at $z \sim 17$, with a global minimum of $ \sim -110 \, \mathrm{mK}$. Compared to the Pop-II only case, whose cosmic dawn signal turns on at $z \sim 20$, our fiducial model turns on 80 million years earlier at $z \sim 30$. Both the Pop II and fiducial model turn into emission at $z\sim 12-13$, as the low-$z$ SFRD is dominated by Pop II stars (cf.~Fig.~\ref{fig:FIG3_sfrd_j21_vs_z}). It is important to note that the exact shape of the global signal is governed by our fiducial choice of X-ray efficiency $L_{40}$ and Lyman-$\alpha$ SEDs, which control the ratio of Lyman-$\alpha$ to X-ray photons. Such parameters are dependent on the assumed astrophysical characteristics of Pop III stars and can be modified by the user.

\subsection{The {\tt Zeus21} Power Spectrum}
    
Using an SFRD constructed with $\mathcal{D}$-blocks lognormal in density and $\mathcal{V}$-blocks log-$\chi^2_3$ in velocity, we computed the fluctuations of the early Lyman-$\alpha$ and X-ray backgrounds in Sec.~\ref{sec:DandVtogether}. Here, we translate these astrophysical into 21-cm fluctuations. Throughout this section, we will show results assuming spherical RSDs $\mu_\mathrm{RSD} = 0.6$ unless otherwise specified.

The statistics of 21-cm brightness temperature anisotropies can then be quantified to first order through the 21-cm \textit{power spectrum} $P_{21}(\mathbf{k}, z)$, defined through
\begin{equation}
    \left\langle\delta T_{21}(\mathbf{k},z) \delta T^{*}_{21}\left(\mathbf{k}^{\prime},z\right)\right\rangle=(2 \pi)^3 \delta_D\left(\mathbf{k}+\mathbf{k}^{\prime}\right) P_{21}(\mathbf{k},z), \nonumber
    \end{equation}
where $\delta T_{21}(\mathbf{k},z)$ is the Fourier transform of $T_{21}\left(\textbf{x},z\right) - \overline{T}_{21}\left(\textbf{x},z\right)$.
For convenience, we use the reduced 21-cm power spectrum
    \begin{equation}
    \Delta_{21}^2(\textbf{k},z)=\frac{k^3 P_{21}(\textbf{k},z)}{2 \pi^2} \, \left[\mathrm{mK}^2\right].
    \end{equation}

To linear order, one can expand spatial perturbations of $T_{21}(\mathbf{x}, z)$ as a sum of weighted terms \citep{barkana05, pritchard07}
    \begin{equation}
        \frac{\delta T_{21}(\mathbf{x}, z)}{\overline{T}_{21}(z)} = \beta_\Delta \delta \Delta + \sum_{i \in \mathrm{II,III}} \beta_\alpha \delta x_\alpha^{(i)} + \beta_X  \delta T_X^{(i)},
    \end{equation}
whose weights $\beta_q$ are defined as
    \begin{align}
        \beta_\Delta &\equiv \frac{1}{\overline{T}_{21}} \frac{\partial T_{21}}{\partial \Delta}  = 1 + \beta_{T_\mathrm{ad}} + \mu_\mathrm{RSD}^2, \\
        \beta_{T_\mathrm{ad}} &\equiv -\frac{2}{3} \frac{(1+z)^2}{D(z)} \int_z^\infty \frac{T_k(z)}{(1+z')^2} \frac{dD}{dz} \beta_X(z)\, dz', \\
        \beta_X &\equiv  \frac{1}{\overline{T}_{21}} \frac{\partial T_{21}}{\partial T_c}\frac{\partial T_c}{\partial T_k} \frac{\partial T_k}{\partial T_X} \approx \frac{T_\mathrm{cmb}}{T_X \left(T_c - T_\mathrm{cmb}\right)}, \\
        \beta_\alpha &\equiv \frac{1}{\overline{T}_{21}} \frac{\partial T_{21}}{\partial x_\alpha} = \frac{1}{\overline{x}_\alpha \left( 1 + \overline{x}_\alpha \right)},
    \end{align}
where $D(z)$ is the cosmological growth factor, and the expression for $\beta_{T_\mathrm{ad}}$ is found through the effective adiabatic index calculation in Sec. IV.2.C of \citet{munoz23}. In the last step, we approximate $\partial T_c / \partial T_k = T_c / T_k - g_\mathrm{col} T_c^2  T_k^3$ by dropping the last subtractive term.

We can then write the (configuration-space) correlation function of $T_{21}$ as
    \begin{align}
        \xi_{21}(r, z)  &= \overline{T}_{21}^2(z) \Big( \beta_\Delta^2 \xi_\Delta + \beta_\alpha^2 \xi_\alpha + \beta_X^2 \xi_X  + 2 \beta_\alpha \beta_X \xi_{\alpha X} \nonumber \\
        &+ \sum_{i \in \mathrm{II,III}} 2 \beta_\Delta \beta_\alpha \xi_{\Delta, \alpha}^{(i)} + 2 \beta_\Delta \beta_X \xi_{\Delta, X}^{(i)} \Big), 
    \end{align}
and Fourier transform to obtain $P_{21}$.

In the lower left panel of Fig.~\ref{fig:FIG11_T21}, we show the evolution of 21-cm power spectrum over redshift at $k = 0.3 \, \mathrm{Mpc}^{-1}$. The 21-cm power spectrum shows the same characteristic double-peak shape from high to low redshift, as it grows at high z when WF coupling turns on, decreases in the transition to X-ray heating, and increases yet again as the gas is heated and $T_{21}$ turns into emission. 
If we had assumed no RSDs ($\mu=0$), or LoS ($\mu=1$), the peaks and troughs in the 21-cm power specrtum would have shifted. Each model plotted in Fig.~\ref{fig:FIG11_T21} assumes different populations of stars and different feedback effects. The inclusion of III stars will spread out the characteristic double-peak shape of the 21-cm power spectrum towards higher redshifts, while the inclusion of different feedback effects will bring it closer to the Pop II case. Thus, both the $k$ and $z$ behavior of the 21-cm power spectrum trace the complex astrophysics taking place during cosmic dawn.

The rightmost panel of Fig.~\ref{fig:FIG11_T21} shows the 21-cm power spectrum against $k$ at $z=16$, which marks the beginning of Pop II dominance in the SFRD (and is around the global minimum of the Pop-II only 21-cm global signal). 
The Pop-II only model exhibits a steep rise in power towards small scales. We attribute this to the negative contribution of the cross-terms between Lyman-$\alpha$ and X-rays, which suppresses large-scale power. 
The models that include Pop III stars are, however, well within the X-ray heating phase, so they exhibit more power at large scales. Yet, they differ from each other. Going from no feedback (blue) to LW only (green), we see a suppression of power at large scales and an increase at small scales. 
This is because of the inhomogeneous distribution of LW photons, which can travel large distances $R$.
We model this through a (negative) correction to the effective biases $\gamma_R^\mathrm{III}$ in Eq.~\eqref{eq:LWcorrectionToGamma1},
which suppress large-scale SFRD fluctuations, and thus $\Delta^2_{21}$.

The Pop II + III model without feedback exhibits lower power than that which includes $v_\mathrm{cb}$ (orange). 
This is to be expected, as Pop III SFRD show increased variance when including relative velocities (see Fig.~\ref{fig:FIG7SFRDRealizations}), which leads to higher overall power on Lyman-$\alpha$ and X-ray fluctuations (through Eqs.~\ref{eq:xiAlpha2} and \ref{eq:xi_X2}). 
This added power is, moreover, in the form of acoustic oscillations, which we now describe.

    \begin{figure}[t!]
        \centering
        \includegraphics[width=\columnwidth]{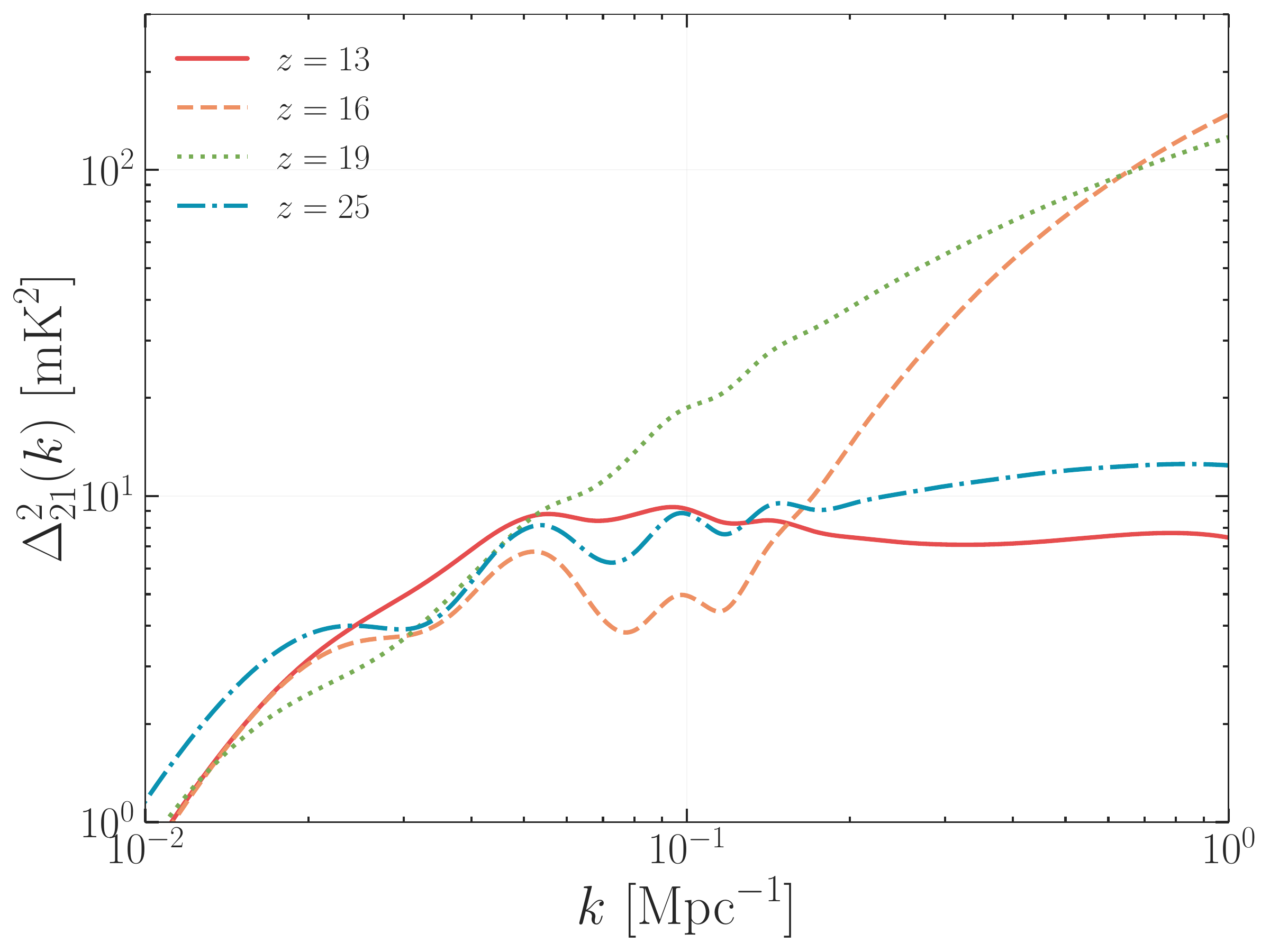}
        \caption{Evolution of the 21-cm power spectrum of our fiducial model (Pop II + III stars with both LW and $v_\mathrm{cb}$ feedback). We highlight the presence of velocity-induced acoustic oscillations (VAOs), which manifest as wiggles in $\Delta_{21}^2$ with three to four discernable peaks at high redshifts $z \gtrsim 25$ and two at intermediate redshifts $13 \lesssim z \lesssim 16$. While the position of the acoustic peaks are only dependent on cosmology, their amplitude depends on the astrophysics of  Lyman-$\alpha$ coupling or X-ray heating. }
      \label{fig:FIG12_21cmPS_differentZ}
    \end{figure}

\subsection{Velocity-Induced Acoustic Oscillations} \label{sec:VAOs}

    \begin{figure*}[t!]
        \centering
        \includegraphics[width=\textwidth]{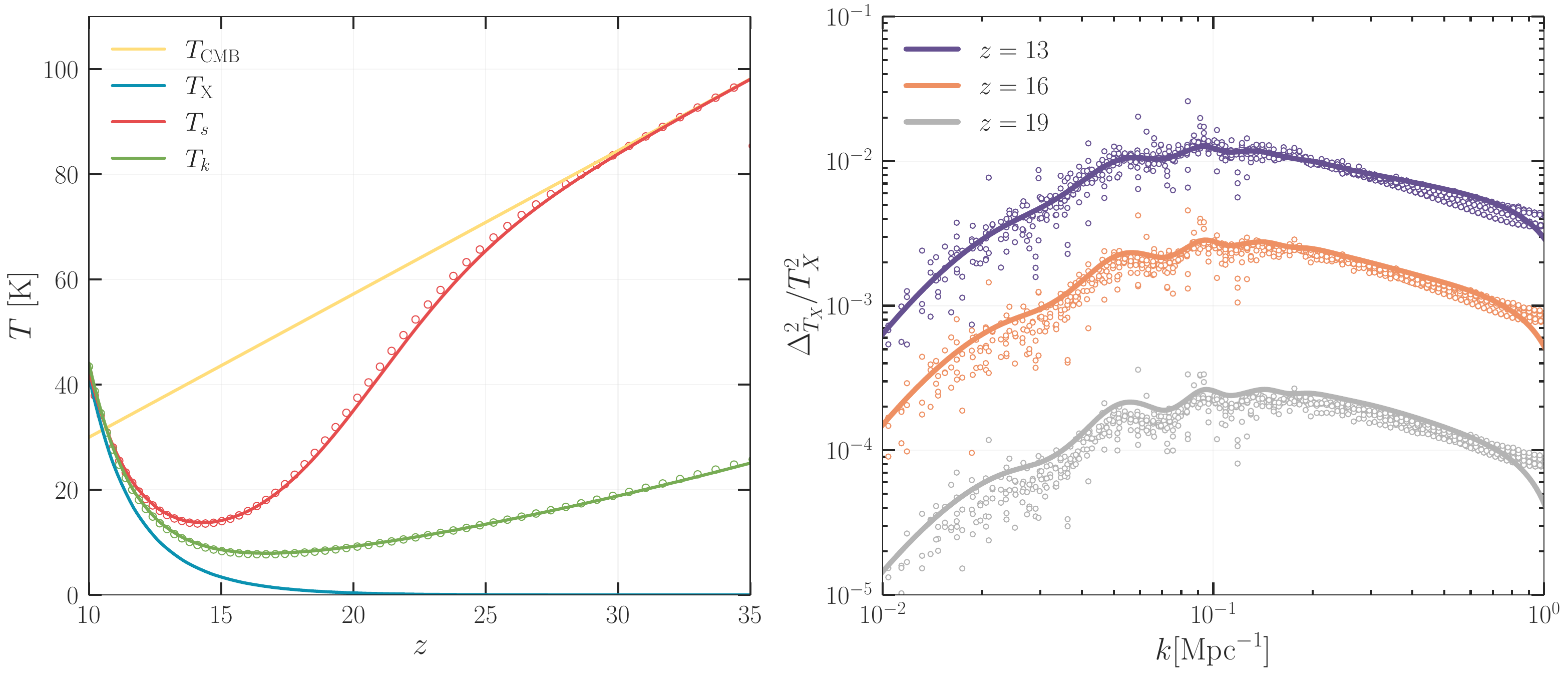}
        \caption{
        A plot of the average and anisotropies of different temperatures predicted by both {\tt Zeus21} (solid lines) and {\tt 21cmFAST} (hollow points). For the latter, we conduct nine simulations of {\tt 21cmFAST} across three different box sizes each with three sets of initial conditions, as detailed in Sec.~\ref{sec:comparisonsWith21cmFAST}. (\textbf{Left}): We show the spin temperature $T_s$ in red, the kinetic temperature $T_k$ in green, the CMB temperature $T_\mathrm{CMB}$ in yellow, and the X-ray heating contribution to the kinetic temperature $T_X$ in blue. The first two quantities are compared against the outputs from {\tt 21cmFAST}, showing remarkable agreement. \textbf{Right:} Power spectrum of the X-ray contribution $T_X$ to the kinetic temperature versus scale $k$, at redshifts $z=13$ (violet), $z=16$ (orange), and $z=19$ (gray). Our {\tt Zeus21} model agrees with the {\tt 21cmFAST} results at the $10\%$ level.}
      \label{fig:FIG13_Tk_21cf}
    \end{figure*}
    
During radiation domination, the radiation pressure on the photon-baryon fluid give rise to baryon acoustic oscillations (BAOs). These manifest as both tessellated cold spots and hot rings in the CMB temperature maps \citep{planck18}, and as overdense rings in the distribution of galaxies in the large scale structure \citep{eisenstein98}, both preferentially correlated by the sound-horizon scale $r_s$, commonly used as standard ruler. Because BAOs also source the relative velocities that suppress star formation, the fluctuations in $v_\mathrm{cb}$ are also correlated by $r_s$ due to their acoustic origin \citep{tseliakhovich10}. These acoustic features manifest as wiggles in the $v_\mathrm{cb}$ power spectrum; which is inherited by the SFRD and early radiation backgrounds, giving rise to velocity-induced acoustic oscillations (VAOs) in the 21-cm power spectrum. \citet{munoz19b} found the position of these peaks to be an astrophysics-robust standard ruler at cosmic dawn, which can be used to infer the cosmic expansion rate and geometry. 

In Fig.~\ref{fig:FIG12_21cmPS_differentZ} we plot $\Delta_{21}^2$ against $k$ at a few redshifts. The power spectra show marked acoustic oscillations across scales $0.05 \, \mathrm{Mpc}^{-1} \lesssim k \lesssim 0.5 \, \mathrm{Mpc}^{-1}$ inherited from the $v_\mathrm{cb}$ fluctuations. At the specific redshifts $z=13, 16,$ and 19, the VAOs increase the power by $\mathcal{O}(1)$, showing their inclusion is critical to modeling Pop III star formation~\cite{munoz22}. At $z = 19$, the VAOs are difficult to distinguish by eye. This redshift is near the transition from the Lyman-$\alpha$ coupling and X-ray heating epochs. The large-scale power (both density- and velocity-sourced) cancels in this intermediate period, as the relative velocity effect on UV and X-rays have counteracting effects on the 21-cm line. During the coupling epoch, areas of larger $v_\mathrm{cb}$ will produce fewer Lyman-$\alpha$ photons, leading to weaker coupling and a more positive $T_{21}$. In contrast, relative velocities will produce fewer X-ray photons, which will lead to less heating at certain scales and contribute a more negative $T_{21}$. These two components must cancel at some redshift, which occurs near $z \sim 19$ for our fiducial model.
    
We can discern four VAO peaks at early times ($z \gtrsim 25$), while only two are distinguishable later on (at $z \sim 13 - 16$). During the epoch of Lyman-$\alpha$ coupling, Lyman-band photons travel distances around $\lambda_\mathrm{mfp} \sim 10^2 \, \mathrm{Mpc}$ \citep{dalal10}, while X-rays travel $\lambda_\mathrm{MFP} \sim 10^0 - 10^1 \, \mathrm{Mpc}$ by the middle of the X-ray heating epoch \citep{mcquinn12, pacucci14, das17}. The more cosmic time has passed, the longer distances these photons travel, and the more washed out the VAO signal becomes. At the end of the Lyman-$\alpha$ coupling epoch at $z \sim 19$, the long distances traveled by Lyman-series photons results in dampened VAOs. A similar effect occurs at the end of the X-ray heating epoch at $z\sim 10$, from the long distances traveled by X-ray photons. While only Lyman-series photons render four noticeable VAO peaks at $z \gtrsim 25$, the combined impact of Lyman-series and X-ray photons results in only two noticeable peaks at $13 \lesssim z \lesssim 16$.

Prior to our study, VAO modeling of the 21-cm power spectrum relied on either perturbation theory~\cite{dalal10,alihamoud11} or semi-numerical simulations \citep{visbal12,munoz19b}. With {\tt Zeus21}, we achieve the first nonlinear fully analytical computation of VAOs in the 21-cm power spectrum. As advanced in Fig.~\ref{fig:FIG8_PS_SFRD}, the linear approach of Ref.~\cite{dalal10} does not fully capture the shape of the VAOs. Specifically, the nonlinear approach estimates a factor of $75-100\%$ more VAO power at $k \gtrsim 0.3 \, \mathrm{Mpc}^{-1}$. While galaxy formation and stellar properties determine the amplitude of the 21-cm power spectrum, and the prominence of the velocity-induced acoustic peaks, cosmology sets the characteristic correlation scale of velocities, and therefore the position of the peaks in $k$. Thus, the position of the VAO peaks are still a robust standard ruler independent of the complex astrophysics at play during cosmic dawn.

    \begin{figure*}[t!]
        \centering
        \includegraphics[width=\textwidth]{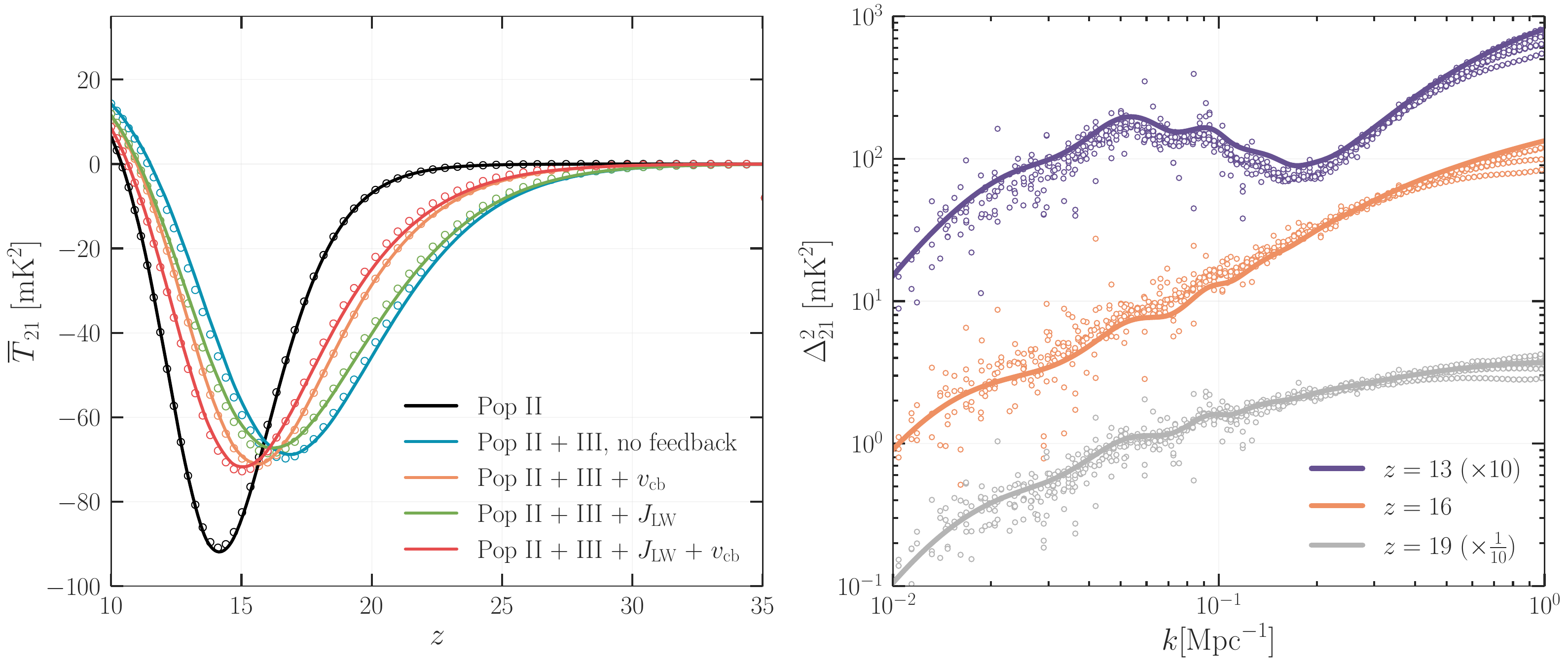}
        \caption{Comparison of the {\tt 21cmFAST} (empty circles) and {\tt Zeus21} (solid lines) 21-cm predictions. \textbf{Left}: Average $\overline{T}_{21}$ versus $z$ for all the population and feedback models we consider: Pop II only (black), Pop II + III without feedback (blue), which moves the signal to higher $z$ and makes it shallower; with$v_\mathrm{cb}$ and LW feedback (orange and green, respectively), and our fiducial Pop II + III + all feedback model  (red). The {\tt Zeus21} prediction matches that of {\tt 21cmFAST} to high fidelity. \textbf{Right:} We show the 21-cm power spectrum across scale $k$ at three different redshifts $z = 13, 16,$ and $ 19$ (violet, orange, and gray, respectively, rescaled for visual clarity). 
        The modeled {\tt Zeus21} power spectra agree with the {\tt 21cmFAST} outputs within $\sim 7\%$. The slight $10\%$ departure at $z = 13$ close to the inflection scale was already present in the Pop II-only model of Ref.~\cite{munoz23}, as we discuss at the end of Sec.~\ref{sec:comparisonsWith21cmFAST}. \hac{Fortunately, these departures remain comfortably within the $\sim 20\%$ theoretical error expected of semi-numerical simulations \citep{park19}.}}
      \label{fig:FIG14_T21_21cf}
    \end{figure*}

\subsection{Comparisons with {\tt 21cmFAST}} \label{sec:comparisonsWith21cmFAST}

As a cross-check of our model, we compare our results against the output of the well-known semi-numerical simulations from {\tt 21cmFAST} \citep{mesinger11, munoz22}. To this end, we compare our results with nine sets of {\tt 21cmFAST} simulations, employing three different resolutions and three different seed values (and therefore different initial conditions.) We run one set of three simulations within a $1350 \, \mathrm{Mpc}$ box with a $3\, \mathrm{Mpc}$ resolution, one set within a $750\, \mathrm{Mpc}$ box with a $2\, \mathrm{Mpc}$ resolution, and one set within a $300\, \mathrm{Mpc}$ box with a $1\, \mathrm{Mpc}$ resolution. While our treatment of the X-ray and Lyman-$\alpha$ backgrounds in {\tt Zeus21} is similar to the excursion-set approach employed in {\tt 21cmFAST}, the parameterization of the SFRDs in terms of the underlying cosmology is fairly different. To make meaningful comparisons, we implement the same astrophysical model within {\tt Zeus21} that the user can activate with the {\tt Flag\_emulate\_21cmfast} flag. Switching on this flag includes fixing $x_e = 2 \times 10^{-4}$ as both codes use different methods to model the leftover free electrons from recombination. \hac{Additionally, since {\tt Zeus21} is valid at redshifts above $z \gtrsim 10$, we turn off reionization by setting $f_\mathrm{esc} = 0$ for both stellar populations and neglect photo-heating feedback \citep{sobacchi14} in both codes.} We set $N_\mathrm{LW}^\mathrm{II} = 3030$ and $N_\mathrm{LW}^\mathrm{III} = 560$ from the Lyman-series SEDs employed by {\tt 21cmFAST.} Lastly, we turn off RSDs and evolve densities linearly to best match the EPS algorithm used to simulate SFRDs in {\tt 21cmFAST}. We refer the interested reader to Table~\ref{tab:tableParams} for the full range of parameters used to compare both codes, and Secs. 6.1 and 6.2 in Ref.~\cite{munoz23} for the detailed prescription used to mimic {\tt 21cmFAST} within {\tt Zeus21}.

We first compare the properties of the IGM across the two codes, which we show in Fig.~\ref{fig:FIG13_Tk_21cf}. The average {\tt 21cmFAST} spin and kinetic temperatures ($T_s$ and $T_k$) are nearly identical to that of {\tt Zeus21}, showing the two codes agree on the overall evolution of the IGM. We also compare the kinetic-temperature power spectra $\Delta^2_{T_k}$ at three redshifts of interest: at $z = 16$ during the peak of the coupling era and close to the global minimum of the $\overline{T}_21$ absorption signal, at $z = 13$ during the ramp-up of X-ray heating, and at $z = 10$ when the 21-cm signal is expected to be in emission. Our nonlinear treatment of X-rays successfully captures the behavior of the kinetic temperature power spectrum fluctuations across redshifts and scales. 
    
Finally, we show our comparison of the predicted 21-cm global signal and power spectrum in Fig.~\ref{fig:FIG14_T21_21cf}. Both {\tt Zeus21} and {\tt 21cmFAST} show remarkably similar global behavior, with an absorption trough at $z \sim 15$ and a fully heated IGM by $z \sim 10$. We plot $\Delta_{21}^2(k)$ across scale $k$ for $z=13, 16, 19$, corresponding to the heating, minimum absorption trough, and Lyman-$\alpha$ coupling epochs. The output of the two codes agree at the $7\%$ level throughout these redshifts. 
In all cases there seems to be a slight downturn in the simulated power at $k\gtrsim 0.6$ Mpc$^{-1}$, which we find to be a resolution-dependent artifact of the realization boxes. Across scales both large and small, the {\tt Zeus21} prediction agree well within the expected theoretical uncertainty of {\tt 21cmFAST}~\cite{zahn11}.  
There is a slightly deviation at $z \approx 13$, where the power in the simulations is $10\%$ smaller than that predicted by our analytic code around $k \sim 0,2 \, \mathrm{Mpc}^{-1}$. 
This may be due to the $k$-binning in simulations, our simplified assumptions for the Lyman-series SED, or perhaps to some other assumption in {\tt 21cmFAST} that may not have been translated into {\tt Zeus21} (for instance, ignoring the initial adiabatic temperature fluctuations~\cite{munoz23}). 
We note that this higher $z\sim 13$ power in {\tt Zeus21} than {\tt 21cmFAST} was already present in the Pop II-only comparison of Ref.~\cite{munoz23}, so it is likely not an artifact of our Pop III modeling. \hac{Fortunately, these departures still remain within the $\sim 20\%$ theoretical error ascribed to semi-numerical simulations \citep{park19, mason23b, hutter23}. Despite these differences, recent analyses using {\tt\string Zeus21} with only Pop II stars \citep{verwohlt2024} recovered highly similar results of previous parameter inferences employing {\tt\string 21cmFAST}. As upcoming data arrives, we expect that improved astrophysical modeling will mitigate these discrepancies.}

In summary, {\tt Zeus21} recovers the nonlinearities and nonlocalities in the 21-cm fluctuations at the level of $10\%$ precision with {\tt 21cmFAST} across cosmic dawn. A subtle point is that in {\tt 21cmFAST} the feedback threshold $M_\mathrm{mol}(z)$ in Eq.~\eqref{eq:Mmol} is not evolved backwards when computing the Pop III SFRD. As such, {\tt 21cmFAST} slightly overpredicts the amount of feedback by using the lower-$z$ $M_\mathrm{mol}$, influenced by a larger LW flux. We have matched this assumption in {\tt Zeus21} when the {\tt Flag\_emulate\_21cmfast} flag is activated, but we show in Appendix~\ref{sec:acausalMmol} how it affects the amplitude of the 21-cm power spectrum up to a factor of 20\%.

\section{Discussion and Conclusions} \label{sec:VII_conclusions}

In this work, we present a fast, fully analytical, effective model for the earliest epochs of star formation. We incorporate the effects of successive populations of stars, including the inhomogeneous feedback that affects their formation, and analytically model the 21-cm signal (both global and fluctuations) resulting from Pop II + III stars  during cosmic dawn. 
The basis of our approach rests in a decomposition of early star formation rate densities (SFRDs) into their population II and III constituents, each of which are further decomposed into their dependence on the local density $\delta$ and relative velocity $v_{\rm cb}$. We show that, to an excellent approximation, the SFRD of Pop III stars traces the underlying density field as an exponential in smoothed fractional density $\delta_R$, and is suppressed as the sum of two exponentials in relative velocity $v_{\mathrm{cb},R}^2$. Thus, the SFRD can be treated as a lognormal random field in density and a log-$\chi^2_3$ random field in relative velocity. This parameterization allows us to analytically find the anisotropies of the SFRD, and thus of the Lyman-$\alpha$ ($J_\alpha$) and X-ray ($J_X$) fluxes, which determine the 21-cm signal.

In our model, molecular-cooling galaxies (MCGs) are the hosts of Pop III stars, and atomic-cooling galaxies form the metal-enriched Pop II. We find that including MCGs with Pop III stars results in earlier cosmic-dawn landmarks, including a shift in the 21-cm absorption signal towards higher redshifts. 
Lyman-Werner (LW) feedback, which halts star formation in MCGs, brings the signal back towards later times, and moreover yields a more homogeneous SFRD.
In our formalism this can be encoded as a shallower dependence of SFRD against density, or a negative correction to the effective biases $\gamma_R^\mathrm{III}$ of Eqs.~\ref{eq:LWcorrectionToGamma1} and \ref{eq:LWcorrectionToGamma2}. 
Given that LW photons propagate significant distances ($R\sim 100$ Mpc) during cosmic dawn, this correction gives rise to a new scale dependence in the effective biases $\gamma_R^\mathrm{III}$, and thus on the 21-cm signal at cosmic dawn.

Confirming previous work, we find that relative-velocity feedback produces an acoustic modulation on 21-cm power spectra.
We perform a fully analytic recalculation of the shape of these so-called velocity-induced acoustic oscillations (VAOs) using our log-$\chi^2_3$ approximation, and find that past analytic work (which assumed a linear approximation) misestimated the VAO amplitude --- and scale dependence (see Fig.~\ref{fig:FIG8_PS_SFRD}).
Of course, the peak structure of the VAOs is not altered, as previously shown in simulations, so they remain a robust standard ruler to cosmic dawn.

Our effective model can predict the 21-cm global signal and power spectrum from the combined impact of both stellar populations. 
We compare the {\tt Zeus21} results against {\tt 21cmFAST}, finding excellent agreement, with deviations in power spectra across all redshifts and scales at the $\sim10\%$ level. Agreement beyond this level is challenging, given the inherent limitations of analytic and semi-numeric models, and would require a dedicated code-comparison project. Unlike {\tt 21cmFAST}, however, a single execution of {\tt Zeus21} runs on the order of seconds on a laptop, three orders of magnitude faster than a typical simulation run on an external computing cluster. That is because {\tt Zeus21} does not create and evolve realizations of the 21-cm brightness temperature, and related quantities, but instead relies on our analytic model to compute observables. While we do not yet compute higher order summary statistics beyond the power spectrum, higher-order correlations can be easily written in terms of two-point functions with our formalism lognormal in density and log-$\chi^2$ in relative velocity \citep{coles91, xavier16}. Provided with redshift probability distributions, cross-correlations with galaxies or other tracers of large scale structure can also be easily computed \citep{bellomo2020, bernal2020}, which we leave for future work. In this new version, {\tt Zeus21} presents a comprehensive and computationally economical calculation of the 21-cm signal, inclusive of the nonlinearities of the local density and velocity fields and the nonlocalities of early radiation backgrounds. 

In sum, here we present an upgrade to {\tt Zeus21}, a public python package that now includes our effective model for Pop III stars and their 21-cm signal. {\tt Zeus21} interfaces with the Boltzmann solver {\tt CLASS}, allowing the user to efficiently test changes to cosmology and astrophysics, including different dark matter models, parameterizations of the halo-galaxy connection, as well as alternative UV and X-ray galaxy SEDs. Our updated {\tt Zeus21} can predict the 21-cm signal in agreement with state-of-the-art simulations at $z\gtrsim 10$ in mere seconds from an ordinary laptop with negligible memory requirements. This takes us a meaningful step closer towards fast and accurate 21-cm data analysis, which is critical to interpret the flurry of upcoming observations and allows us to  compute the first billion years in seconds.

\acknowledgments

We thank Volker Bromm, Jordan Flitter, Steven Furlanetto, Nick Kokron, Andrei Mesinger, Jordan Mirocha, and Steven Murray for useful discussions. HAC was supported by the National Science Foundation Graduate Research Fellowship under Grant No.\ DGE2139757. 
This work was supported at UT Austin by NSF Grants AST-2307354 and AST-2408637, and at Johns Hopkins by NSF Grant No.\ 2112699, the Simons Foundation, and the Templeton Foundation. NS was supported by a Horizon Fellowship from Johns Hopkins University. Part of this work was carried out at the Advanced Research Computing at Hopkins (ARCH) core facility  (arch.jhu.edu), which is supported by the National Science Foundation (NSF) grant number OAC1920103.
The authors thank Nordita for their hospitality during part of this work.



\appendix

\section{Correlation Function of the $\mathcal{V}$ Building Block} \label{sec:logchisquareCorrFunc}

In Sec.~\ref{sec:effectiveLNCHImodel}, we describe the suppressive effects of relative velocities on the SFRD with a fit using the sum of two exponentials in $\eta$. In this Appendix, we find the functional connection between our model of SFRD anisotropies to velocity anisotropies. To resummarize, we found in Eq.~\eqref{eq:velBuildingBlock} that the Pop III SFRD suppression from velocities is best modeled by
    \begin{equation*}
        \mathcal{V}\left( z | v_\mathrm{cb}^2 \right) \approx \frac{\Lambda_R e^{-\lambda_R \tilde{\eta}_R} + \Omega_R e^{-\omega_R \tilde{\eta}_R}}{\Lambda_R(1+2\lambda_R)^{-3/2} + \Omega_R(1+2\omega_R)^{-3/2}},
    \end{equation*}
where $\Lambda_R, \Omega_R, \lambda_R, \omega_R$ are all fitted constants. We will first find the correlation function between two fields described by single exponentials,
    \begin{align*}
        U_1(\mathbf{r}) &= \Lambda_{R_1} \exp \left(-\lambda_{R_1} \tilde{\eta}_R(\mathbf{r}) \right), \\
        U_2(\mathbf{r}) &= \Lambda_{R_2} \exp \left(-\lambda_{R_2} \tilde{\eta}_R(\mathbf{r}) \right),
    \end{align*}
whose correlation function is $\xi_U(r)$. Later, we will derive the full expression of correlations of the sums of two exponentials. The task at hand simplifies then to find $\xi_{U}(r)$ in terms of $\xi_{\eta}(r)$, and then write the correlation function of the velocity building block $\xi_\mathcal{V}(r)$ in terms of $\xi_{U}(r)$.

We will make frequent use of the \citet{Peebles:1980yev} convention in Eq.~\eqref{eq:peeblesCorrFunc1} defining the two-point correlation function of homogenous and isotropic random fields $U_1(\mathbf{r})$ and $U_2(\mathbf{r})$ as simply the spatial covariance between both fields normalized by their means. Before we proceed with the derivation, we note two important remarks. Firstly, by consequence of Eq.~\eqref{eq:peeblesCorrFunc1}, we note that variables that differ by multiplicative and subtractive constants exhibit identical correlation functions (i.e. $\xi_{\eta}(r) = \xi_{\tilde{\eta}}(r)$). Secondly, as was the case with densities, computations of the $\dot{\rho}_*(z|\delta_R, v_{\mathrm{cb},R}^2)$ are computed using shells of the smoothed relative velocity field whose shell radius and smoothing radius are the same $R$. Thus, we can treat correlations between smoothed velocity fields $v^2_{\mathrm{cb},R_1}$ and $v^2_{\mathrm{cb},R_2}$ as functionally identical to convolving the velocity power spectrum $P_\eta(k)$ directly with window functions $W_{R_1}(k)$ and $W_{R_2}(k)$. For notational simplicity, we drop subscripts specifying smoothing radii until the end.

\subsection{Correlations of Single Exponentials}

Our first task is to express the correlation function of exponentiated velocities $\xi_{U}(r)$ in terms of correlations of the velocities themselves $\xi_{\eta}(r)$. Doing so requires an exercise into the statistics behind the velocity field itself.

\subsubsection{Correlations of $\chi^2_3$ Random Variables}
For simplicity, consider the case of correlated random variables. Let $v_1 = (v_{1x}, v_{1y}, v_{1z})$ and $v_2 = (v_{2x}, v_{2y}, v_{2z})$ be two three-dimensional vectors whose elements are Gaussian random variables with zero mean, shared variance (i.e. $v_{ij} \sim \mathcal{N}(0, \sigma_i^2)$), and identical covariance (i.e. $\langle v_{1j} v_{2j'}\rangle = \rho \sigma_1 \sigma_2$.) The variables $\tilde{\eta}_1 = 3v_1^2 / \sigma_1^2$ and $\tilde{\eta}_2 = 3v_2^2 / \sigma_2^2$, where $v_i^2 = v_{ix}^2 + v_{iy}^2 + v_{iz}^2$, are therefore correlated $\chi^2_{n=3}$ variables with three degrees of freedom. The joint probability distribution function (PDF) of $\tilde{\eta}_1$ and $\tilde{\eta}_2$ with $n$ degrees of freedom is (see Appendix A of \citep{givans23}):
    \begin{equation} \label{eq:chi2pdf}
    \begin{aligned}
    p\left(\tilde{\eta}_1, \tilde{\eta}_2\right)= & \frac{\left(1-\rho^2\right)^{-\frac{n}{2}}\left(\tilde{\eta}_1 \tilde{\eta}_2\right)^{\frac{n-2}{2}}}{2^n \sqrt{\pi} \Gamma\left(\frac{n}{2}\right) \Gamma\left(\frac{n-1}{2}\right)} \exp \left(-\frac{\tilde{\eta}_1+\tilde{\eta}_2}{2\left(1-\rho^2\right)}\right) \\
    & \times \int_{-1}^1\left(1-\mu^2\right)^{(n-3) / 2} \exp \left(\mu \frac{\rho  \sqrt{\tilde{\eta}_1 \tilde{\eta}_2}}{1-\rho^2}\right) d \mu .
    \end{aligned}
    \end{equation}
Note that the integral over $\mu$ can also be recast into
    \begin{equation}
        \int_{-1}^1 \left(1 - \mu^2\right)^{\frac{n-3}{2}} e^{\mu \alpha} d\mu = \sqrt{\pi} \frac{\Gamma\left(\frac{n-1}{2}\right)}{\Gamma\left(\frac{n}{2}\right)}{}_0F_1\left[;\frac{n}{2}, \frac{\alpha^2}{4}\right],
    \end{equation}
in terms of a confluent hypergeometric function, defined as
    \begin{equation}
        {}_0F_1[;a, z] = \sum_{k=0}^{\infty} \frac{1}{(a)_k} \frac{z^k}{k!}, \qquad (a)_k = \frac{\Gamma(a+k)}{\Gamma(a)},
    \end{equation}
which can facilitate numerical evaluation when using Mathematica. The first moment of $\tilde{\eta}_1$ and $\tilde{\eta}_2$ is given by
    \begin{equation}
        \langle \tilde{\eta}_i \rangle \equiv \iint \tilde{\eta}_i p(\tilde{\eta}_1,\tilde{\eta}_2) \ d\tilde{\eta}_1 d\tilde{\eta}_2 = n = 3,
    \end{equation}
and their mean-normalized covariance using the convention in Eq.~\eqref{eq:peeblesCorrFunc1} is given by
    \begin{align}
        \xi_{\eta} &\equiv \frac{\big\langle (\tilde{\eta}_1 - \langle \tilde{\eta}_1 \rangle )  (\tilde{\eta}_2 - \langle \tilde{\eta}_2 \rangle )\big\rangle}{\langle \tilde{\eta}_1 \rangle\langle \tilde{\eta}_2 \rangle} \nonumber \\
        &= \frac{\langle \tilde{\eta}_1 \tilde{\eta}_2\rangle - \langle \tilde{\eta}_1 \rangle \langle \tilde{\eta}_2 \rangle}{\langle \tilde{\eta}_1 \rangle \langle \tilde{\eta}_2 \rangle} \nonumber \\
        &= \frac{\iint \tilde{\eta}_1 \tilde{\eta}_2 \ p(\tilde{\eta}_1\tilde{\eta}_2) \ d\tilde{\eta}_1 d\tilde{\eta}_2}{n^2} - 1 \nonumber \\
        &= \frac{n(n+2\rho^2)}{n^2} - 1 = \frac{2}{3}\rho^2.
    \end{align}
Now, we define two random variables
\begin{equation}
    U_1 = \Lambda_1 e^{\lambda_1 \tilde{\eta}_1}, \qquad U_2 = \Lambda_2 e^{\lambda_2 \tilde{\eta}_2},
\end{equation}
where $\Lambda_1, \Lambda_2, \lambda_1, \lambda_2$ are constants. Their first moments are given by
\begin{equation}
    \left\langle U_i \right\rangle  \equiv \iint U_i p(\tilde{\eta}_1,\tilde{\eta}_2) \ d\tilde{\eta}_1 d\tilde{\eta}_2 = \left(1+2\lambda_i\right)^{-\frac{n}{2}},
\end{equation}
and their mean-normalized covariance is given by
\begin{align} \label{eq:corrFuncDerivation}
    \xi_{U} &\equiv \frac{\big\langle (U_1- \langle U_1 \rangle )  (U_2 - \langle U_2 \rangle )\big\rangle}{\langle U_1 \rangle\langle U_2 \rangle} \nonumber \\
    &= \frac{\big\langle U_1 U_2\big\rangle - \left\langle U_1\right\rangle\left\langle U_2\right\rangle}{\left\langle U_1\right\rangle\left\langle U_2\right\rangle} \nonumber \\
    &= \frac{\iint U_1 U_2 \ p(\tilde{\eta}_1,\tilde{\eta}_2) \ d\tilde{\eta}_1 d\tilde{\eta}_2}{\left(1+2\lambda_1\right)^{-n/2}\left(1+2\lambda_2\right)^{-n/2}} - 1 \nonumber \\
    &= \frac{1}{\left(1 - \frac{2n\lambda_1\lambda_2}{(1+2\lambda_1)(1+2\lambda_2)} \frac{2\rho^2}{n}\right)^{\frac{n}{2}}} - 1 \nonumber \\
    &= \frac{1}{\left(1 - \frac{6\lambda_1\lambda_2}{(1+2\lambda_1)(1+2\lambda_2)} \xi_{\eta}\right)^{\frac{3}{2}}} - 1.
\end{align}
The last expression yields the desired relationship between correlation functions.

\subsubsection{Correlations of $\chi^2_3$ Random Fields}

Little modification is needed when generalizing this result from random variables to random fields. The CDM-baryon relative velocity field $\mathbf{v_\mathrm{cb}}(\mathbf{r}) = (v_x(\mathbf{r}), v_y(\mathbf{r}), v_z(\mathbf{r}))$ is a three-dimensional vector field whose magnitude squared exhibits a spatial average of $\langle v_\mathrm{cb}^2(\mathbf{r}) \rangle = \sigma^2_\mathrm{cb}$. Each spatial component of the velocity is a Gaussian random field with zero mean and equipartitioned variance, i.e. $v_i(\mathbf{r}) \sim \mathcal{N}(0, \sigma_\mathrm{cb}^2 / 3)$. The quantity $\tilde{\eta}(\mathbf{r}) = 3 v_\mathrm{cb}^2/\sigma_\mathrm{cb}^2 $ is therefore a $\chi^2_{3}$ random field with three degrees of freedom. 

One can proceed with the same exercises as in the previous subsection, with the amendment that $U_1, U_2 \rightarrow U(\mathbf{r}_1), U(\mathbf{r}_2)$ become random fields dependent on single exponentials in $\tilde{\eta}(\mathbf{r})$, and that Eq.~\eqref{eq:chi2pdf} would become a joint-probability distribution functional of the form
    \begin{align}
        \mathcal{P}\left[\tilde{\eta}(\mathbf{r}_1),\tilde{\eta}(\mathbf{r}_2) \right] &= \frac{ (\tilde{\eta}(\mathbf{r}_1) \tilde{\eta}(\mathbf{r}_2))^{\frac{n-2}{2}}}{2^n \Gamma^2\left( \frac{n}{2}\right)\left(1-\rho^2\right)^{\frac{n}{2}}}  \exp \left(-\frac{\tilde{\eta}(\mathbf{r}_1)+\tilde{\eta}(\mathbf{r}_2)}{2(1- \rho^2)}\right) \nonumber \\
        &\times { }_0 F_1 \left[;\frac{n}{2} ; \frac{\rho^2 \tilde{\eta}(\mathbf{r}_1) \tilde{\eta}(\mathbf{r}_2)}{4\left(1- \rho^2\right)^2}\right]. 
    \end{align}
Moments and covariances of field quantities $q(\mathbf{r})$ would then be computed by a path integral over all field values $\tilde{\eta}(\mathbf{r})$, such that 
\begin{equation}
    \xi_q(r) = \frac{\iint q(\mathbf{r}_1)q(\mathbf{r}_2) \mathcal{P}\left[\tilde{\eta}(\mathbf{r}_1),\tilde{\eta}(\mathbf{r}_2) \right]  \ \mathcal{D}\tilde{\eta}(\mathbf{r}_1) \mathcal{D}\tilde{\eta}(\mathbf{r}_2)}{\left\langle q(\mathbf{r}_1)\right\rangle\left\langle q(\mathbf{r}_2)\right\rangle} - 1.
\end{equation}
The above prescription yields the same functional form as Eq.~\eqref{eq:corrFuncDerivation} despite the necessary modifications for fields. Thus, the spatial two-point correlation function for two log-$\chi^2_3$-distributed random fields is given by
\begin{equation} \label{eq:xi_expeta}
    \xi_{U}(r, \lambda_1, \lambda_2) = \frac{1}{\left(1 - \frac{6\lambda_1\lambda_2}{(1+2\lambda_1)(1+2\lambda_2)} \xi^{R_1, R_2}_{\eta}(r)\right)^{\frac{3}{2}}} - 1.
\end{equation}

\subsection{Full Expression for $\xi_\mathcal{V}$}\label{sec:correlationsInV}

Given that the velocity building block denoted in Eq.~\eqref{eq:velBuildingBlock} outlines $\mathcal{V}(z | v_{\mathrm{cb},R}^2)$ as the sum of two exponential terms in $\tilde{\eta}$, correlations in $\mathcal{V}$ are simply---but quite verbosely---defined as the sum of four terms each proportional to $\xi_U$ correlation function as in Eq.~\eqref{eq:xi_expeta}. Expanding the terms in Eq. \eqref{eq:xi_velBlock} yields the full expression of the two-point function of the velocity building block:
    \begin{align} \label{eq:fullVCorrelation}
        \xi_{\mathcal{V}}(r) &= \frac{1}{\Lambda_1(1+2\lambda_{R_1})^{-\frac{3}{2}} + \Omega_1(1+2\omega_{R_1})^{-\frac{3}{2}}}  \\
        & \times \frac{1}{\Lambda_2(1+2\lambda_{R_2})^{-\frac{3}{2}} + \Omega_2(1+2\omega_{R_2})^{-\frac{3}{2}}} \nonumber \\
        & \times \Big[ \Lambda_1 \Lambda_2 (1+2\lambda_{R_1})^{-\frac{3}{2}} (1+2\lambda_{R_2})^{-\frac{3}{2}} \xi_{U}(r, \lambda_{R_1}, \lambda_{R_2}) \nonumber \\
        & +  \Lambda_1 \Omega_2 (1+2\lambda_{R_1})^{-\frac{3}{2}} (1+2\omega_{R_2})^{-\frac{3}{2}} \xi_{U}(r, \lambda_{R_1}, \omega_{R_2}) \nonumber \\
        & + \Omega_1 \Lambda_2 (1+2\omega_{R_1})^{-\frac{3}{2}} (1+2\lambda_{R_2})^{-\frac{3}{2}} \xi_{U}(r, \omega_{R_1}, \lambda_{R_2}) \nonumber \\
        &+ \Omega_1 \Omega_2 (1+2\omega_{R_1})^{-\frac{3}{2}} (1+2\omega_{R_2})^{-\frac{3}{2}} \xi_{U}(r, \omega_{R_1}, \omega_{R_2}) \Big] \nonumber.
    \end{align}

\section{Correlations of the Total Pop II + III SFRD} \label{sec:apx_sfrdCorrelation}

As described in Sec.~\ref{sec:DandVtogether}, the full perturbed SFRD of a shell centered at redshift $z$ with shell and smoothing radius $R$ with both Pop. II and III stars is given by
\begin{align} \label{eq:sfrdFullAppendix}
    \dot{\rho}_*(z| \delta_R, v_{\mathrm{cb}, R}^2) &= \dot{\bar{\rho}}_*^\mathrm{II}(z) \mathcal{D}^\mathrm{II} \left(z|\delta_R\right)\\
    &+ \dot{\bar{\rho}}_*^\mathrm{III}(z) \mathcal{V} \left(z|v_{\mathrm{cb}, R}^2\right) \mathcal{D}^\mathrm{III} \left(z|\delta_R\right) , \nonumber
\end{align} 
where both the density and velocity building blocks $\langle \mathcal{D}^\mathrm{II} \rangle = \langle \mathcal{D}^\mathrm{III} \rangle = \langle \mathcal{V} \rangle = 1$ are normalized. In this appendix, we write out the full functional form of correlation functions of the total SFRD by using the using the \citet{Peebles:1980yev} convention, in terms of all stellar populations and feedback effects. For brevity, we drop the functional arguments of each term and rewrite out the the ``unitfull'' (i.e. not mean-normalized) covariance of the total SFRD in Eq.~\eqref{eq:sfrdFullAppendix} as
\begin{align}
    \xi_{\dot{\rho}_*}(r) &= \left\langle\left(\rho_1^\mathrm{II} \mathcal{D}_1^\mathrm{II} + \rho_1^\mathrm{III} \mathcal{V}_1\mathcal{D}_1^\mathrm{III}\right)\left(\rho_2^\mathrm{II} \mathcal{D}_2^\mathrm{II} + \rho_2^\mathrm{III} \mathcal{V}_2\mathcal{D}_2^\mathrm{III} \right)\right\rangle \\
    &- \langle \rho_1^\mathrm{II} \mathcal{D}_1^\mathrm{II} + \rho_1^\mathrm{III} \mathcal{V}_1\mathcal{D}_1^\mathrm{III} \rangle \langle \rho_2^\mathrm{II} \mathcal{D}_2^\mathrm{II} + \rho_2^\mathrm{III} \mathcal{V}_2 \mathcal{D}_2^\mathrm{III} \rangle \nonumber.
\end{align}
Covariances of linear combinations of random variables can be simplified into their component terms, yielding
\begin{align}
    & \xi_{\dot{\rho}_*}(r) = \rho_1^\mathrm{II}\rho_2^\mathrm{II} \left( \langle \mathcal{D}_1^\mathrm{II} \mathcal{D}_2^\mathrm{II} \rangle - \langle \mathcal{D}_1^\mathrm{II} \rangle\langle \mathcal{D}_2^\mathrm{II} \rangle \right)  &\textrm{Term 1} \nonumber \\
    &+ \rho_1^\mathrm{II}\rho_2^\mathrm{III} \left( \langle \mathcal{D}_1^\mathrm{II}\mathcal{V}_2 \mathcal{D}^\mathrm{III}_2 \rangle - \langle \mathcal{D}_1^\mathrm{II}\rangle\langle \mathcal{V}_2 \mathcal{D}^\mathrm{III}_2 \rangle \right)  &\textrm{Term 2}  \nonumber \\
    &+ \rho_1^\mathrm{III}\rho_2^\mathrm{II} \left( \langle \mathcal{V}_1\mathcal{D}^\mathrm{III}_1\mathcal{D}^\mathrm{II}_2 \rangle - \langle \mathcal{V}_1\mathcal{D}^\mathrm{III}_1\rangle\langle \mathcal{D}^\mathrm{II}_2 \rangle \right)  &\textrm{Term 3}  \nonumber \\
    &+ \rho_1^\mathrm{III} \rho_2^\mathrm{III} \left( \langle \mathcal{V}_1\mathcal{D}^\mathrm{III}_1 \mathcal{V}_2 \mathcal{D}^\mathrm{III}_2 \rangle - \langle \mathcal{V}_1\mathcal{D}^\mathrm{III}_1\rangle\langle \mathcal{V}_2 \mathcal{D}^\mathrm{III}_2 \rangle \right) .   &\textrm{Term 4} \nonumber 
\end{align}
The goal is then to collect terms of covariances of two variables, and then express as spatially-dependent correlation functions $\xi(r)$. Term 1 is already in such form:
\begin{equation*}
    \langle \mathcal{D}^\mathrm{II}_1\mathcal{D}^\mathrm{II}_2 \rangle - \langle \mathcal{D}^\mathrm{II}_1\rangle\langle \mathcal{D}^\mathrm{II}_2 \rangle = \langle \mathcal{D}^\mathrm{II}_1\rangle\langle \mathcal{D}^\mathrm{II}_2 \rangle \xi_{\mathcal{D}}^\mathrm{II}.
\end{equation*}
Terms 2 and 3 share the same form. Assuming $\mathcal{V}_2$ is uncorrelated with either $\mathcal{D}^\mathrm{II}_1$ or $\mathcal{D}^\mathrm{III}_2$, and $\mathcal{V}_1$ is uncorrelated with either $\mathcal{D}^\mathrm{III}_1$ or $\mathcal{D}^\mathrm{II}_2$ (a safe assumption, elaborated upon in App.~\ref{sec:apx_DVcross}), these terms reduce to
\begin{align*}
    \langle \mathcal{D}^\mathrm{II}_1\mathcal{V}_2\mathcal{D}^\mathrm{III}_2 \rangle - \langle \mathcal{D}^\mathrm{II}_1\rangle\langle \mathcal{V}_2\mathcal{D}^\mathrm{III}_2 \rangle &= \langle \mathcal{D}^\mathrm{II}_1\rangle\langle \mathcal{V}_2 \rangle \langle \mathcal{D}^\mathrm{III}_2 \rangle \xi_{\mathcal{D}}^{\mathrm{II}\times\mathrm{III}}, \\
    \langle \mathcal{V}_1\mathcal{D}^\mathrm{III}_1\mathcal{D}^\mathrm{II}_2 \rangle - \langle \mathcal{V}_1\mathcal{D}^\mathrm{III}_1\rangle\langle \mathcal{D}^\mathrm{II}_2 \rangle &= \langle \mathcal{V}_1\rangle\langle \mathcal{D}^\mathrm{III}_1 \rangle \langle \mathcal{D}^\mathrm{II}_2 \rangle \xi_{\mathcal{D}}^{\mathrm{III}\times\mathrm{II}}.
\end{align*}
Term 4 can be simplified into products of covariances by assuming that all other permutations of pairs are uncorrelated except for $\mathcal{V}_1,\mathcal{V}_2$ and $\mathcal{D}^\mathrm{III}_1,\mathcal{D}^\mathrm{III}_2$. Such assumptions yield:
\begin{align*}
    &\langle \mathcal{V}_1\mathcal{D}^\mathrm{III}_1\mathcal{V}_2\mathcal{D}^\mathrm{III}_2 \rangle - \langle \mathcal{V}_1\mathcal{D}^\mathrm{III}_1\rangle\langle \mathcal{V}_2\mathcal{D}^\mathrm{III}_2 \rangle \\
    &=\Big[ \left(\langle \mathcal{V}_1\mathcal{V}_2\rangle - \langle \mathcal{V}_1\rangle \langle \mathcal{V}_2\rangle  \right) \left(\langle \mathcal{D}^\mathrm{III}_1\mathcal{D}^\mathrm{III}_2\rangle - \langle \mathcal{D}^\mathrm{III}_1 \rangle \langle \mathcal{D}^\mathrm{III}_2 \rangle  \right) \\
    &\quad+ \langle \mathcal{V}_1 \rangle \langle \mathcal{V}_2 \rangle \left(\langle \mathcal{D}^\mathrm{III}_1\mathcal{D}^\mathrm{III}_2 \rangle -\langle \mathcal{D}^\mathrm{III}_1\rangle \langle \mathcal{D}^\mathrm{III}_2 \rangle \right) \\
    &\quad+ \langle \mathcal{D}^\mathrm{III}_1 \rangle \langle \mathcal{D}^\mathrm{III}_2 \rangle \left(\langle \mathcal{V}_1\mathcal{V}_2\rangle - \langle \mathcal{V}_1 \rangle \langle \mathcal{V}_2\rangle  \right)\Big] \\
    &= \langle \mathcal{V}_1 \rangle \langle \mathcal{V}_2 \rangle \langle \mathcal{D}^\mathrm{III}_1 \rangle \langle \mathcal{D}^\mathrm{III}_2 \rangle \left[ \xi_{\mathcal{V}} \xi_{\mathcal{D}}^\mathrm{III} + \xi_{\mathcal{V}} + \xi_{\mathcal{D}}^\mathrm{III}\right].
\end{align*}
Note that the expectations $\langle \mathcal{D}^\mathrm{II}_1 \rangle = \langle \mathcal{V}_1 \rangle = \langle \mathcal{D}^\mathrm{III}_1 \rangle = \langle \mathcal{D}^\mathrm{II}_2 \rangle = \langle \mathcal{V}_2 \rangle = \langle \mathcal{D}^\mathrm{III}_2 \rangle = 1$ are unity due to normalization. After plugging in for the recasted variables, one recovers the following relation for the correlation function for the full perturbed SFRD given by
\begin{align}
    \xi_{\dot{\rho}_*}(r) &=  \dot{\bar{\rho}}_*^\mathrm{II}(z_1) \dot{\bar{\rho}}_*^\mathrm{II}(z_2) \xi_{\mathcal{D}}^\mathrm{II}(r) \\
    &+ \dot{\bar{\rho}}_*^\mathrm{II}(z_1)\dot{\bar{\rho}}_*^\mathrm{III}(z_2) \xi_{\mathcal{D}}^\mathrm{II \times III}(r) \nonumber \\
    &+ \dot{\bar{\rho}}_*^\mathrm{III}(z_1)\dot{\bar{\rho}}_*^\mathrm{II}(z_2) \xi_{\mathcal{D}}^\mathrm{III \times II}(r) \nonumber \\
    &+ \dot{\bar{\rho}}_*^\mathrm{III}(z_1)\dot{\bar{\rho}}_*^\mathrm{III}(z_2) \left[  \xi_{\mathcal{V}}(r) \xi_{\mathcal{D}}^\mathrm{III}(r) + \xi_{\mathcal{V}}(r) + \xi_{\mathcal{D}}^\mathrm{III}(r)\right], \nonumber
\end{align}
whose individual correlation functions are given by
\begin{align}
    \xi_{\mathcal{D}}^\mathrm{II}(r) &= \exp \left(\gamma^\mathrm{II}_{R_1}\gamma^\mathrm{II}_{R_2} \xi^{R_1, R_2}_\delta(r) \right) - 1,  \\
    \xi_{\mathcal{D}}^\mathrm{II \times III}(r) &= \exp \left(\gamma^\mathrm{II}_{R_1}\gamma^\mathrm{III}_{R_2} \xi^{R_1, R_2}_\delta(r) \right) - 1, \nonumber \\
    \xi_{\mathcal{D}}^\mathrm{III \times II}(r) &= \exp \left(\gamma^\mathrm{III}_{R_1}\gamma^\mathrm{II}_{R_2} \xi^{R_1, R_2}_\delta(r) \right) - 1, \nonumber \\
    \xi_{\mathcal{D}}^\mathrm{III}(r) &= \exp \left(\gamma^\mathrm{III}_{R_1}\gamma^\mathrm{III}_{R_2} \xi^{R_1, R_2}_\delta(r) \right) - 1, \nonumber 
\end{align}
with $\xi_\mathcal{V}(r)$ given by Eq.~\eqref{eq:fullVCorrelation}.

\begin{figure}[t!]
        \centering
        \includegraphics[width=\columnwidth]{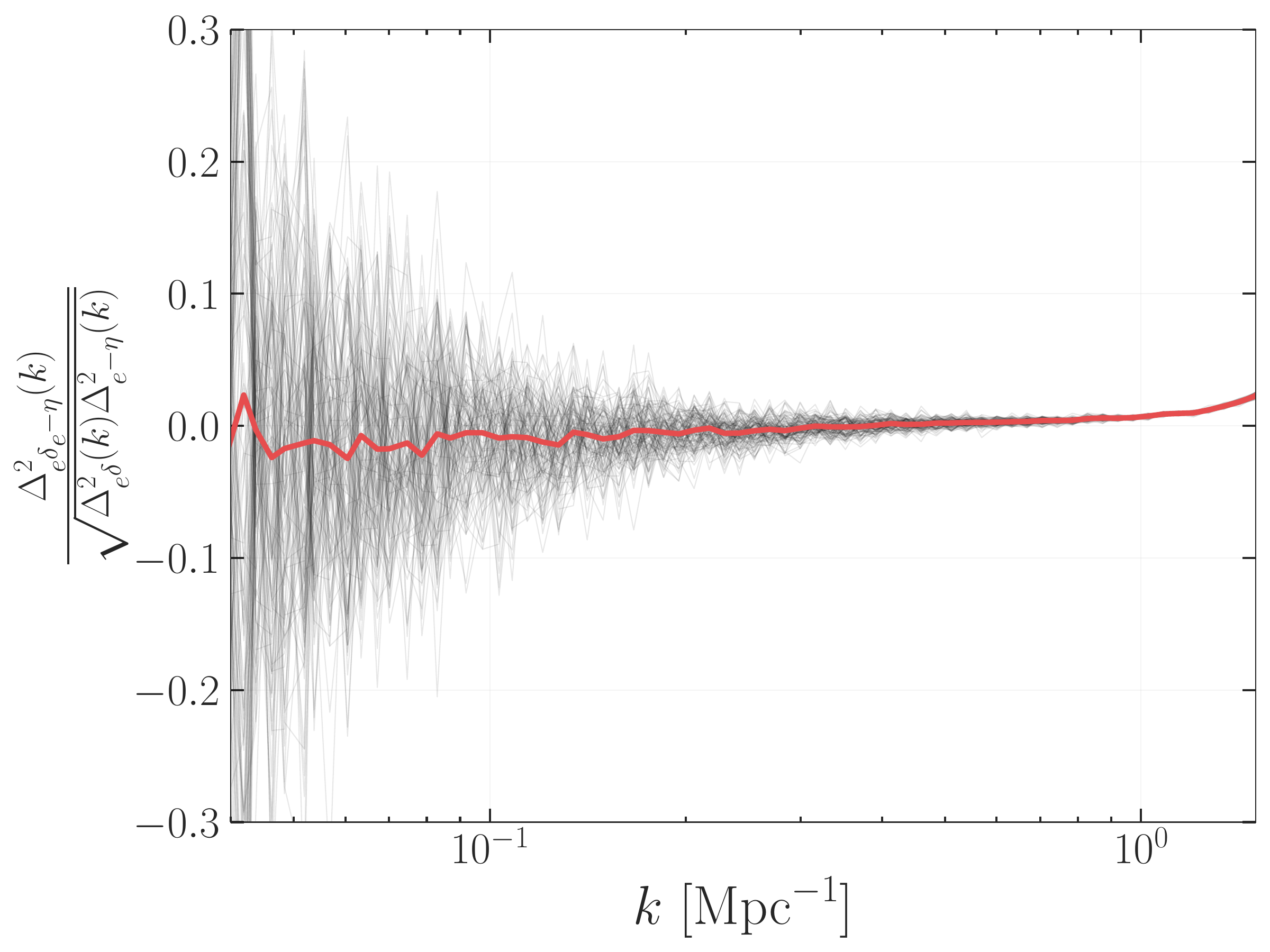}
        \caption{A plot of the cross power spectrum between a realization exponential in density $e^\delta$ and a realization exponential in velocity $e^{-\eta}$. To minimize the relative variances of each realization, we normalize the cross-power with the square root of the product of the auto power spectrum of each realization. In black lines are the power spectra of 100 different realization boxes and seed values of side length $600 \, \mathrm{Mpc}$ and resolution $2 \, \mathrm{Mpc}$. In red, we compute the average power spectrum from all 100 realizations, finding a result close to zero power. Due to the negligible power of the averaged cross power spectrum and ``white noise'' features from each individual realization, we conclude that $e^\delta$ and $e^\eta$ are very approximately independent and uncorrelated, as was the case for their linear counterparts $\delta$ and $\eta$. }
      \label{fig:FIG_APP_crossPowerDV}
    \end{figure}

    \begin{figure*}[hbt!]
        \centering
        \includegraphics[width=\textwidth]{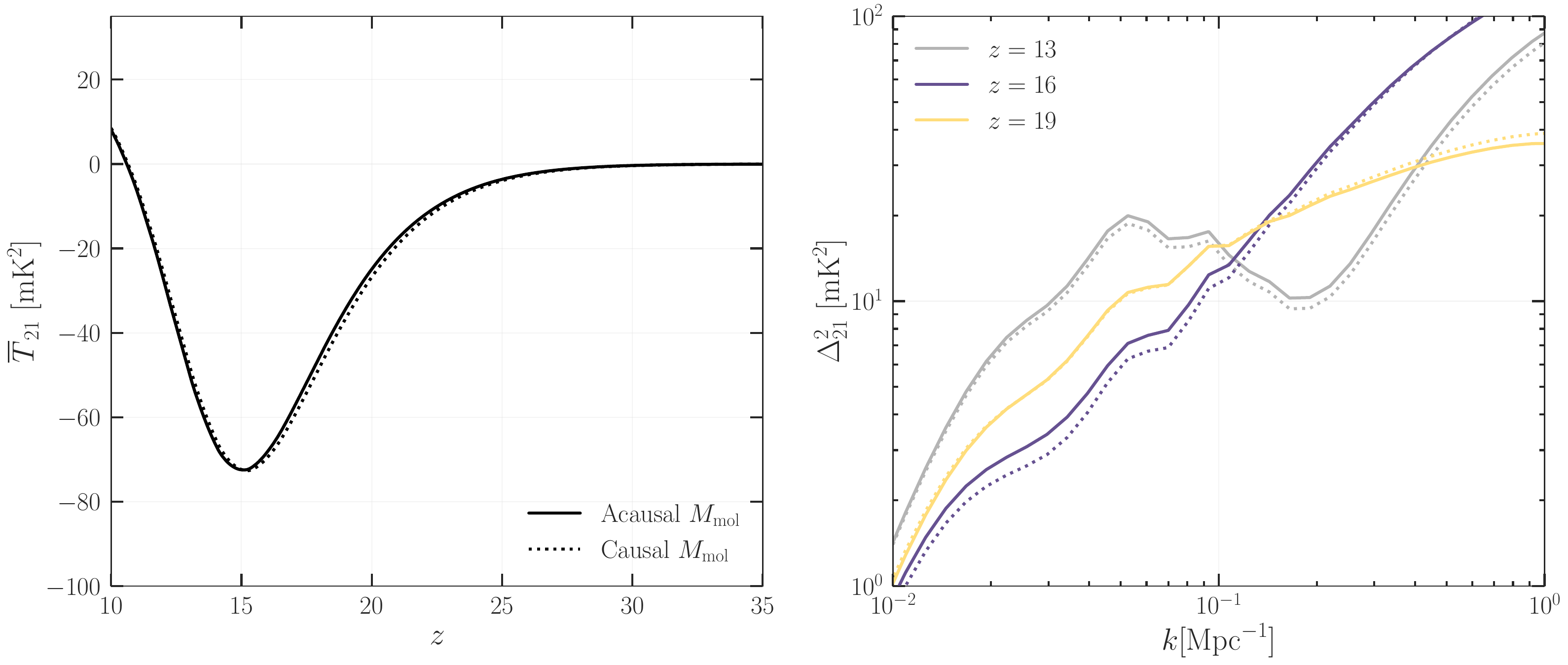}
        \caption{A demonstration of the effects of acausal assumptions in $M_\mathrm{mol}$ as described in App.~\ref{sec:acausalMmol}. (\textbf{Left}): A plot of the $\overline T_{21}$ global signal while the right panel plots 21-cm power spectra against $k$, at a few different redshifts. Without backward-evolving the Pop III molecular cooling mass threshold, the prediction of the 21-cm global signal will be shifted by $\Delta z = 0.15$ towards lower redshifts, as the LW flux will tend to be overestimated. (\textbf{Right}): A plot of the 21-cm power spectrum at redshifts $z = 13, 16$, and $19$. Assuming an acausal mass threshold will also lead to a $15-20\%$ prediction in power spectra across all redshifts. }
      \label{fig:FIG_APP_ACAUSAL}
    \end{figure*}
    
\section{Independent $\delta$-$v_\mathrm{cb}^2$ Cross Correlations} \label{sec:apx_DVcross}

Throughout our prescription for SFRD correlation functions, we have assumed that the cross term between $\exp(\gamma_R^\mathrm{III} \tilde{\delta}_R)$ and $\exp(-\lambda_R \tilde{\eta}_R)$ is negligible. In this appendix, we justify our assumption. In Fig.~\ref{fig:FIG_APP_crossPowerDV}, we plot the cross-power spectrum $\Delta_{e^\delta e^{-\eta}}^2$ of 100 different realizations of a field exponential in density and velocity, each realization a cube of length $600 \, \mathrm{Mpc}$ with $2\, \mathrm{Mpc}$ resolution. To understand the size of this cross spectrum,, we normalize it by the square root of the product of the individual auto power spectra of $e^\delta$ and $e^{-\eta}$. 
We find that the cross spectra appear as white noise, with an average cross spectrum (of all 100 realizations) consistent with zero, and a scatter across power spectra highly dependent on the initial seed value that determines initial conditions.
Thus, we can rest assured that the cross-terms yield negligible power contribution to overall radiative and 21-cm anisotropies.
    
\section{Full Correlation Functions of the IGM} \label{sec:allCorrelations}
In this Appendix, we write out the full functional form for the correlation functions that describe the IGM in Sec.~\ref{sec:V_IGM}.

The correlation functions that describe Lyman-$\alpha$ coupling anisotropies in Eqs.~\eqref{eq:xiAlpha1} and \eqref{eq:xiAlpha2} are
    \begin{align}
        \xi^\mathrm{II}_\alpha(r, z) &= c_{1, \alpha}(z)^2 \sum_{R_1, R_2} c_{2, \alpha}^\mathrm{II}(z, R_1) c_{2, \alpha}^\mathrm{II}(z, R_2) \xi_\mathcal{D}^{\mathrm{II}} (r, z), \nonumber \\
        \xi^\mathrm{II \times III}_\alpha(r, z) &= c_{1, \alpha}(z)^2 \sum_{R_1, R_2} c_{2, \alpha}^\mathrm{II}(z, R_1) c_{2, \alpha}^\mathrm{III}(z, R_2) \xi_{\mathcal{D}}^{\mathrm{II, III}} (r, z), \nonumber \\
        \xi^\mathrm{III}_\alpha(r, z) &= c_{1, \alpha}(z)^2 \sum_{R_1, R_2} c_{2, \alpha}^\mathrm{III}(z, R_1) c_{2, \alpha}^\mathrm{III}(z, R_2) \nonumber \\
        &\times \left[  \xi_{\mathcal{V}}(r) \xi_{\mathcal{D}}^\mathrm{III}(r,z) + \xi_{\mathcal{V}}(r) + \xi_{\mathcal{D}}^\mathrm{III}(r,z)\right]. \nonumber 
    \end{align}
which uses our lognormal $\mathcal{D}$-density and log-$\chi^2_3$ $\mathcal{V}$-velocity building blocks. By symmetry, $\xi_\alpha^{\mathrm{II}\times\mathrm{III}} = \xi_\alpha^{\mathrm{III}\times\mathrm{II}}$. Similarly, the correlation functions that describe the X-ray contribution to the kinetic temperature in Eqs.~\eqref{eq:xi_X1} and \eqref{eq:xi_X2} are given by
    \begin{align} 
        \xi^\mathrm{II}_X(r, z) &= (1+z)^4 \sum_{z_1', z_2' \geq z }c_{1, X}(z_1')c_{1, X}(z_2') \\
        & \times \sum_{R_1, R_2} c_{2, X}^\mathrm{II}(z_1', R_1) c_{2, X}^\mathrm{II}(z_2', R_2) \xi_\mathcal{D}^{\mathrm{II}} (r, z), \nonumber \\
        \xi^\mathrm{II\times III}_X(r, z) &= (1+z)^4 \sum_{z_1', z_2' \geq z }c_{1, X}(z_1')c_{1, X}(z_2') \nonumber \\
        & \times \sum_{R_1, R_2} c_{2, X}^\mathrm{II}(z_1', R_1) c_{2, X}^\mathrm{III}(z_2', R_2) \xi_\mathcal{D}^{\mathrm{II \times III}} (r, z), \nonumber \\
        \xi^\mathrm{III}_X(r, z) &= (1+z)^4 \sum_{z_1', z_2' \geq z }c_{1, X}(z_1')c_{1, X}(z_2') \nonumber \\
        & \times \sum_{R_1, R_2} c_{2, X}^\mathrm{III}(z_1', R_1) c_{2, X}^\mathrm{III}(z_2', R_2) \nonumber \\
        & \times \left[  \xi_{\mathcal{V}}(r) \xi_{\mathcal{D}}^\mathrm{III}(r,z) + \xi_{\mathcal{V}}(r) + \xi_{\mathcal{D}}^\mathrm{III}(r,z)\right]. \nonumber 
    \end{align}
where again by symmetry, $\xi_X^{\mathrm{II}\times\mathrm{III}} = \xi_X^{\mathrm{III}\times\mathrm{II}}$. 
Lastly, their cross-correlation is expressed as
    \begin{align}
        \xi^\mathrm{II}_{\alpha X}(r, z) &= (1+z)^2 c_{1, \alpha}(z)\sum_{z_2' \geq z }c_{1, X}(z_2') \\
        & \times \sum_{R_1, R_2} c_{2, \alpha}^\mathrm{II}(z, R_1) c_{2, X}^\mathrm{II}(z_2', R_2) \xi_\mathcal{D}^{\mathrm{II}} (r, z), \nonumber \\
        \xi^\mathrm{II\times III}_{\alpha X}(r, z) &= (1+z)^2 c_{1, \alpha}(z)\sum_{z_2' \geq z }c_{1, X}(z_2') \nonumber \\
        & \times \sum_{R_1, R_2} c_{2, \alpha}^\mathrm{II}(z, R_1) c_{2, X}^\mathrm{III}(z_2', R_2) \xi_\mathcal{D}^{\mathrm{II \times III}} (r, z), \nonumber \\
        \xi^\mathrm{III\times II}_{\alpha X}(r, z) &= (1+z)^2 c_{1, \alpha}(z)\sum_{z_2' \geq z }c_{1, X}(z_2') \nonumber \\
        & \times \sum_{R_1, R_2} c_{2, \alpha}^\mathrm{III}(z, R_1) c_{2, X}^\mathrm{II}(z_2', R_2) \xi_\mathcal{D}^{\mathrm{III \times II}} (r, z), \nonumber \\
        \xi^\mathrm{III}_{\alpha X}(r, z) &= (1+z)^2 c_{1, \alpha}(z)\sum_{z_2' \geq z }c_{1, X}(z_2') \nonumber \\
        & \times \sum_{R_1, R_2} c_{2, \alpha}^\mathrm{III}(z, R_1) c_{2, X}^\mathrm{III}(z_2', R_2) \nonumber \\
        & \times \left[  \xi_{\mathcal{V}}(r) \xi_{\mathcal{D}}^\mathrm{III}(r,z) + \xi_{\mathcal{V}}(r) + \xi_{\mathcal{D}}^\mathrm{III}(r,z)\right], \nonumber 
    \end{align}
which does not exhibit the same symmetry between the Pop II and III cross terms.

\section{Acausal Evolution of $M_\mathrm{mol}(z)$ in {\tt 21cmFAST}} \label{sec:acausalMmol}

One modeling assumption in {\tt 21cmFAST} is that the molecular-cooling threshold $M_{\rm turn}^{\rm III} (\mathbf x, z)$ is not evolved backwards when computing the Pop III SFRD. That is, when computing radiative fields (say $J_{\alpha},J_X$, and even $J_{\rm LW}$), {\tt 21cmFAST} reads a realization of $M_{\rm turn}^{\rm III}$ at the current redshift $z$ and  integrates the SFRD over the past lightcone smoothing $\log_{10} M_{\rm turn}^{\rm III}$ over a radius $R$, but does not de-evolve $M_{\rm turn}^{\rm III}$ to $z'(R) > z$, where $z'(R)$ is the redshift of emission of a source at a distance $R$ away from redshift $z$.
This is an acausal approximation, as the LW flux from $z$, which sets $M_{\rm turn}^{\rm III}$, is being used at earlier times $z'$ than it was emitted.
In {\tt Zeus21}, instead, we directly compute $M_{\rm turn}^{\rm III}$ at all $z'(R)$, since it is analytic and thus extremely fast.
In this Appendix we compare these two approaches to find what degree of uncertainty it causes.

Through the text we have mimicked the acausal {\tt 21cmFAST} behavior in $M_{\rm turn}^{\rm III}$ when setting the {\tt Flag\_emulate\_21cmfast} on, so as to fairly compare results. 
However, we can easily turn on and off the acausal $M_{\rm turn}^{\rm III}$ behavior within {\tt Zeus21}. 
In Fig.~\ref{fig:FIG_APP_ACAUSAL}, we show the effect of this assumption. 
We see how ignoring the backwards evolution of $M_{\rm turn}^{\rm III}$ (as in {\tt 21cmFAST}) changes the 21-cm global signal and power spectrum at the $\sim 10\%$ level. In particular, for our fiducial parameters, it shifts the timing of cosmic dawn by $\Delta z\approx 0.1-0.2$, and the amplitude of the 21-cm power spectrum by $15-20\%$. This is within the typical $\sim 20\%$ ``theoretical uncertainty'' assumed in semi-numerical models of cosmic dawn.
We suggest a simple fix, by interpolating between different saved $M_{\rm turn}^{\rm III}$ boxes in {\tt 21cmFAST}.

\newpage
\bibliography{refs}

\end{document}